\documentclass[journal]{IEEEtran}
 \usepackage{color} % Colored texts

\usepackage{graphicx}
 \usepackage{cite}
 \providecommand{\shortcite}[1]{\cite{#1}}

  \usepackage{hyperref}       % hyperlinks
\usepackage{url}            % simple URL typesetting
\usepackage{booktabs}       % professional-quality tables
\usepackage{amsfonts}       % blackboard math symbols
 \usepackage{microtype}      % microtypography

\usepackage{amsmath}
\usepackage{amsthm}
\usepackage{algorithm}
 \usepackage{cases}

\usepackage{amsmath,amssymb,amsfonts}
\usepackage{algorithmic}
\usepackage{graphicx}
\usepackage{textcomp}
\usepackage{xcolor}

 \usepackage{hyperref}

\usepackage{url}  %Required
\usepackage{graphicx}  %Required
 \newcommand{\dr}{\,\mbox{d}}

\usepackage{caption}

\newcommand{\fr}[1]{{\mathbb{F}}\left[{#1}\right]}

\usepackage{url}
 \usepackage{epsfig,endnotes}
 \usepackage{floatpag}
 \usepackage{enumitem}
\setlist{leftmargin=13pt}
\usepackage{hyphenat}
\usepackage{relsize}
 \usepackage{tikz}

\newcommand{\con}{\boldsymbol{\mid}}

\usepackage{multirow}
\usepackage{setspace}
\usepackage{mathrsfs}

\DeclareMathOperator*{\erf}{erf}

\DeclareMathOperator*{\erfc}{erfc}

\DeclareMathOperator*{\inverf}{inverf}

\DeclareMathOperator*{\inverfc}{inverfc}

\usepackage{bbm}
\usepackage{pifont}

\newcommand{\qeda}{\hfill\ensuremath{\blacksquare}}

\usepackage{tabularx}
\usepackage{listings}
\usepackage{graphicx}
\usepackage{amssymb,booktabs}
\usepackage{array}
\usepackage{cases}

 \usepackage{supertabular}
\usepackage{psfrag}
\usepackage{bbding}

\usepackage{float}

\usepackage{amsbsy}

\makeatletter
\providecommand{\leftsquigarrow}{%
  \mathrel{\mathpalette\reflect@squig\relax}%
}
\newcommand{\reflect@squig}[2]{%
  \reflectbox{$\m@th#1\rightsquigarrow$}%
}
\makeatother

\usepackage{algorithm}
 \usepackage{algorithmic}

\makeatletter
\newcommand{\newalgname}[1]{%
  \renewcommand{\ALG@name}{#1}%
}

\newcommand{\bp}[1]{{\mathbb{P}}\left[{#1}\right]}

\newcommand{\bfu}[1]{{\mathbb{F}}\left[{#1}\right]}

\newcommand{\fsquare}{\vrule height6pt width7pt depth1pt}   % filled square
\newcommand{\pfe}{\hfill\fsquare \\}             % end of proof

\def\centerhack#1{\hbox to 0pt{\hss\footnotesize #1\hss}}
\def\centerhackn#1{\hbox to 0pt{\hss #1\hss}}
\def\dchack#1{\vbox to 0pt{\vss{\hbox to 0pt{\hss#1\hss}}\vss}}
\newcommand{\pr}[1]{{\mathbb{P}}\left[{#1}\right]} % probability measure

\usepackage{etoolbox}
\AtBeginEnvironment{align}{\setcounter{subeqn}{0}}% Reset subequation number at start of align
\newcounter{subeqn} %
\usetikzlibrary{positioning}

\newcounter{mysub}

\newtheorem{defn}{Definition}

\newtheorem{lem}{Lemma}
\newtheorem{thm}{\textbf{Theorem}}

\newtheorem{rem}{Remark}

\newtheorem{prop}{Proposition}
\newtheorem*{proposition1.1}{Proposition 1.1}
\newtheorem*{proposition1.2}{Proposition 1.2}
\newtheorem*{proposition1.3}{Proposition 1.3}
\newtheorem*{proposition2.1}{Proposition 2.1}
\newtheorem*{proposition2.2}{Proposition 2.2}

\usepackage{subfigure}

\begin{document}

% \title{Computation Offloading for
% Edge Computing
% %Mobile-Edge Cloud Computing
% in %Emerging
%  Wireless Networks with Intelligent Surfaces}

\title{Reviewing and Improving the Gaussian Mechanism for Differential Privacy}
%\title{K-Connectivity in Secure Wireless Sensor Networks under an ON/OFF Channel Model}

%(the details are given in Section VI of the full version \cite{full} due to space limitation)

%
%where ``for any sufficiently large $n$'' means ``for any $n \geq
%N_0$, where $N_0$ is a selected constant''


\author{\mbox{\fontsize{10.7}{10.7}\selectfont Jun Zhao, Teng Wang, Tao Bai, Kwok-Yan Lam, Zhiying Xu, Shuyu Shi, Xuebin Ren, Xinyu Yang, Yang Liu, Han Yu}% <-this % stops a space
\IEEEcompsocitemizethanks{Jun Zhao, Teng Wang, Tao Bai,  Kwok-Yan Lam, and Han Yu are with Nanyang Technological University, Singapore %639798
 (Emails:  \mbox{junzhao@ntu.edu.sg}, \mbox{N1805892E@e.ntu.edu.sg}, \mbox{bait0002@e.ntu.edu.sg}, \mbox{kwokyan.lam@ntu.edu.sg}, \mbox{han.yu@ntu.edu.sg}).\newline \indent
Zhiying Xu and Shuyu Shi are with Nanjing University, China (Emails:  \mbox{zyxu@smail.nju.edu.cn}, \mbox{ssy@nju.edu.cn}).\newline \indent
Xuebin Ren and Xinyu Yang are with Xi'an Jiaotong University, China (Emails:  \mbox{xuebinren@mail.xjtu.edu.cn}, \mbox{yxyphd@mail.xjtu.edu.cn}).\newline \indent Yang Liu is with WeBank Co Ltd, China (Email:  \mbox{yangliu@webank.com}).}}

\markboth{}%
{}



\maketitle

\begin{abstract}
\mbox{Differential privacy provides a rigorous framework} to quantify data privacy, and has received considerable interest recently. A randomized mechanism satisfying \mbox{$(\epsilon, \delta)$-differential privacy} (DP) roughly means that, except with a small probability~$\delta$, altering a record in a dataset cannot change the probability that an output is seen by more than a multiplicative factor $e^{\epsilon} $. A well-known solution to \mbox{$(\epsilon, \delta)$-DP} is the Gaussian mechanism initiated by Dwork~\textit{et~al.}~\cite{dwork2006our} in 2006 with an improvement by Dwork and Roth~\cite{dwork2014algorithmic} in 2014, where a Gaussian noise amount $\sqrt{2\ln \frac{2}{\delta}} \times \frac{\Delta}{\epsilon}$ of~\cite{dwork2006our} or $\sqrt{2\ln \frac{1.25}{\delta}} \times \frac{\Delta}{\epsilon}$ of~\cite{dwork2014algorithmic} is added independently to each dimension of the query result, for a query with $\ell_2$-sensitivity $\Delta$. Although both classical Gaussian mechanisms~\cite{dwork2006our,dwork2014algorithmic} explicitly assume $0 < \epsilon \leq 1$ only, our review finds that many studies in the literature have used the classical Gaussian mechanisms under values of $\epsilon$ and $\delta$ where we show the added noise amounts of~\cite{dwork2006our,dwork2014algorithmic} do \textit{not} achieve $(\epsilon,\delta)$-DP. We obtain such result by analyzing the optimal (i.e., least) Gaussian noise amount $ \sigma_{\texttt{\upshape DP-OPT}}$ for $(\epsilon,\delta)$-DP and   identifying the set of $\epsilon$ and $\delta$ where the noise amounts of classical Gaussian mechanisms are even less than $ \sigma_{\texttt{\upshape DP-OPT}}$. The inapplicability of mechanisms of~\cite{dwork2006our,dwork2014algorithmic} to large $\epsilon$ can also be seen from our result that
$ \sigma_{\texttt{\upshape DP-OPT}}$ for large $\epsilon$ can be written as $ \Theta\big(\frac{1}{\sqrt{\epsilon}}\big)$\vspace{2pt}, but not $\Theta\left(\frac{1}{\epsilon}\right)$.

Since $ \sigma_{\texttt{\upshape DP-OPT}}$ has no \mbox{closed-form} expression and needs to be approximated in an iterative manner, we propose  Gaussian mechanisms by deriving \mbox{closed-form} upper bounds for $\sigma_{\texttt{\upshape DP-OPT}}$. Our mechanisms achieve $(\epsilon,\delta)$-DP for \textit{any} $\epsilon$, while the classical Gaussian mechanisms~\cite{dwork2006our,dwork2014algorithmic} do \emph{not} achieve $(\epsilon,\delta)$-DP for large $\epsilon$ given $\delta$. Moreover, the utilities of our proposed Gaussian mechanisms improve those of the classical Gaussian mechanisms~\cite{dwork2006our,dwork2014algorithmic} and are close to that of the optimal yet more computationally expensive Gaussian mechanism.

Since most mechanisms proposed in the literature for \mbox{$(\epsilon,\delta)$-DP}  are obtained by ensuring a condition called $(\epsilon,\delta)$-probabilistic differential privacy (pDP), we also present an extensive discussion of $(\epsilon,\delta)$-pDP including deriving Gaussian noise amounts to achieve it.

% We also prove that the optimal Gaussian noise amount $ \sigma_{\texttt{\upshape DP-OPT}}$ for $(\epsilon,\delta)$-DP with any $ \epsilon >0 $ and any $0<\delta<1$ is always less than $ \frac{\Delta}{2\sqrt{2} \cdot \inverf(\delta)} $, where $\inverf()$ denotes the inverse of the error function. This is in contrast to the classical Gaussian mechanisms' noise amounts which scale with $\frac{1}{\epsilon}$ and hence tend to $\infty$ as $ \epsilon \to 0 $.

% Since most mechanisms proposed in the literature for \mbox{$(\epsilon,\delta)$-DP} are obtained by ensuring a condition called $(\epsilon,\delta)$-probabilistic differential privacy (pDP), we also investigate the difference/relationship between $(\epsilon,\delta)$-DP and $(\epsilon,\delta)$-pDP, and analyze the optimal Gaussian mechanism for $(\epsilon,\delta)$-pDP. Since this optimal noise amount has no \mbox{closed-form} expression, we also provide closed-form Gaussian mechanisms for $(\epsilon,\delta)$-pDP which are not optimal but achieve utilities close to that of the optimal one.

% Moreover, we present  analyses for the composition of Gaussian mechanisms to achieve $(\epsilon,\delta)$-DP or $(\epsilon,\delta)$-pDP. Finally, we conduct both numerical and experimental studies to show the superiority of our proposed Gaussian mechanisms over the classical Gaussian mechanisms~\cite{dwork2006our,dwork2014algorithmic}.

 To summarize, our paper fixes the literature's long-time misuse of  Gaussian mechanism~\cite{dwork2006our,dwork2014algorithmic} for $(\epsilon, \delta)$-differential privacy and provides a comprehensive study for the Gaussian mechanisms.
\end{abstract}

%computation offloading for mobile-edge server computing in emerging  wireless networks with Reconfigurable intelligent surfaces

\begin{IEEEkeywords}
Differential privacy, Gaussian mechanism, probabilistic differential privacy,
 %composition,
 data analysis. %\vspace{-2pt} deep learning
\end{IEEEkeywords}

\section{Introduction}%\cite{dwork2006differential,dwork2006calibrating}

\textbf{Differential privacy.}
Differential privacy~\cite{dwork2006calibrating} has received considerable interest~\cite{dwork2006our,KairouzIT2017,song2013stochastic,wang2016online,jalko2016differentially,heikkila2017differentially,dwork2016concentrated,bun2016concentrated,meiser2018tight} since it provides a rigorous framework to quantify data privacy. Roughly speaking, a randomized mechanism achieving \mbox{$(\epsilon,\delta)$-differential privacy} (DP) means that, except with a (typically small)  probability $\delta$, altering a record in a dataset \mbox{cannot}  change the probability that an output is seen by more than a multiplicative factor $e^{\epsilon} $. Formally, for $D$ and $D'$ iterating through all pairs of neighboring datasets which differ by  one record, and for $\mathcal{Y}$ iterating through all subsets of the output range of a randomized mechanism $Y$, the mechanism $Y$ achieves \mbox{{$(\epsilon, \delta)$-DP}}   if {$\pr{Y(D) \in \mathcal{Y}} \leq e^{\epsilon} \pr{Y(D')\in \mathcal{Y}} + \delta,$} where $\pr{\cdot}$ denotes the probability, and the probability space is over the coin flips of the randomized mechanism $Y$. If $ \delta = 0$, the notion of $(\epsilon,\delta)$-DP becomes~\mbox{{$\epsilon$-DP}}.

% In several studies~\mbox{\cite{dwork2016concentrated,bun2016concentrated,geng2016optimal,dwork2015generalization,bun2015differentially}}, $(\epsilon, \delta)$-DP and {$\epsilon$-DP} are also referred to as \textit{approximate} DP and \textit{pure} DP, respectively.

\textbf{Classical Gaussian mechanisms~\cite{dwork2006our,dwork2014algorithmic} to achieve \mbox{$(\epsilon,\delta)$-differential privacy.}}
Among various mechanisms to achieve DP, the \textit{Gaussian mechanism} for real-valued queries initiated by~\cite{dwork2006our} has received much attention, where a certain amount of zero-mean Gaussian noise is added independently to each dimension of the query result.  Below, for a Gaussian mechanism with parameter $\sigma$, we mean that $\sigma$ is the standard deviation of the Gaussian noise.

As shown in~\cite{dwork2006our,dwork2014algorithmic}, the noise amount in the Gaussian mechanism scales with the $\ell_2$-sensitivity $\Delta$ of a query, which is defined as the maximal $\ell_2$ distance between the true query results for any two neighboring datasets $D$ and $D'$ that differ in one record; i.e., \mbox{$\Delta  = \max_{\textrm{neighboring $D,D'$}} \|Q(D) - Q(D')\|_{2}$}. We will elaborate the notion of neighboring datasets in Remark~\ref{rem-neighboring-datasets} on Page~\pageref{rem-neighboring-datasets}.
For a query with $\ell_2$-sensitivity\footnote{For $p=1,2,\ldots$, the $\ell_p$-sensitivity of a query $Q$ is defined as the maximal $\ell_p$ distance between the outputs for two neighboring datasets $D$ and $D'$ that differ in one record: \mbox{$\Delta_{Q,p} = \max_{\textrm{neighboring $D,D'$}} \|Q(D) - Q(D')\|_{p}$}.\label{footnotelp}} $\Delta$, the noise amount in the first Gaussian mechanism proposed by Dwork~\textit{et~al.}~\cite{dwork2006our} in 2006 to achieve $(\epsilon,\delta)$-DP, denoted by \texttt{\upshape Dwork-2006}, is given by
\begin{align}
 \textstyle{\sigma_{\texttt{\upshape Dwork-2006}}: =\sqrt{2\ln \frac{2}{\delta}} \times \frac{\Delta}{\epsilon}.}    \label{dwork-2006}
\end{align}
Improving \texttt{\upshape Dwork-2006} via a smaller amount of noise addition, the   Gaussian mechanism by Dwork and Roth~\cite{dwork2014algorithmic} in 2014, denoted by \texttt{\upshape Dwork-2014}, adds Gaussian noise with standard deviation
%  \footnote{The statement of Theorem A.1 in~\cite{dwork2014algorithmic} specifies the noise amount of the Gaussian mechanism as \mbox{$c  {\Delta} /\epsilon $} for $c>\sqrt{2\ln \frac{1.25}{\delta}}$, but the proof actually requires only $c \geq \sqrt{2\ln \frac{1.25}{\delta}}$, so we consider~\cite{dwork2014algorithmic}'s noise amount as $\sqrt{2\ln \frac{1.25}{\delta}} \times \frac{\Delta}{\epsilon}$.}
\begin{align}
  \textstyle{\sigma_{\texttt{\upshape Dwork-2014}}: = \sqrt{2\ln \frac{1.25}{\delta}} \times \frac{\Delta}{\epsilon}.}    \label{dwork-2014}
\end{align}

Both Page 6 in~\cite{dwork2006our} for \texttt{\upshape Dwork-2006} and Theorem A.1 on Page 261 in~\cite{dwork2014algorithmic} for \texttt{\upshape Dwork-2014} consider
%  \footnote{The statement of Theorem A.1 in~\cite{dwork2014algorithmic} notes $ \epsilon < 1$, but the proof actually applies to $ \epsilon \leq 1$.}
 $ \epsilon \leq 1$.
We will formally prove that \texttt{\upshape Dwork-2006} and \texttt{\upshape Dwork-2014} fail to achieve $(\epsilon,\delta)$-DP for large $\epsilon$ given $\delta$. Moreover, we will show in Section~\ref{sec-main-review-Gaussian} that many studies~\cite{imtiaz2018distributed,liu2018privacy,wang2018private,ermis2017differentially,liu2018adaptive,imtiaz2017differentially,jalko2016differentially,heikkila2017differentially,imtiaz2018differentially,pyrgelis2017knock,jain2014near,wang2015privacy} %,gong2016optimal,yang2017approximate
 applying \texttt{\upshape Dwork-2006} and \texttt{\upshape Dwork-2014} neglect the condition $ \epsilon \leq 1$, and use \texttt{\upshape Dwork-2006} or \texttt{\upshape Dwork-2014} under values of  $\epsilon$ and $\delta$ where the added Gaussian noise amount actually does \textbf{not} achieve $(\epsilon,\delta)$-DP. This renders their obtained results  inaccurate.

One may wonder why we consider both mechanisms since clearly it holds that
\begin{align} \label{eq-res0}
\text{$\sigma_{\texttt{\upshape Dwork-2014}}$ in Eq.~(\ref{dwork-2014})} <
 \text{$\sigma_{\texttt{\upshape Dwork-2006}}$ in Eq.~(\ref{dwork-2006})} .
\end{align}
The reason is as follows. Although \texttt{\upshape Dwork-2014} achieves higher utility than that of \texttt{\upshape Dwork-2006} for the set of   $\epsilon$ and $\delta$ under which they both achieve $(\epsilon,\delta)$-DP, \texttt{\upshape Dwork-2006} has wider applicability than \texttt{\upshape Dwork-2014}; i.e., the set of   $\epsilon$ and $\delta$ where \texttt{\upshape Dwork-2014} achieves $(\epsilon,\delta)$-DP is a strict subset of the set of   $\epsilon$ and $\delta$ where \texttt{\upshape Dwork-2006} achieves $(\epsilon,\delta)$-DP. Given the above, we discuss both mechanisms.

\textbf{Our contributions.} We make the following contributions in this paper.
\begin{enumerate}
    \item[\textbf{1)}] \textbf{Failures of classical Gaussian mechanisms for large~$\epsilon$.} We    prove (in Theorem~\ref{thm-Dwork-2014-not-work} on Page~\pageref{thm-Dwork-2014-not-work}) that the classical Gaussian mechanisms  \texttt{\upshape Dwork-2006} of~\cite{dwork2006our} and \texttt{\upshape Dwork-2014} of~\cite{dwork2014algorithmic}  \textbf{fail} to achieve $(\epsilon,\delta)$-DP for large $\epsilon$ given $\delta$. In fact, we prove that for any Gaussian mechanism with noise amount $F(\delta) \times \frac{\Delta}{\epsilon}$ for some function $F(\delta)$,
 there exists a positive function $G(\delta)$ for \mbox{\textbf{any} $0<\delta<1$} such that the above Gaussian mechanism does~\textbf{not} achieve \mbox{$(\epsilon,\delta)$-DP}  for \textbf{any} $\epsilon >  G(\delta)$. The above result applies to \texttt{\upshape Dwork-2006} and \texttt{\upshape Dwork-2014},
where the former specifies $F(\delta)$ as $ \sqrt{2\ln \frac{2}{\delta}} $ and the latter specifies $F(\delta)$ as $ \sqrt{2\ln \frac{1.25}{\delta}} $. \vspace{3pt}
    \item[\textbf{2)}] \textbf{The literature's misuse of classical Gaussian mechanisms for large $\epsilon$.} After a literature review (in Table~\ref{tab:misuse-GM} on Page~\pageref{tab:misuse-GM}), we find that many papers~\cite{imtiaz2018distributed,liu2018privacy,wang2018private,ermis2017differentially,liu2018adaptive,imtiaz2017differentially,jalko2016differentially,heikkila2017differentially,imtiaz2018differentially,pyrgelis2017knock,jain2014near,wang2015privacy} %,gong2016optimal,yang2017approximate
  use the classical Gaussian mechanism \texttt{\upshape Dwork-2006} or \texttt{\upshape Dwork-2014} under values of  $\epsilon$ and $\delta$ where the added noise amount actually does~\textbf{not}  achieve $(\epsilon,\delta)$-DP. This makes their obtained results inaccurate.\vspace{3pt}
    \item[\textbf{3)}] \textbf{An $\epsilon$-independent upper bound and asympotics of the optimal Gaussian noise amount for $(\epsilon,\delta)$-DP.} We prove (in Theorem~\ref{thm-DP-OPT-asymptotics} on Page~\pageref{thm-DP-OPT-asymptotics}) that  the optimal (i.e., least) Gaussian noise amount $ \sigma_{\texttt{\upshape DP-OPT}}$ for $(\epsilon,\delta)$-DP is always less than $ \frac{\Delta}{2\sqrt{2} \cdot \inverf(\delta)} $, which \textbf{does} not depend on  $\epsilon$, where $\inverf()$ denotes the inverse of the error function. This is in contrast to the  classical  Gaussian  mechanisms' noise amounts $\sigma_{\texttt{\upshape Dwork-2006}}$ in Eq.~(\ref{dwork-2006}) and $\sigma_{\texttt{\upshape Dwork-2014}}$ in Eq.~(\ref{dwork-2014}) which scale with $\frac{1}{\epsilon}$ and   tend to $\infty$ as $  \epsilon \to 0 $. In fact, we prove that $ \sigma_{\texttt{\upshape DP-OPT}}$ given
     a fixed
      $\delta$ converges to its upper bound $\frac{\Delta}{2\sqrt{2} \cdot \inverf(\delta)} $ as \mbox{$\epsilon \to 0$}, and is\footnote{A positive sequence $x$ can be written as $\Theta\left(y\right)$ for a positive sequence $y$ if $\liminf \frac{x}{y} $ and $\limsup \frac{x}{y} $ are greater than $0$ and smaller than $\infty$.} $\Theta\left(\frac{1}{\sqrt{\epsilon}}\right)$ as \mbox{$\epsilon \to \infty$}. Also,  we show that $\sigma_{\texttt{\upshape DP-OPT}}$ given
     a fixed
     $\epsilon  $ is $\Theta\left( \sqrt{\ln \frac{1}{\delta}} \hspace{1.5pt}\right)$ as $\delta \to 0$.\vspace{3pt} %, where $\inverfc()$ denotes the inverse of the complementary error function
\item[\textbf{4)}] \textbf{Our Gaussian mechanisms  for $(\epsilon,\delta)$-differential privacy with \mbox{closed-form} expressions.} Although the optimal Gaussian mechanism  for $(\epsilon,\delta)$-DP has been proposed in a very recent work~\cite{balle2018improving}, its   noise amount $ \sigma_{\texttt{\upshape DP-OPT}}$  has no \mbox{closed-form} expression and needs to be approximated in an iterative manner. Hence, we propose new Gaussian mechanisms (\texttt{\upshape Mechanism~1} and \texttt{\upshape Mechanism~2} in Theorems~\ref{thm-Mechanism-1} and~\ref{thm-Mechanism-2} on Page~\pageref{thm-Mechanism-1}) by deriving \mbox{closed-form} upper bounds for $\sigma_{\texttt{\upshape DP-OPT}}$.\vspace{3pt}
%  As two examples, our \texttt{\upshape Mechanism-1-Variant-2} adds Gaussian noise amount $\frac{\big( b_{\#}+\sqrt{{b_{\#}}^2+\epsilon}\hspace{1.5pt} \big) \cdot  \Delta }{\epsilon\sqrt{2}} $ for $b_{\#} : =  \inverfc\left(2 \delta\right)$, while our \texttt{\upshape Mechanism 2} adds Gaussian noise amount $\frac{\left( c+\sqrt{c^2+\epsilon}\hspace{1.5pt} \right) \cdot  \Delta }{\epsilon\sqrt{2}}$ for $c : = \sqrt{\ln \frac{2}{\sqrt{16\delta+1}-1}}$.
 We summarize the advantages of our Gaussian mechanisms as follows.
    \begin{itemize}%[leftmargin=10pt]
        \item[i)] As discussed, our Gaussian mechanisms have \mbox{closed-form}  expressions and are  computationally efficient than \cite{balle2018improving}'s optimal Gaussian noise amount, which has no \mbox{closed-form} expression and needs to be approximated in an iterative manner. In addition, both numerical and experimental studies show that the utilities of our Gaussian mechanisms are close to that of the optimal yet more computationally expensive Gaussian mechanism by~\cite{balle2018improving}.
               \item[ii)] Our Gaussian mechanisms all achieve $(\epsilon,\delta)$-DP  for any $\epsilon$, while the classical Gaussian mechanisms  \texttt{\upshape Dwork-2006} of~\cite{dwork2006our} and \texttt{\upshape Dwork-2014} of~\cite{dwork2014algorithmic} were proposed for only $0<\epsilon \leq 1$ and we show that they do \textbf{not} achieve $(\epsilon,\delta)$-DP for large $\epsilon$ given $\delta$, as noted in Contribution 1) above.
               \item[iii)] We \textit{prove} (in Inequality~(\ref{compare-eq-all}) on Page~\pageref{compare-eq-all}) that
the noise amounts of our Gaussian mechanisms are   less than that of  {\texttt{\upshape Dwork-2014}} (and hence also less than that of \texttt{\upshape Dwork-2006}), for $0<\epsilon \leq 1$ where the proofs of \texttt{\upshape Dwork-2006} of~\cite{dwork2006our} and \texttt{\upshape Dwork-2014} of~\cite{dwork2014algorithmic}  require.
               \item[iv)] For a subset of $  \epsilon > 1$ where {\texttt{\upshape Dwork-2014}} \textit{happens} to work ({\texttt{\upshape Dwork-2014}}'s original proof requires $  \epsilon \leq 1$), experiments (in Figure~\ref{fig:noiseamounts} on Page~\pageref{fig:noiseamounts}) show that our \texttt{\upshape Mechanism~1} often adds noise amount less than that of {\texttt{\upshape Dwork-2014}}.
    \end{itemize}
\item[\textbf{5)}] \textbf{$(\epsilon,\delta)$-Differential privacy versus $(\epsilon,\delta)$-probabilistic differential privacy.} Since most mechanisms proposed in the literature for $(\epsilon,\delta)$-differential privacy (DP) are obtained by ensuring a notion called $(\epsilon,\delta)$-probabilistic differential privacy (pDP), which requires the privacy loss random variable to fall in the interval $[-\epsilon, \epsilon]$ with probability at least $1- \delta$, we also investigate $(\epsilon,\delta)$-pDP, and show its difference/relationship with $(\epsilon,\delta)$-DP (in Section~\ref{sec-pDP-optimal-ours} on Page~\pageref{sec-pDP-optimal-ours}). In particular, the minimal Gaussian noise amount to achieve $(\epsilon,\delta)$-pDP given $\delta$ scales with $\frac{1}{\epsilon}$ as \mbox{$\epsilon \to 0$} (from Theorem~\ref{thm-pDP-OPT-asymptotics} on Page~\pageref{thm-pDP-OPT-asymptotics}), while the minimal Gaussian noise amount to achieve $(\epsilon,\delta)$-DP given $\delta$ converges to its upper bound $\frac{\Delta}{2\sqrt{2} \cdot \inverf(\delta)} $ as \mbox{$\epsilon \to 0$} (from Theorem~\ref{thm-DP-OPT-asymptotics} on Page~\pageref{thm-DP-OPT-asymptotics}). Moreover, while  clearly $(\epsilon,\delta)$-pDP implies $(\epsilon,\delta)$-DP, we also prove that $(\epsilon,\delta)$-DP implies $(\epsilon_*,  \frac{\delta (1+e^{-\epsilon_*})}{1-e^{\epsilon-\epsilon_*}} )$-pDP for any $\epsilon_* > \epsilon$.\vspace{3pt}
\item[\textbf{6)}] \textbf{Gaussian mechanisms for $(\epsilon,\delta)$-probabilistic differential privacy.} For  $(\epsilon,\delta)$-pDP, we also derive the optimal Gaussian mechanism (in Theorem~\ref{thm-pDP-OPT} on Page~\pageref{thm-pDP-OPT}) which adds the least amount of Gaussian noise (denoted by $\sigma_{\texttt{\upshape pDP-OPT}}$). However, since $\sigma_{\texttt{\upshape pDP-OPT}}$   has no \mbox{closed-form} expression and needs to be approximated in an iterative manner, we propose Gaussian mechanisms for   $(\epsilon,\delta)$-pDP (\texttt{\upshape Mechanism~3} and \texttt{\upshape Mechanism~4} in Theorems~\ref{thm-Mechanism-3} and~\ref{thm-Mechanism-4} on Page~\pageref{thm-Mechanism-3}) by deriving more computationally efficient upper bounds (in \mbox{closed-form} expressions) for $\sigma_{\texttt{\upshape pDP-OPT}}$.
% \item[\textbf{7)}] \textbf{Composition of Gaussian mechanisms.} We present  analyses (in Section~\ref{sec-main-composition} on Page~\pageref{sec-main-composition}) for the composition of Gaussian mechanisms to achieve $(\epsilon,\delta)$-DP or $(\epsilon,\delta)$-pDP.
% \item[\textbf{8)}] \textbf{Experiments.} In addition to numerical studies, we conduct   experiments on various data analytics tasks including  deep learning, to show the superiority of  our proposed Gaussian mechanisms.
\end{enumerate}

% Although variants of differential privacy, including mean-concentrated differential privacy (mCDP) \cite{dwork2016concentrated}, zero-concentrated differential privacy (zCDP) \cite{bun2016concentrated}, R\'{e}nyi differential privacy \cite{mironov2017renyi} (RDP), and truncated concentrated differential privacy (tCDP) \cite{bun2018composable} have been recently proposed as alternatives to $(\epsilon, \delta)$-differential privacy, we anticipate that $(\epsilon, \delta)$-differential privacy will still be used in many applications. Moreover, we show in Appendix~\ref{sec-DP-CDP-Related} of this supplementary file that achieving $(\epsilon,\delta)$-DP by first ensuring one of these privacy definitions (mCDP, zCDP, RDP, and tCDP)   give Gaussian mechanisms worse than ours.

\textbf{Organization.} The rest of the paper is organized as follows.
%  Section~\ref{sec-main-related} surveys related work.
%      In Section~\ref{sec-main-review-Gaussian}, we elaborate $(\epsilon,\delta)$-differential privacy (DP) and review the literature's misuse of classical Gaussian mechanisms.
%       In Section~\ref{sec-DP-optimal-our}, we discuss the optimal Gaussian mechanism for $(\epsilon,\delta)$-DP, where the noise amount has no \mbox{closed-form} expression.
%      Section~\ref{sec-DP-ours} presents our Gaussian mechanisms for $(\epsilon,\delta)$-DP with \mbox{closed-form} expressions of noise amounts. We conduct experiments on real datasets in Section~\ref{sec-main-experiments}. We conclude the paper in Section~\ref{sec-main-conclusion}.
\begin{itemize}
\item Section~\ref{sec-main-related} surveys related work.
    \item In Section~\ref{sec-main-review-Gaussian}, we elaborate $(\epsilon,\delta)$-differential privacy (DP) and review the literature's misuse of classical Gaussian mechanisms.
    \item  In Section~\ref{sec-DP-optimal-our}, we discuss the optimal Gaussian mechanism for $(\epsilon,\delta)$-DP, where the noise amount has no \mbox{closed-form} expression.
    \item Section~\ref{sec-DP-ours} presents our Gaussian mechanisms for $(\epsilon,\delta)$-DP with \mbox{closed-form} expressions of noise amounts.
    \item Since most mechanisms proposed in the literature for $(\epsilon,\delta)$-DP are obtained by ensuring a notion called $(\epsilon,\delta)$-probabilistic differential privacy (pDP), Section~\ref{sec-pDP-optimal-ours} is devoted to $(\epsilon,\delta)$-pDP, where we discuss the difference/relationship between $(\epsilon,\delta)$-pDP and $(\epsilon,\delta)$-DP, and derive the optimal Gaussian mechanism for $(\epsilon,\delta)$-pDP, where the noise amount has no \mbox{closed-form} expression. Then we propose  Gaussian mechanisms for $(\epsilon,\delta)$-pDP with \mbox{closed-form} expressions of noise amounts.
    \item In view that concentrated differential privacy~\cite{dwork2016concentrated} and related notions~\cite{bun2016concentrated,mironov2017renyi,bun2018composable} have recently been proposed as variants of differential privacy, we show in Section~\ref{sec-DP-CDP-Related} that achieving $(\epsilon,\delta)$-DP by  ensuring one of these privacy definitions gives Gaussian mechanisms worse than ours.
    % \item Section~\ref{sec-Comparison-Gaussian} provides numerical studies to show the superiority of  our proposed Gaussian mechanisms for $(\epsilon,\delta)$-DP and $(\epsilon,\delta)$-pDP.
    \item Section~\ref{sec-main-experiments} presents experimental results.
        \item We conclude the paper in Section~\ref{sec-main-conclusion}.
\end{itemize}

Due to the space limitation, additional details including the proofs are provided in the appendices of this supplementary file.

\textbf{Notation.} \label{paraNotation} Throughout the paper, $\pr{\cdot}$ denotes the probability, and $\fr{\cdot}$ stands for the probability density function. The error function is denoted by ${\erf}()$, and its complement is ${\erfc}()$; i.e., ${\erf}(x):=\frac{2}{\sqrt{\pi}}\int_{0}^{x}e^{-{t}^2} \hspace{1pt} \emph{d} t$ and ${\erfc}(x):=1-{\erf}(x)$. In addition, $\inverf()$ is the inverse of the error function, and $\inverfc()$ is the inverse of the complementary error function. %\vspace{-5pt}

\section{Related Work }  \label{sec-main-related}

%https://scholar.google.com.sg/scholar?as_q="gaussian+mechanism"&as_epq=differential+privacy

\textbf{Differential privacy.}
The notion of differential privacy (DP)~\cite{dwork2006calibrating} has received much attention~\cite{shokri2015privacy,abadi2016deep,dwork2014analyze,nikolov2013geometry,hsu2016jointly,elahi2014privex} since it provides a rigorous framework to quantify data privacy. The Gaussian mechanism to achieve DP has been investigated in~\cite{dwork2006our,dwork2014algorithmic}, while the Laplace mechanism is introduced in~\cite{dwork2006calibrating} and the exponential mechanism  is proposed in~\cite{mcsherry2007mechanism}. The Gaussian (resp., Laplace) mechanism adds independent Gaussian (resp., Laplace) noise to each dimension of the query result, while the exponential mechanism can address non-numeric queries. Recently, the following mechanisms to achieve DP have been proposed: the truncated Laplacian mechanism~\cite{geng2018truncated}, the staircase mechanism~\cite{geng2014optimal,geng2015staircase}, and the Podium mechanism~\cite{pihur2019podium}. Compared with these mechanisms, the Gaussian mechanism is more friendly for composition analysis since the privacy loss random variable (defined in Section~\ref{sec-pDP-optimal-ours} on Page~\pageref{sec-pDP-optimal-ours}) after composing independent Gaussian mechanisms follows a Gaussian distribution, whereas the privacy loss after composing independent truncated Laplacian mechanisms (staircase mechanisms, or podium mechanisms) has a complicated probability distribution.

\textbf{Use of Gaussian mechanism.} The Gaussian mechanism has been used by Dwork~\mbox{\emph{et~al.}}~\cite{dwork2014analyze} to design algorithms for privacy-preserving principal
component analysis.
%  , where the goal is to  find a subspace that captures the covariance
% of a set of data
% records.
 Nikolov~\mbox{\emph{et~al.}}~\cite{nikolov2013geometry} leverage the Gaussian mechanism for differentially private release of a $k$-way marginal query.
%   , which answers the fraction of a dataset's records with a given value in each of a given set of up to $k$ columns.
 The Gaussian mechanism is also used by Hsu~\mbox{\emph{et~al.}}~\cite{hsu2016jointly} for enabling multiple parties to distributedly solve convex optimization problems in a privacy-preserving and distributed manner. Gilbert and McMillan
~\cite{gilbert2017local} apply the Gaussian mechanism to   differentially private recovery of  heat source location.
 Bun~\mbox{\emph{et~al.}}~\cite{bun2014fingerprinting} employ the Gaussian mechanism to derive a lower bound on the length of a combinatorial object called a fingerprinting code, proposed
by Boneh and Shaw~\cite{boneh1998collusion} for watermarking copyrighted content.
%  They further utilize the obtained bound to analyze the sample complexity of $(\epsilon, \delta)$-differentially private algorithms for answering a large number of queries.
 Abadi~\mbox{\emph{et~al.}}~\cite{abadi2016deep} apply $(\epsilon, \delta)$-DP to stochastic gradient descent of deep learning, where the Gaussian
mechanism is used for adding noise to the gradient. Recently,
Liu~\cite{liu2018generalized} have presented a generalized Gaussian
mechanism based on the $\ell_p$-sensitivity$^{\text{\ref{footnotelp}}}$.

\textbf{Probabilistic differential privacy.}
Most mechanisms proposed in the literature for $(\epsilon,\delta)$-DP are obtained by ensuring a notion called $(\epsilon,\delta)$-probabilistic differential privacy (pDP)~\cite{machanavajjhala2008privacy}, which requires the privacy loss random variable to fall in the interval $[-\epsilon, \epsilon]$ with probability at least $1- \delta$. For the formal definition and results discussed below, see  Section~\ref{sec-pDP-optimal-ours} for details, where we present i) relations between $(\epsilon,\delta)$-DP and $(\epsilon,\delta)$-pDP, ii) an analytical but not \mbox{closed-form} expression for the optimal Gaussian mechanism (denoted by Mechanism \texttt{\upshape pDP-OPT}) to achieve $(\epsilon,\delta)$-pDP, and iii) Gaussian mechanisms for $(\epsilon,\delta)$-pDP, denoted by \texttt{\upshape Mechanism~3} and \texttt{\upshape Mechanism~4}, respectively.

% For completeness, figures in this paper include these pDP mechanisms.

\textbf{Other variants of differential privacy.} Different variants of differential privacy have been proposed in the literature recently, including mean-concentrated differential privacy (mCDP)~\cite{dwork2016concentrated}, zero-concentrated differential privacy (zCDP)~\cite{bun2016concentrated}, R\'{e}nyi differential privacy~\cite{mironov2017renyi} (RDP), and truncated concentrated differential privacy (tCDP)~\cite{bun2018composable}. These notions are more complex than $(\epsilon, \delta)$-DP, so we  believe that $(\epsilon, \delta)$-DP will still be used in many applications. Therefore, any issue concerning the classical Gaussian mechanism for $(\epsilon, \delta)$-DP is worthy of serious discussions in the research community. Moreover, we show in Section~\ref{sec-DP-CDP-Related} on Page~\pageref{sec-DP-CDP-Related}  that achieving $(\epsilon,\delta)$-DP by  ensuring one of these privacy definitions (i.e., mCDP, zCDP, RDP, and tCDP)   gives Gaussian mechanisms worse than ours.

\textbf{Composition.} One of the appealing properties of differential privacy is the composition property~\cite{dwork2014algorithmic}, meaning that the composition of differentially private algorithms satisfies a certain level of differential privacy. In Appendix~\ref{sec-main-composition} of this supplementary file, we provide   analyses for the composition of Gaussian mechanisms. Our result is that for $m$ queries $Q_1, Q_2, \ldots, Q_m$ with
$\ell_2$-sensitivity   $\Delta_1, \Delta_2, \ldots, \Delta_m$, if the query result of $Q_i$ is added with independent Gaussian noise of amount (i.e., standard deviation) $\sigma_i$, then the differential privacy (DP) level for the composition of the $m$ noisy answers is the same as that of a Gaussian mechanism with noise amount $\sigma_{*}:= \left(\sum_{i=1}^m \frac{  {\Delta_i}^2 }{{\sigma_i}^2}\right)^{-1/2}$ for a query with $\ell_2$-sensitivity   $1$. Let ${\sigma_{\epsilon,\delta}^{\textup{DP}}}$ be a Gaussian noise amount which achieves $(\epsilon, \delta)$-DP for a query with $\ell_2$-sensitivity   $1$, where the expression of ${\sigma_{\epsilon,\delta}^{\textup{DP}}}$ can follow from classical \texttt{\upshape Dwork-2006} and  \texttt{\upshape Dwork-2014} of~\cite{dwork2006our,dwork2014algorithmic} (when $ \epsilon \leq 1$), the optimal one (i.e., \texttt{\upshape DP-OPT}), or  our proposed mechanisms (i.e., \texttt{\upshape Mechanism~1} and \texttt{\upshape Mechanism~2}). Then the above composition satisfies $(\epsilon,\delta)$-DP for $\epsilon$ and $\delta$ satisfying $\sigma_{*}\geq {\sigma_{\epsilon,\delta}^{\textup{DP}}}$ with $\sigma_{*}$ defined above.

\section{$(\epsilon,\delta)$-Differential Privacy and Usage of the Gaussian Mechanism}\label{sec-main-review-Gaussian}

% \subsection{$(\epsilon,\delta)$-Differential privacy}

% $(\epsilon,\delta)$-Differential privacy?

The formal definition of
$(\epsilon,\delta)$-differential privacy~\cite{dwork2006our} is as follows.

\begin{defn}[\textbf{$(\epsilon,\delta)$-Differential privacy}~\cite{dwork2006our}] \label{defn-DP}
A randomized algorithm $Y$ satisfies $(\epsilon,\delta)$-differential privacy, if for any two neighboring datasets $D$ and $D'$ that differ only in
one record, and for any possible subset of outputs $\mathcal{Y}$ of $Y$, we have
\begin{align} \label{eqn:defn-DP}
\bp{ Y(D) \in \mathcal{Y} } \le e^{\epsilon} \cdot \bp{Y(D') \in  \mathcal{Y}}+\delta.
\end{align}
where $\bp{ \cdot }$ denotes the probability of an event. If $\delta=0$, $Y$ is said to satisfy $\epsilon$-differential privacy.
\end{defn}

\begin{rem}[\textbf{Notion of neighboring datasets}] \label{rem-neighboring-datasets}

Two datasets $D$ and $D'$ are called neighboring if they differ only in one record. There are still variants about this. In the first case, the size of $D$ and $D'$ differ by one so that $D'$ is obtained by adding one record to $D$ or deleting one record from $D$. In the second case, $D$ and $D'$ have the same size (say~$n$), and have different records at only one of the $n$ positions. Finally, the notion of neighboring datasets can also be defined to include both cases above. Our results in this paper do not rely on how neighboring datasets are specifically defined. In a  differential privacy application, after the notion of neighboring datasets is defined, what we need is just the $\ell_2$-sensitivity $\Delta$ of a query $Q$ with respect to neighboring datasets: \mbox{$\Delta  = \max_{\textrm{neighboring datasets $D,D'$}} \|Q(D) - Q(D')\|_{2}$.}

 \end{rem}

%and $Y$ satisfies $\epsilon$-differential privacy when $\delta=0$.

% In what follows, we say that two datasets are {\it neighboring} if one dataset can be obtained by adding a record to the other dataset. Hence, the numbers of records in our notion of neighboring datasets differ by $1$.

% \subsection{Failures of the classical Gaussian mechanisms}

Theorem~\ref{thm-Dwork-2014-not-work} below shows failures of the classical Gaussian mechanisms~\cite{dwork2014algorithmic,dwork2006our} for large $\epsilon$.

% \textbf{Failures of the classical Gaussian mechanisms.} Theorem~\ref{thm-Dwork-2014-not-work} shows that any Gaussian mechanism with the standard deviation of noise in the form of $F(\delta) \times \frac{\Delta}{\epsilon}$ for a function $F(\delta)$ will \textbf{not}~satisfy $(\epsilon,\delta)$-differential privacy for sufficiently large $\epsilon $ given any $\delta$.

% \begin{thm}[\textbf{Failure of the classical Gaussian mechanisms of Dwork and Roth~\cite{dwork2014algorithmic} in 2014 and of Dwork~\textit{et~al.}~\cite{dwork2006our} to achieve  $(\epsilon,\delta)$-differential
% privacy for slightly large $\epsilon$}] \label{thm-Dwork-2014-not-work}
% For \textbf{any} $0<\delta<1$, there exists a positive function $F(\delta)$ such that the Gaussian mechanism of Dwork and Roth~\cite{dwork2014algorithmic}, which adds Gaussian noise with standard deviation $\sqrt{2\ln \frac{1.25}{\delta}} \times \frac{1}{\epsilon}$ to each dimension of the query, \textbf{does~\emph{not}} achieve $(\epsilon,\delta)$-differential
% privacy for \textbf{any} $\epsilon  \geq  F(\delta)$.

% % \\
% % \textbullet~$\epsilon \geq 6.78$ and $\delta = 10^{-2}$;~~~
% % \textbullet~$\epsilon  \geq  7.47$ and $\delta = 10^{-3}$;\\
% % \textbullet~$\epsilon  \geq  8.00$ and $\delta = 10^{-4}$;~~~
% % \textbullet~$\epsilon  \geq  8.43$ and $\delta = 10^{-5}$;\\
% % \textbullet~$\epsilon  \geq  8.79$ and $\delta = 10^{-6}$;~~~
% % \textbullet~$\epsilon  \geq  9.11$ and $\delta = 10^{-7}$.
% \end{thm}

\begin{thm}[\textbf{Failures of the classical Gaussian mechanisms of Dwork and Roth~\shortcite{dwork2014algorithmic} and of Dwork~\textit{et~al.}~\shortcite{dwork2006our} to achieve  $(\epsilon,\delta)$-differential privacy for large $\epsilon$}] \label{thm-Dwork-2014-not-work}

% Let $c$ be a positive constant, where $c$ the Gaussian mechanism of Dwork and Roth~\cite{dwork2014algorithmic}.

For a positive function $F(\delta)$,
consider a Gaussian mechanism which adds Gaussian noise with standard deviation $F(\delta) \times \frac{\Delta}{\epsilon}$ to each dimension of a query with $\ell_2$-sensitivity $\Delta$.
With an \textbf{arbitrarily} fixed $0<\delta<1$, as $\epsilon$ increases, the above Gaussian mechanism \textbf{does~\emph{not}} achieve $(\epsilon,\delta)$-differential privacy for large enough $\epsilon$ (specifically, for \textbf{any} $\epsilon >  G(\delta)$ with $G(\delta)$ being some positive function). This result applies to the classical Gaussian mechanism \texttt{\upshape Dwork-2014} of Dwork and Roth~\shortcite{dwork2014algorithmic} and mechanism \texttt{\upshape Dwork-2006} of Dwork~\textit{et~al.}~\shortcite{dwork2006our},
where the former specifies $F(\delta)$ as $ \sqrt{2\ln \frac{1.25}{\delta}} $ and the latter specifies $F(\delta)$ as $ \sqrt{2\ln \frac{2}{\delta}} $.
% \\
% \textbullet~$\epsilon \geq 6.78$ and $\delta = 10^{-2}$;~~~
% \textbullet~$\epsilon  \geq  7.47$ and $\delta = 10^{-3}$;\\
% \textbullet~$\epsilon  \geq  8.00$ and $\delta = 10^{-4}$;~~~
% \textbullet~$\epsilon  \geq  8.43$ and $\delta = 10^{-5}$;\\
% \textbullet~$\epsilon  \geq  8.79$ and $\delta = 10^{-6}$;~~~
% \textbullet~$\epsilon  \geq  9.11$ and $\delta = 10^{-7}$.
\end{thm}

We formally prove Theorem~\ref{thm-Dwork-2014-not-work} in Appendix~\ref{secprf-thm-Dwork-2014-not-work}.

\begin{figure*}
\centering
\footnotesize
\begin{tabular}{cc}
\hspace{0mm}\includegraphics[width=0.3\textwidth]{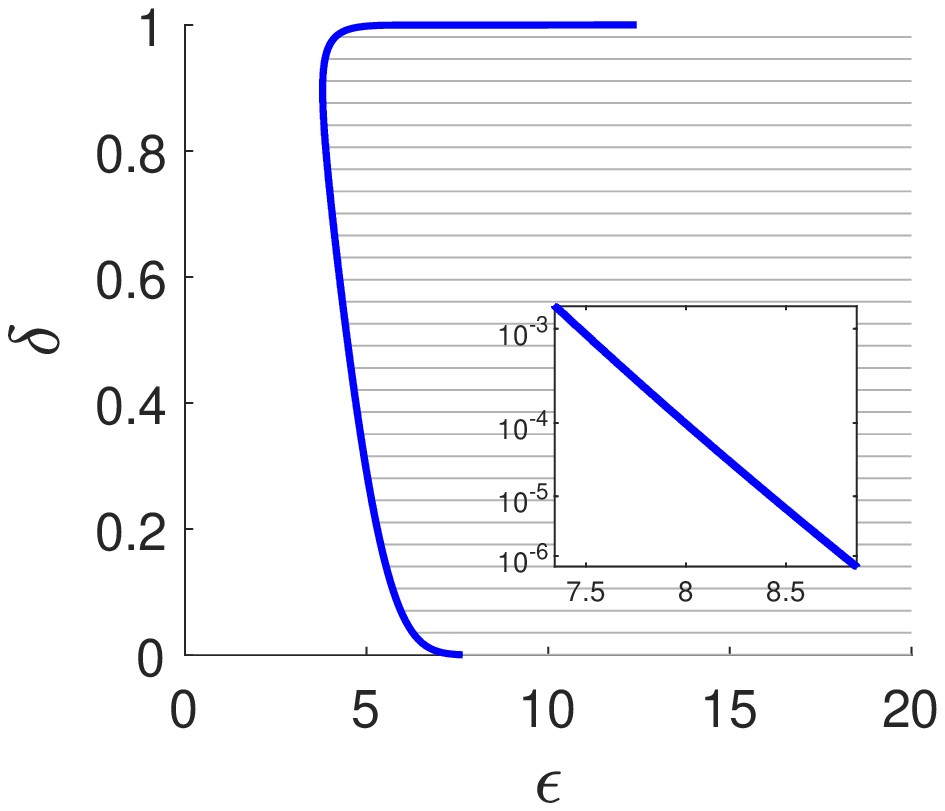} &
\hspace{-2mm}\includegraphics[width=0.3\textwidth]{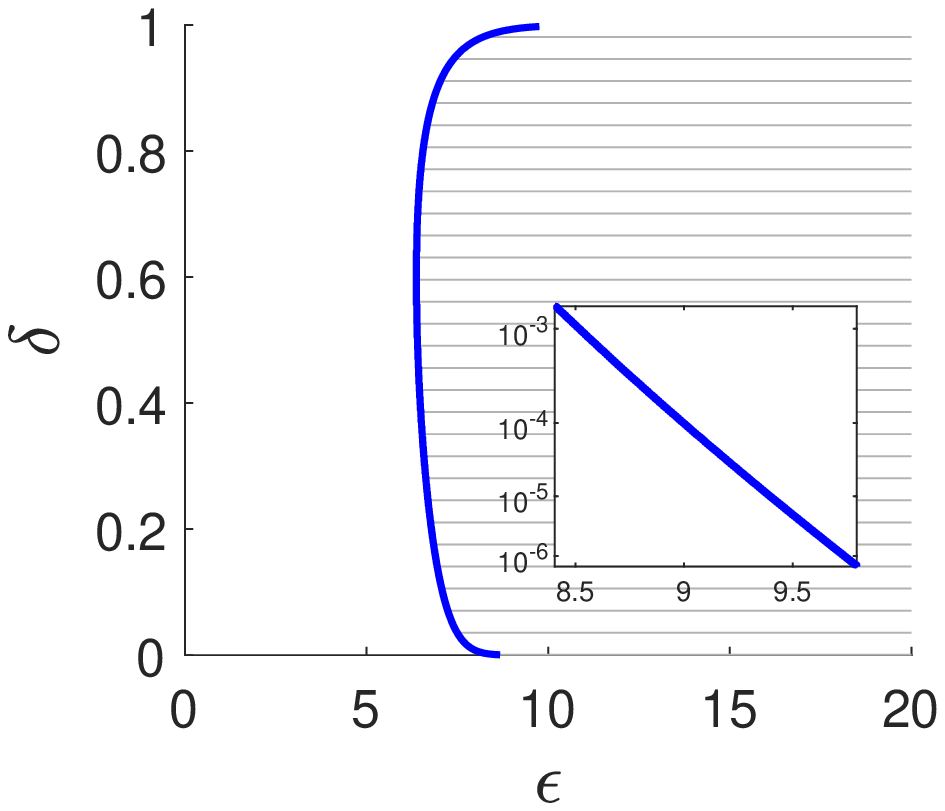}\\[-6pt]
(i) Mechanism \texttt{\upshape Dwork-2014} of Dwork and Roth~\shortcite{dwork2014algorithmic}  & (ii) Mechanism \texttt{\upshape Dwork-2006} of  Dwork~\textit{et~al.}~\shortcite{dwork2006our}
\end{tabular}
 \vspace{-6pt} \caption{The shaded area in each subfigure represents the set of $(\epsilon,\delta)$ where Mechanism \texttt{\upshape Dwork-2014} (resp.,~\texttt{\upshape Dwork-2006}) does \textbf{not} achieve \mbox{$(\epsilon,\delta)$-differential privacy}. \vspace{0mm}  }
\label{fig:region-dp}
\end{figure*}

\begin{rem}\label{rem-Dwork-2014-not-work}
For the Gaussian mechanism \texttt{\upshape Dwork-2014} of~\cite{dwork2014algorithmic}, the blue line in Figure~\ref{fig:region-dp}(i) on Page~\pageref{fig:region-dp} illustrates all points $(\delta, G_*(\delta))$ such that \texttt{\upshape Dwork-2014} {does~\emph{not}} achieve $(\epsilon,\delta)$-differential privacy for $\epsilon >  G_*(\delta)$; e.g., $G_*(10^{-3}) = 7.47$, $G_*(10^{-4}) = 8.00$, $G_*(10^{-5}) = 8.43$, and $G_*(10^{-6}) = 8.79$. For the Gaussian mechanism \texttt{\upshape Dwork-2006} of~\cite{dwork2006our}, the blue line in Figure~\ref{fig:region-dp}(ii) on Page~\pageref{fig:region-dp} illustrates all points $(\delta, G_{\#}(\delta))$ such that \texttt{\upshape Dwork-2006} {does~\emph{not}} achieve $(\epsilon,\delta)$-differential privacy for $\epsilon >  G_{\#}(\delta)$; e.g.,
% \textbullet~$G(10^{-2}) = 6.78$;~~~~~~~~~\textbullet~$G(10^{-3}) = 7.47$;\\ \textbullet~$G(10^{-4}) = 8.00$;~~~~~~~~~\textbullet~$G(10^{-5}) = 8.43$;\\ \textbullet~$G(10^{-6}) = 8.79$;~~~~~~~~~\textbullet~$G(10^{-7}) =9.11$.
$G_{\#}(10^{-3}) = 8.51$, $G_{\#}(10^{-4}) = 8.99$, $G_{\#}(10^{-5}) = 9.39$, and $G_{\#}(10^{-6}) = 9.73$.
\end{rem}

% \subsection{The literature's misuse of the classical Gaussian mechanisms} \label{subsec-Literature-Misuse}

\textbf{The literature's misuse of the classical Gaussian mechanisms.}
After a literature review, we find that many papers~\cite{imtiaz2018distributed,liu2018privacy,wang2018private,ermis2017differentially,liu2018adaptive,imtiaz2017differentially,jalko2016differentially,heikkila2017differentially,imtiaz2018differentially,pyrgelis2017knock,jain2014near,wang2015privacy} use the classical Gaussian mechanism \texttt{\upshape Dwork-2006} (resp., \texttt{\upshape Dwork-2014}) under values of  $\epsilon$ and $\delta$ where \texttt{\upshape Dwork-2006}  (resp., \texttt{\upshape Dwork-2014}) actually does not achieve $(\epsilon,\delta)$-differential privacy.
%  This renders their obtained results inaccurate.
Table~\ref{tab:misuse-GM} on Page~\pageref{tab:misuse-GM} summarizes   selected papers which misuse the classical Gaussian mechanism \texttt{\upshape Dwork-2006} or \texttt{\upshape Dwork-2014}.

% We note that the recent work~\shortcite{balle2018improving} of Balle and Wang also pointed out that ``the rate $\sigma = \Theta(1/\epsilon)$ provided by the classical
% Gaussian mechanism cannot be extended beyond $\epsilon \in (0,1)$.''

\textbf{Usage of $\epsilon>1$.} Although $\epsilon\leq1$ is preferred in practical applications, there are still  cases where $\epsilon>1$ is used, so it is necessary to have Gaussian mechanisms which apply to not only $\epsilon\leq1$ but also $\epsilon>1$. We discuss usage of $\epsilon>1$ as follows. First, the references~\cite{imtiaz2018distributed,liu2018privacy,wang2018private,ermis2017differentially,liu2018adaptive,imtiaz2017differentially,jalko2016differentially,heikkila2017differentially,imtiaz2018differentially,pyrgelis2017knock,jain2014near,wang2015privacy} in Table~\ref{tab:misuse-GM} have used $\epsilon>1$. Second,
the Differential Privacy Synthetic Data Challenge organized by the National Institute of Standards and Technology
 (NIST)~\cite{NIST2} included experiments of $\epsilon$ as $10$.
Third, for a variant of differential privacy called local differential privacy~\cite{duchi2013local} which is implemented in several industrial applications, Apple~\cite{tang2017privacy,Apple2} and Google~\cite{erlingsson2014rappor} have adopted $\epsilon>1$.

% \begin{thm}[\textbf{Failure of the classical Gaussian mechanism of Dwork~\textit{et~al.}~\cite{dwork2006our} in 2006 to achieve  $(\epsilon,\delta)$-differential
% privacy for slightly large $\epsilon$}] \label{thm-Dwork-2006-not-work}
% For \textbf{any} $0<\delta<1$, there exists a positive function $G(\delta)$ such that the Gaussian mechanism of Dwork and Roth~\cite{dwork2006our}, which adds Gaussian noise with standard deviation $\sqrt{2\ln \frac{2}{\delta}} \times \frac{\Delta}{\epsilon}$ to each dimension of a query with $\ell_2$-sensitivity $\Delta$, \textbf{does~\emph{not}} achieve $(\epsilon,\delta)$-differential
% privacy for \textbf{any} $\epsilon  \geq  G(\delta)$.
% \end{thm}

% \begin{rem}\label{rem-Dwork-2006-not-work}
% %We formally prove Theorem~\ref{thm-Dwork-2014-not-work} in Section~\ref{proof-dwork-not-work}. Figure~? illustrates ?.

% \end{rem}

%Mechanism \texttt{\upshape Dwork-2014}:
%$\sigma_{\texttt{\upshape Dwork-2014}}: = \sqrt{2\ln \frac{1.25}{\delta}} \times \frac{\Delta}{\epsilon}$

%Mechanism \texttt{\upshape Dwork-2006}:
%$\sigma_{\texttt{\upshape Dwork-2006}}: =\sqrt{2\ln \frac{2}{\delta}} \times \frac{\Delta}{\epsilon}$

\begin{table*}[t]
\begin{center}
%~\\[-20pt]
\caption{Misuse of the Classical Gaussian Mechanisms in the Literature from 2014 to 2018.\vspace{-6pt}}
\centering
\setlength{\tabcolsep}{1.1mm}
\resizebox*{1\textwidth}{!}{
 \begin{tabular}{c|c|c|c|c|c}
 \hline\hline
 \textbf{Selected papers}& \textbf{Mechanism}&\textbf{$\epsilon$} & \textbf{$\delta$} & \textbf{The resulting noise amounts}& \textbf{The least noise amounts} \\ [0.5ex]
 \hline
 Imtiaz and Sarwate~\shortcite{imtiaz2018distributed}  & \texttt{\upshape Dwork-2014}&10 & 0.01 & 0.3108 & 0.3501 \\
 \hline
 Liu~\textit{et~al.}~\shortcite{liu2018privacy}  & \texttt{\upshape Dwork-2014}&6, 10 & 0.1 & 0.3746, 0.2248 & 0.3813, 0.2818 \\
 \hline
 Wang~\textit{et~al.}~\shortcite{wang2018private}  & \texttt{\upshape Dwork-2014}&8.87, 9.59 & $10^{-5}$ & 0.5462, 0.5052 & 0.5172, 0.5513 \\
 \hline
%  \multirow{5}{*}{Liu~\textit{et~al.}~\shortcite{liu2018privacy} } & \multirow{5}{*}{\texttt{\upshape Dwork-2014}}&6 & \multirow{5}{3.5ex}{0.1} & 0.3746 & 0.3813 \\
%  & & 7 & & 0.3211 & 0.3479 \\
%  & & 8 & & 0.2809 & 0.3215 \\
%  & & 9 & & 0.2497 & 0.2999 \\
%  & & 10 &  & 0.2248 & 0.2818\\
%  \hline
%  & & 8.87 &  & 0.5462 & 0.5513 \\[-1ex]
%  \raisebox{1.5ex}{Wang~\textit{et~al.}~\shortcite{wang2018private} } &\raisebox{1.5ex}{\texttt{\upshape Dwork-2014}} & 9.59 & \raisebox{1.5ex}{$10^{-5}$} & 0.5052 & 0.5172\\
%  \hline
Ermis and Cemgil~\shortcite{ermis2017differentially}  &\texttt{\upshape Dwork-2014}& {10} & $10^{-2}$, $10^{-5}$ & 0.3108, 0.4845 & 0.3501, 0.4999 \\
%  \multirow{4}{*}{Ermis and Cemgil~\shortcite{ermis2017differentially} } &\multirow{4}{*}{\texttt{\upshape Dwork-2014}}&  \multirow{4}{3.5ex}{10} & $10^{-2}$ & 0.3108 & 0.3501 \\
%  & & & $10^{-3}$ & 0.3776 & 0.4061 \\
%  & & & $10^{-4}$& 0.4344 & 0.4553 \\
%  & & & $10^{-5}$ & 0.4845 & 0.4999 \\
 \hline
 Liu~\textit{et~al.}~\shortcite{liu2018adaptive}  & \texttt{\upshape Dwork-2014}& 8 & $10^{-1}$ & 0.2809 &0.3215\\ %[1ex]
%  \hline
%  Yang and Murmann~\shortcite{yang2017approximate}  & \texttt{\upshape Dwork-2014}& 10 & $10^{-5}$ & 0.4845 & 0.4999\\
 \hline
 Imtiaz and Sarwate~\shortcite{imtiaz2017differentially}  & \texttt{\upshape Dwork-2014}& 10 & $10^{-2}$ & 0.3108 & 0.3501\\
 \hline
 J{\"a}lk{\"o}~\textit{et~al.}~\shortcite{jalko2016differentially}  &\texttt{\upshape Dwork-2014}& 10 & $10^{-3}$ & 0.3776 & 0.4061\\
 \hline
%  & &  & & 0.4344 & 0.4553 \\[-1ex]
 {Heikkil{\"a}~\textit{et~al.}~\shortcite{heikkila2017differentially} } & {\texttt{\upshape Dwork-2014}} &10, 31.62 & $10^{-4}$ & 0.4344, 0.1374 & 0.1976, 0.4553\\
 \hline
 Imtiaz and Sarwate~\shortcite{imtiaz2018differentially}  &\texttt{\upshape Dwork-2006}& 10 & $10^{-2}$ & 0.3325 & 0.3501\\
 \hline
 Pyrgelis~\textit{et~al.}~\shortcite{pyrgelis2017knock}  &\texttt{\upshape Dwork-2006}& 10 & $10^{-1}$ & 0.2448 & 0.2818\\
%  \hline
%  Gong~\textit{et~al.}~\shortcite{gong2016optimal}  &\texttt{\upshape Dwork-2006}& 10 & $10^{-1}$ & 0.2448 & 0.2818\\ % delta not appointed
 \hline
 Wang~\textit{et~al.}~\shortcite{wang2015privacy} &\texttt{\upshape Dwork-2006}& 10 & $10^{-1}$ & 0.2448 & 0.2818\\
 \hline
 Jain and Thakurta~\shortcite{jain2014near}&\texttt{\upshape Dwork-2006}& 10 & $10^{-3}$ & 0.3898 & 0.4061 \\
 \hline
\end{tabular}
}
\label{tab:misuse-GM}
\end{center} \vspace{-10pt}
\end{table*}

\begin{table*}[]
\caption{Different mechanisms to achieve~$(\epsilon,\delta)$-differential privacy (DP). \vspace{-6pt}}
\label{table:DP-mechanisms}
\centering
\setlength\tabcolsep{3pt}\begin{tabular}{|l|l|l|}
\hline
DP Mechanisms & Comparison                                    & Common properties                                                  \\ \hline
\begin{tabular}[c]{@{}l@{}}\texttt{\upshape DP-OPT}\\of Theorem~\ref{thm-DP-OPT}\end{tabular} & \begin{tabular}[c]{@{}l@{}}\textbullet~the optimal Gaussian mechanism   to achieve~$(\epsilon,\delta)$-DP,\\ \textbullet~no closed-form expression,\\ \textbullet~computed using the bisection method  \\ ~~with the number of iterations being \\ ~~logarithmic in the given error (i.e., tolerance).\end{tabular} & \multirow{3}{*}{\begin{tabular}[c]{@{}l@{}}Noise amounts $\sigma_{\texttt{\upshape DP-OPT}}$, $\sigma_{\texttt{\upshape Mechanism-1}}$, \\ and $\sigma_{\texttt{\upshape Mechanism-2}}$ are all smaller than \\$\sigma_{\texttt{\upshape Dwork-2014}}$ and $\sigma_{\texttt{\upshape Dwork-2006}}$, \\for $0<\epsilon \leq 1$ which the proofs of\\ {\texttt{\upshape Dwork-2014}} and {\texttt{\upshape Dwork-2006}} require\\ (The proof is in Appendix~\ref{sec-Mechanism-2-Dwork-2014}). \end{tabular}} \\ \cline{1-2}
\begin{tabular}[c]{@{}l@{}}Our \texttt{\upshape Mechanism~1}\\of Theorem~\ref{thm-Mechanism-1}\end{tabular} & \begin{tabular}[c]{@{}l@{}}\textbullet~closed-form expression involving \\ ~~the complementary error function \\ ~~$\erfc()$ and its inverse $\inverfc()$,\\ \textbullet~computational complexity: dependent on\\ ~~$\erfc()$ \& $\inverfc()$ implementations and often very efficient, \\ \textbullet~$\sigma_{\texttt{\upshape Mechanism-1}}$ is slightly greater than $\sigma_{\texttt{\upshape DP-OPT}}$. \end{tabular} &                                                                    \\ \cline{1-2}
\begin{tabular}[c]{@{}l@{}}Our \texttt{\upshape Mechanism~2}\\of Theorem~\ref{thm-Mechanism-2}\end{tabular} & \begin{tabular}[c]{@{}l@{}}\textbullet~closed-form expression involving \\ ~~only elementary functions,\\ \textbullet~computed in constant amount of time, \\ \textbullet~$\sigma_{\texttt{\upshape Mechanism-2}}$ is slightly greater than $\sigma_{\texttt{\upshape Mechanism-1}}$. \end{tabular} &                                                                    \\ \hline
\end{tabular}
 \vspace{-6pt}\end{table*}

\section{The Optimal  Gaussian Mechanism for $(\epsilon,\delta)$-Differential Privacy}\label{sec-DP-optimal-our}

% The Optimal (Yet more computationally expensive) Gaussian Mechanism for $(\epsilon,\delta)$-Differential Privacy

%\newpage
%In Mechanism \texttt{\upshape DP-OPT},

% \subsection{Analytical but not \mbox{closed-form} expressions for the optimal Gaussian mechanism of $(\epsilon,\delta)$-differential privacy}

A recent work~\cite{balle2018improving} of Balle and Wang in ICML 2018 analyzed the optimal Gaussian mechanism for $(\epsilon,\delta)$-differential privacy, where ``optimal'' means that the noise amount is the least among Gaussian mechanisms. This optimal Gaussian mechanism is also analyzed by Sommer~\emph{et~al.}~\cite{sommer2019privacy}, where the
shape of the privacy loss is also discussed. Based on~\cite{balle2018improving}, we present Theorem~\ref{thm-DP-OPT} below.

% , where analytical rather than \mbox{closed-form} expressions are given. We present and rewrite their result in Theorem~\ref{thm-DP-OPT} below.

% Balle and Wang~\cite{balle2018improving} also revisited Gaussian mechanism and showed that the classical Gaussian mechanisms have several limitations. Then, they have developed an optimal Gaussian mechanism concluded as follows.

\begin{thm}[\textbf{Optimal Gaussian mechanism for  $(\epsilon,\delta)$-differential privacy}] \label{thm-DP-OPT}
The optimal Gaussian mechanism for $(\epsilon,\delta)$-differential privacy, denoted by Mechanism \texttt{\upshape DP-OPT}, adds Gaussian noise with standard deviation $\sigma_{\texttt{\upshape DP-OPT}}$ specified below to each dimension of a query with $\ell_2$-sensitivity $\Delta$.
\begin{itemize}
  \setlength\itemsep{0em}
\item[(i)]
We derive $\sigma_{\texttt{\upshape DP-OPT}}$ as follows based on Theorem 8 of Balle and Wang~\shortcite{balle2018improving}:
\begin{align}
& \text{With $a$ satisfying } \erfc\left(a \right)   -  e^{\epsilon} \erfc\left( \sqrt{a^2 + \epsilon} \right)  =  2 \delta,\nonumber \\ & \text{we get~} \textstyle{\sigma_{\texttt{\upshape DP-OPT}}  := \frac{\left(a+\sqrt{a^2+\epsilon} \hspace{1.5pt}\right)  \cdot \Delta }{\epsilon\sqrt{2}} } , \label{eqn-sigma-DP-OPT}
\end{align}
where $\erfc()$ is the complementary error function.
\end{itemize}
For $\epsilon \geq 0.01$ and $0<\delta \leq 0.05$, we prove the following results:
% we show for the first time that there exist positive functions $A(\delta)$, $B(\delta)$, $C(\delta)$, and $D(\delta)$ such that the optimal Gaussian noise amount $\sigma_{\texttt{\upshape DP-OPT}}$ is greater than $\frac{A(\delta)}{\epsilon} + \frac{B(\delta)}{\sqrt{\epsilon}}$ and less than $\frac{C(\delta)}{\epsilon} + \frac{D(\delta)}{\sqrt{\epsilon}}$.
% Specifically, we prove the following results for $\epsilon \geq 1$ and $0< \delta \leq 0.005$:
\begin{itemize}
  \setlength\itemsep{0em}
\item[(ii)]   \mbox{$\sigma_{\texttt{\upshape DP-OPT}} >  \frac{\Delta}{\sqrt{2\epsilon\,}} $.}
% The term $a$ in Eq.~(\ref{eqn-sigma-DP-OPT}) satisfies $\sqrt{a^2 + \epsilon} \geq \sqrt{ \ln \frac{1}{\delta} +  \ln \frac{\epsilon}{ 8 \sqrt{\pi} } - \frac{3}{2} \ln \ln \frac{1}{\delta} } >\sqrt{ \ln \frac{0.0057}{\delta} } $, which implies a lower bound in the form of $\Theta\left(\frac{1}{\sqrt{\epsilon}}\right)$ for $\sigma_{\texttt{\upshape DP-OPT}}$: $\sigma_{\texttt{\upshape DP-OPT}} > \sqrt{ \frac{1}{8} \ln \frac{0.0057}{\delta} } \cdot\frac{\Delta }{\epsilon} + \frac{\Delta}{\sqrt{8\epsilon\,}} $.
\item[(iii)]  \mbox{$\sigma_{\texttt{\upshape DP-OPT}} <  \sqrt{2\ln \frac{1}{2\delta} }\cdot\frac{\Delta}{\epsilon} + \frac{\Delta}{\sqrt{2\epsilon\,}} $.} % The term $a$ in Eq.~(\ref{eqn-sigma-DP-OPT}) satisfies $a < \sqrt{ \ln \frac{1}{2\delta} } $,  which implies an upper bound in the form of $\Theta\left(\frac{1}{\sqrt{\epsilon}}\right)$ for $\sigma_{\texttt{\upshape DP-OPT}}$: $\sigma_{\texttt{\upshape DP-OPT}} <  \sqrt{2\ln \frac{1}{2\delta} }\cdot\frac{\Delta}{\epsilon} + \frac{\Delta}{\sqrt{2\epsilon\,}} $.
\end{itemize}
%so that $\Phi\left( \cdot \right)$.
% Moreover,   $\sigma_{\texttt{\upshape DP-OPT}}$ given $\delta $ is $O\left( \frac{1}{\sqrt{\epsilon}} \right)$ as $\epsilon \to \infty$; specifically, given $\delta $, $ \lim_{\epsilon \to \infty}\sigma_{\texttt{\upshape DP-OPT}} \bigg/ \left( \frac{1}{\sqrt{\epsilon}} \right) = \frac{\Delta }{\sqrt{2}} $.
\end{thm}

\begin{rem}
Results (ii) and (iii) of Theorem~\ref{thm-DP-OPT} mean that $\sigma_{\texttt{\upshape DP-OPT}}$ is in the form of $   \Theta\left(\frac{1}{\sqrt{\epsilon}}\right)$ for large $\epsilon$ (note $\frac{1}{\epsilon}$ is smaller than $\frac{1}{\sqrt{\epsilon}}$ for large $\epsilon$). This further implies the result of Theorem~\ref{thm-Dwork-2014-not-work} for $0<\delta \leq 0.05$ (our direct proof for Theorem~\ref{thm-Dwork-2014-not-work} in Appendix~\ref{secprf-thm-Dwork-2014-not-work} works for any $0< \delta < 1$).
% Again,    $\sigma_{\texttt{\upshape DP-OPT}}$ for large $\epsilon$ can be written as $\Theta\left(\frac{1}{\sqrt{\epsilon}}\right)$, but not $\Theta\left(\frac{1}{\epsilon}\right)$.
\end{rem}

\begin{rem}  \label{rem-Mechanism-DP-OPT-ru}
With $r(u) : = \erfc\left(u \right)   -  e^{\epsilon} \erfc\left( \sqrt{u^2 + \epsilon} \right) $, the term $a$ in Eq.~(\ref{eqn-sigma-DP-OPT}) satisfies $r(a) =   2 \delta  $.
Then $r(u)$ strictly decreases as $u$ increases given the derivative $r'(u)  =  \frac{2}{\sqrt{\pi}} \exp(-u^2) \times \frac{u - \sqrt{u^2 + \epsilon}  }{\sqrt{u^2 + \epsilon}} < 0 $. Based on this and $r(0)  = 1 - e^{\epsilon} \erfc\left( \sqrt{ \epsilon} \right)  $, for $a$ in Eq.~(\ref{eqn-sigma-DP-OPT}), we obtain $a > 0$ if $ 2 \delta < 1 - e^{\epsilon} \erfc\left( \sqrt{ \epsilon} \right) $, and $a \leq 0$ otherwise. More discussions about Remark~\ref{rem-Mechanism-DP-OPT-ru} are presented in Appendix~\ref{subsection-rem-Mechanism-DP-OPT-ru-detailed} of this supplementary file.
% , where we show in Appendix~\ref{subsection-rem-Mechanism-DP-OPT-ru-detailed} of this supplementary file that $e^{\epsilon} \erfc\left( \sqrt{ \epsilon} \right)$ strictly decreases as $\epsilon$ increases.
\end{rem}

\begin{rem} \label{rem-Mechanism-DP-OPT}
Mechanism \texttt{\upshape DP-OPT} is just the optimal \underline{Gaussian} mechanism for $(\epsilon,\delta)$-differential privacy in the sense that it gives the minimal required amount of noise when the noise follows a Gaussian distribution. However, it may not be the optimal mechanism for $(\epsilon,\delta)$-differential privacy, since there may exist other perturbation methods~\cite{pihur2019podium,geng2018truncated,geng2016optimal} which may outperform a Gaussian mechanism under certain utility measure~\cite{geng2014optimal}.
% after defining a utility function to quantify the optimality.
\end{rem}

We prove Theorem~\ref{thm-DP-OPT} in Appendix~\ref{appsecprf-DP-OPT} of this supplementary file.

% Lemma~\ref{lem-DP-OPT-bounds} below provide bounds for $\sigma_{\texttt{\upshape DP-OPT}}$ of Theorem~\ref{thm-DP-OPT}.

Since $\sigma_{\texttt{\upshape DP-OPT}}$ of Theorem~\ref{thm-DP-OPT} has no \mbox{closed-form} expression and needs to be approximated in an iterative manner, we first provide its asympotics in Theorem~\ref{thm-DP-OPT-asymptotics} and present more computationally efficient upper bounds for $\sigma_{\texttt{\upshape DP-OPT}}$ in Section~\ref{sec-DP-ours}.

% $0<\delta<1$ vs $0<\delta<0.5$

%\subsection{An algorithm to approximate the optimal Gaussian noise amount for $(\epsilon,\delta)$-differential privacy}

In Appendix~\ref{appendix-sec-Alg-opt} of this supplementary file, we present
Algorithm~\ref{Alg-opt} to compute $\sigma_{\texttt{\upshape DP-OPT}}$ of Theorem~\ref{thm-DP-OPT}.

We now analyze the asympotics for the optimal Gaussian noise amount $\sigma_{\texttt{\upshape DP-OPT}}$ of $(\epsilon,\delta)$-differential privacy. As a side result, we prove that $\sigma_{\texttt{\upshape DP-OPT}}$ is always less than $ \frac{\Delta}{2\sqrt{2} \cdot \inverf(\delta)} $ and hence bounded even for $ \epsilon \to 0 $. This is in contrast to the classical Gaussian mechanisms' noise amounts $\sigma_{\texttt{\upshape Dwork-2006}}$ and $\sigma_{\texttt{\upshape Dwork-2014}}$ in Eq.~(\ref{dwork-2006}) and~(\ref{dwork-2014}) which scale with $\frac{1}{\epsilon}$ and hence tend to $\infty$ as $ \epsilon \to 0 $.

\begin{thm}[\textbf{An upper bound and asympotics of the optimal Gaussian noise amount for $(\epsilon,\delta)$-differential privacy}] \label{thm-DP-OPT-asymptotics}  ~
% We have the following asympotic results for the optimal Gaussian noise amount $\sigma_{\texttt{\upshape DP-OPT}}$ in Theorem~\ref{thm-DP-OPT}   for $(\epsilon,\delta)$-differential
% privacy.
\begin{itemize}
   \item[\ding{172}] For any $\epsilon >0 $ and $0<\delta<1$, $\sigma_{\texttt{\upshape DP-OPT}}$ is less than $\frac{\Delta}{2\sqrt{2} \cdot \inverf(\delta)} $, which is the optimal Gaussian noise amount to achieve $(0,\delta)$-differential privacy.
    \item[\ding{173}] {Given a fixed $0<\delta<1$, $\sigma_{\texttt{\upshape DP-OPT}}$ converges to its upper bound $\frac{\Delta}{2\sqrt{2} \cdot \inverf(\delta)} $ as $\epsilon \to 0$}.
    % More specifically, given a fixed $0<\delta<1$, we have for $ \lim_{\epsilon \to \infty}\sigma_{\texttt{\upshape DP-OPT}} = \frac{\Delta}{2\sqrt{2} \cdot \inverf(\delta)} $.
    \item[\ding{174}] {Given a fixed $0<\delta<1$, $\sigma_{\texttt{\upshape DP-OPT}}$ is $\Theta\left( \frac{1}{\sqrt{\epsilon}} \right)$ as $\epsilon \to \infty$; specifically,  $ \lim_{\epsilon \to \infty}\sigma_{\texttt{\upshape DP-OPT}} \Big/ \left( \frac{\Delta}{\sqrt{2\epsilon\hspace{1.5pt}}} \right) = 1 $.}
%  .\\ Given a fixed $0<\delta<1$, $\sigma_{\texttt{\upshape DP-OPT}}$ is $\Theta\left( \frac{1}{\epsilon} \right)$ as $\epsilon \to 0$. \\ Given a fixed $0<\delta<1$, $\sigma_{\texttt{\upshape DP-OPT}}$ is $\Theta\left( \frac{1}{\sqrt{\epsilon}} \right)$ as $\epsilon \to 0$. \\
%     More specifically, we have:\\
%   Given a fixed $0<\delta<1$, $ \lim_{\epsilon \to \infty}\sigma_{\texttt{\upshape DP-OPT}} \bigg/ \left( \frac{1}{\sqrt{\epsilon}} \right) = \frac{\Delta }{\sqrt{2}} $.\\
%     Given a fixed $\delta < 0.5 $, $ \lim_{\epsilon \to 0}\sigma_{\texttt{\upshape DP-OPT}} \bigg/ \left( \frac{1}{\epsilon} \right) = \sqrt{2} \cdot \Delta  \cdot \inverfc(2 \delta)  $.\\
%   Given a fixed $\delta \geq 0.5 $, $ \lim_{\epsilon \to 0}\sigma_{\texttt{\upshape DP-OPT}} \bigg/ \left( \frac{1}{\sqrt{\epsilon}} \right) = \frac{\Delta }{\sqrt{2}} $.
  \item[\ding{175}] {Given a fixed $\epsilon >0 $, $\sigma_{\texttt{\upshape DP-OPT}}$ is $\Theta\left( \sqrt{\ln \frac{1}{\delta}} \right)$ as $\delta \to 0$; specifically,  $ \lim_{\delta \to 0}\sigma_{\texttt{\upshape DP-OPT}} \Big/ \left( \frac{\Delta}{\epsilon} \sqrt{2  \ln \frac{1}{\delta}  }\hspace{1.5pt}\right) = 1 $.}
    \end{itemize}
\end{thm}

\textbf{Intuition of Result \ding{172} of Theorem~\ref{thm-DP-OPT-asymptotics} based on Theorem~\ref{thm-DP-OPT}.} With $\delta$ fixed, when $\epsilon$ tends to 0, the quantity $a$ in Eq.~(\ref{eqn-sigma-DP-OPT}) of Theorem~\ref{thm-DP-OPT} is negative and is close to $-\inverfc(1-\delta)$; i.e., $\inverf(\delta)$, where we use $\erfc\left( -a\right)=2-\erfc\left(a \right) $. Then the numerator $(a+\sqrt{a^2+\epsilon})\Delta$ of Eq.~(\ref{eqn-sigma-DP-OPT}) can be written as $\frac{\epsilon\Delta}{-a+\sqrt{a^2+\epsilon}} $ and approaches $\frac{\epsilon\Delta}{(-a)\cdot 2} $ to  scale with $\epsilon$  instead of scaling with $\sqrt{\epsilon}$ as $\epsilon \to 0$. As the numerator and denominator of Eq.~(\ref{eqn-sigma-DP-OPT}) are both $\Theta(\epsilon)$ as $\epsilon \to 0$, $\sigma_{\texttt{\upshape DP-OPT}}$ with fixed $\delta$  does not grow unboundedly as $\epsilon \to 0$.

% Details in Eq(20) on Page13 of the full version show the optimal amount has a finite limit as $\epsilon$ tends to 0.

We prove Theorem~\ref{thm-DP-OPT-asymptotics} in Appendix~\ref{secprf-thm-DP-OPT-asymptotics} of this supplementary file.
Theorem~\ref{thm-DP-OPT-asymptotics} provides the first asymptotic results in the literature on the optimal Gaussian noise amount for $(\epsilon,\delta)$-differential privacy. The proofs delicately bound $\sigma_{\texttt{\upshape DP-OPT}}$ to avoid over-approximation.

%  Results \ding{172} and \ding{173} of Theorem~\ref{thm-DP-OPT-asymptotics} above for $\sigma_{\texttt{\upshape DP-OPT}}$ are in contrast to the  classical  Gaussian  mechanisms' noise amounts $\sigma_{\texttt{\upshape Dwork-2006}}$ and $\sigma_{\texttt{\upshape Dwork-2014}}$ which scale with $\frac{1}{\epsilon}$ and thus tend to $\infty$ as $  \epsilon \to 0 $.

%  In fact, $ \sigma_{\texttt{\upshape DP-OPT}}$ given a fixed $0<\delta<1$ converges to its upper bound $\frac{\Delta}{2\sqrt{2} \cdot \inverf(\delta)} $ as \mbox{$\epsilon \to 0$}, and is $\Theta\left( \frac{1}{\sqrt{\epsilon}} \right)$ as \mbox{$\epsilon \to \infty$}, where $\inverfc()$ denotes the inverse of the complementary error function.

% Moreover, f

For clarification, we note that  Results \ding{173} and \ding{175} of Theorem~\ref{thm-DP-OPT-asymptotics} do not contradict   each other since Result~\ding{173} fixes $0<\delta<1$ and considers $\epsilon \to 0$ so that $\epsilon/\delta \to 0$, while Result \ding{175} fixes $\epsilon > 0$ and considers $\delta \to 0$ so that $\delta/\epsilon \to 0$. More specifically, to bound $\sigma_{\texttt{\upshape DP-OPT}}$ in Result \ding{175}, we consider $\epsilon > f(\delta )$ for some function $f$, which clearly holds given a fixed $\epsilon >0 $ and $\delta \to 0$. With $\epsilon > f(\delta )$, the expression $ \frac{\Delta}{\epsilon} \sqrt{2  \ln \frac{1}{\delta}  }$ in Result \ding{175} is less than $ \frac{\Delta}{f(\delta )} \sqrt{2  \ln \frac{1}{\delta}  }$, which is less than $\frac{\Delta}{2\sqrt{2} \cdot \inverf(\delta)} $ for suitable $f(\delta )$, so Result \ding{173} does not contradict   Result \ding{175}.

\begin{figure*}
  \centering
  \footnotesize
  \begin{tabular}{cccc}\footnotesize
~ & ~ &~  & ~ \\[-25pt] \multirow{2}{*}{\includegraphics[width=0.17\textwidth]{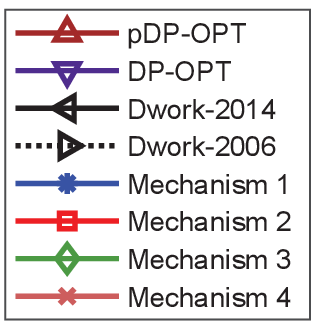}} &
    \hspace{-3mm}\includegraphics[width=0.23\textwidth]{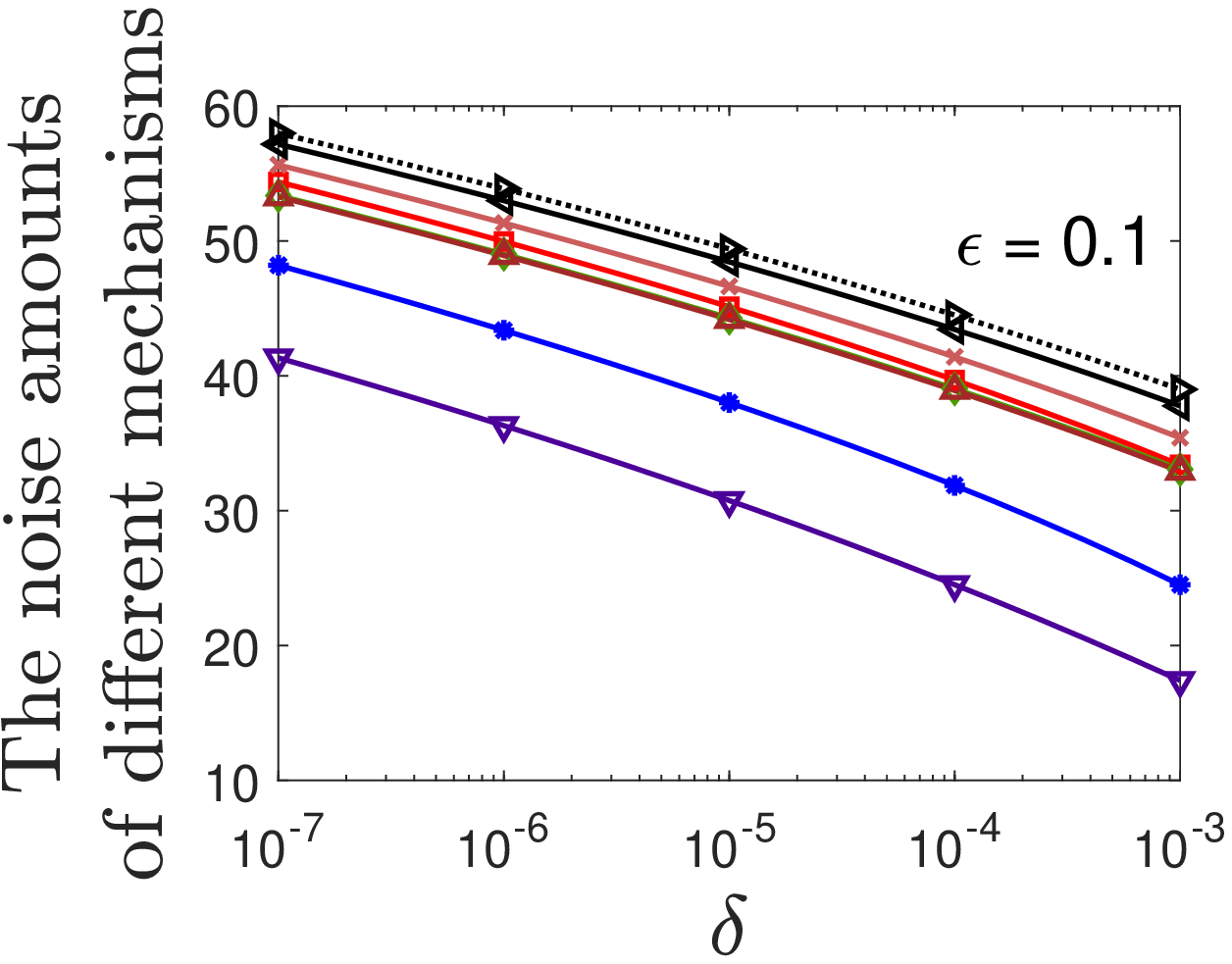} &
    \hspace{-3mm}\includegraphics[width=0.23\textwidth]{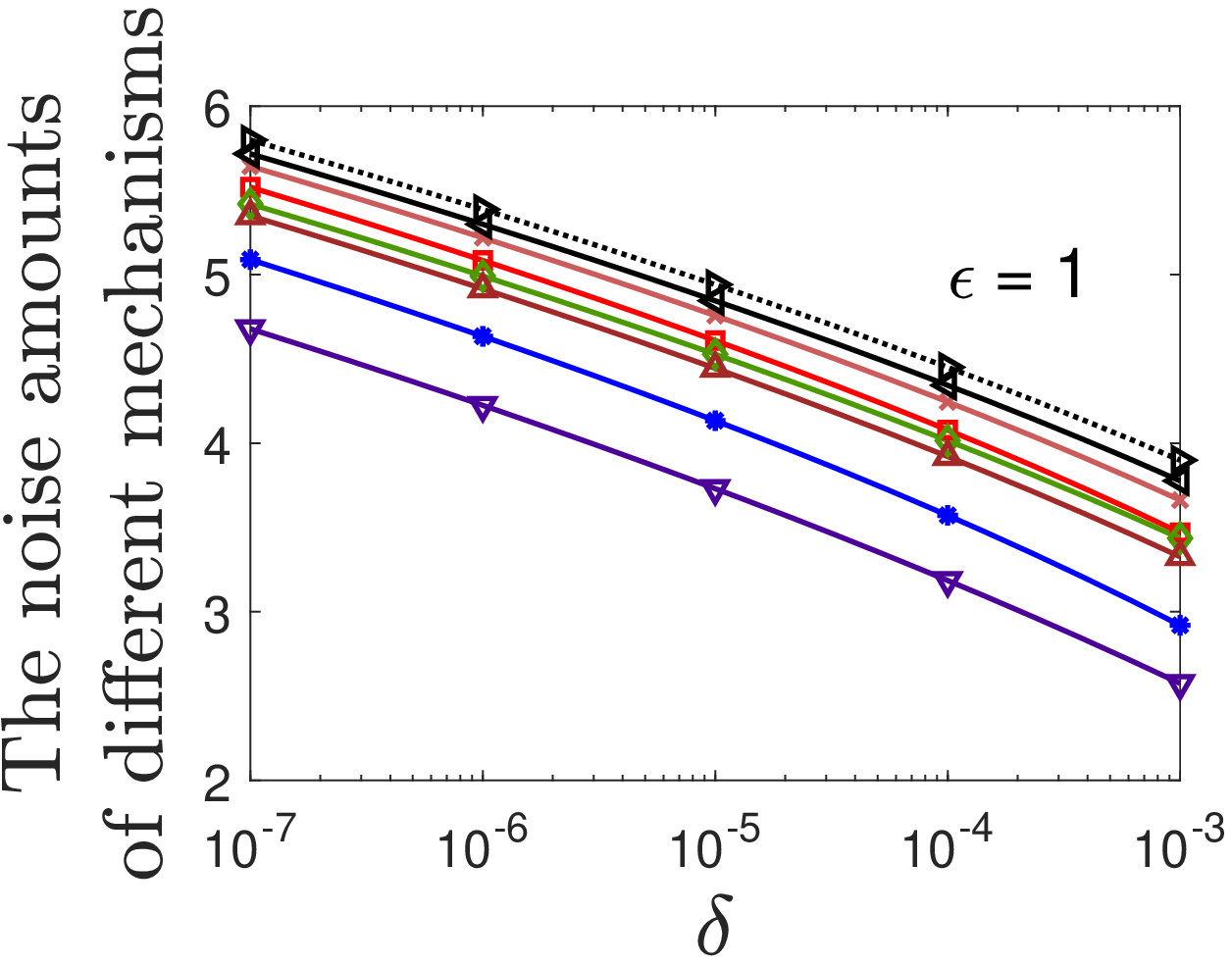} & \hspace{-3mm}\includegraphics[width=0.23\textwidth]{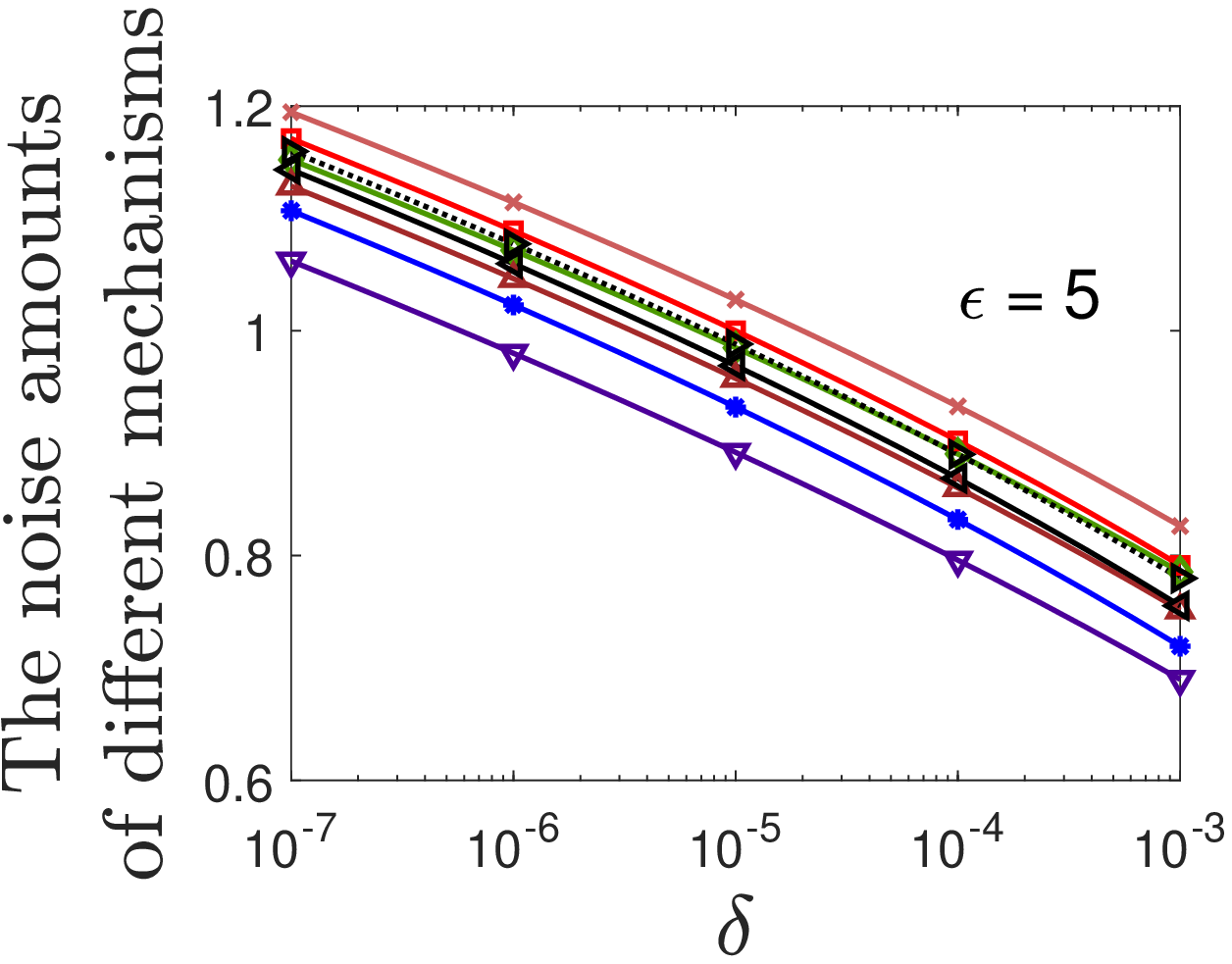} \\
   ~ & {\footnotesize (i)} & (ii)  & (iii)  \\
     ~ &   \hspace{-3mm}\includegraphics[width=0.23\textwidth]{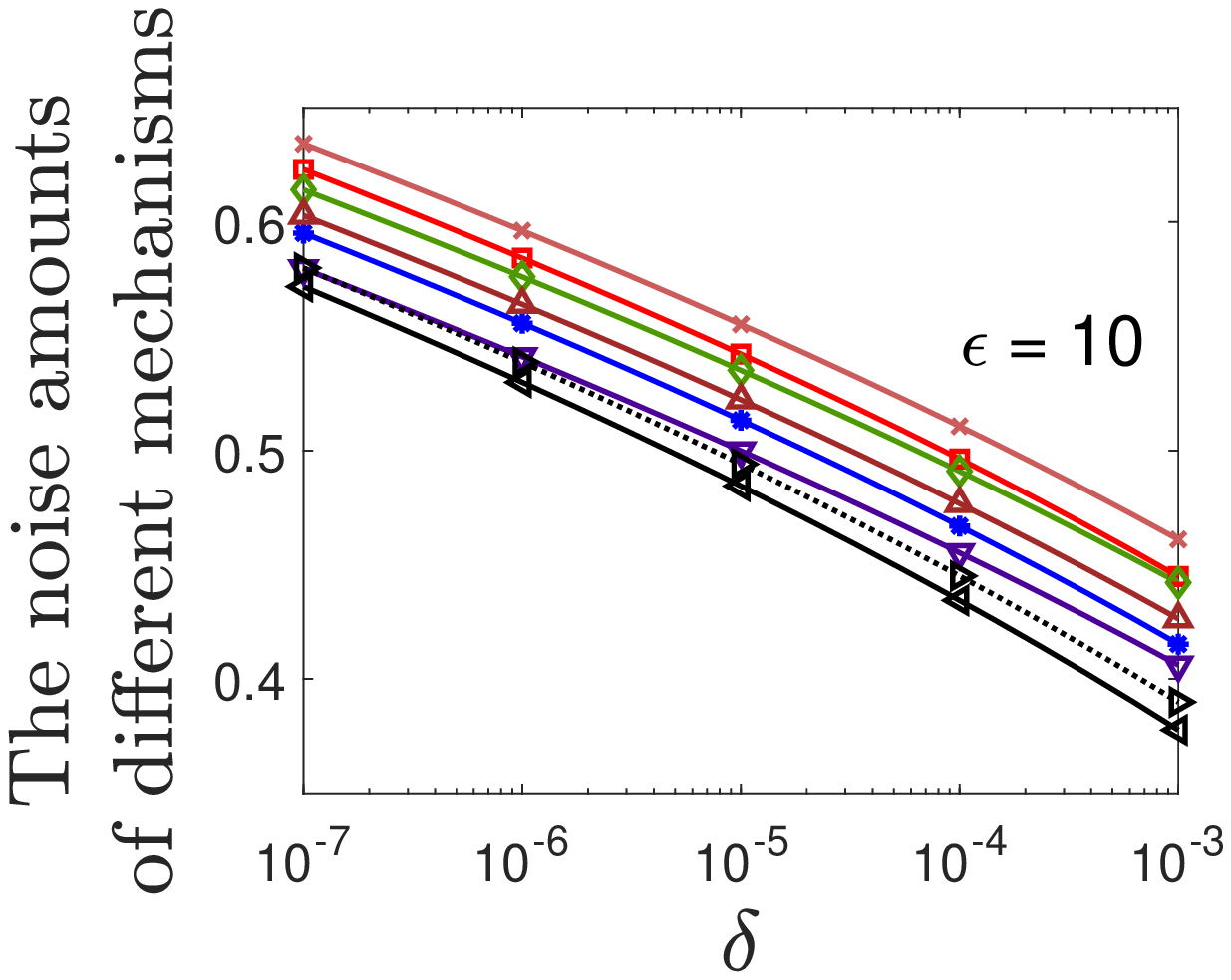}   &   \hspace{-3mm}\includegraphics[width=0.23\textwidth]{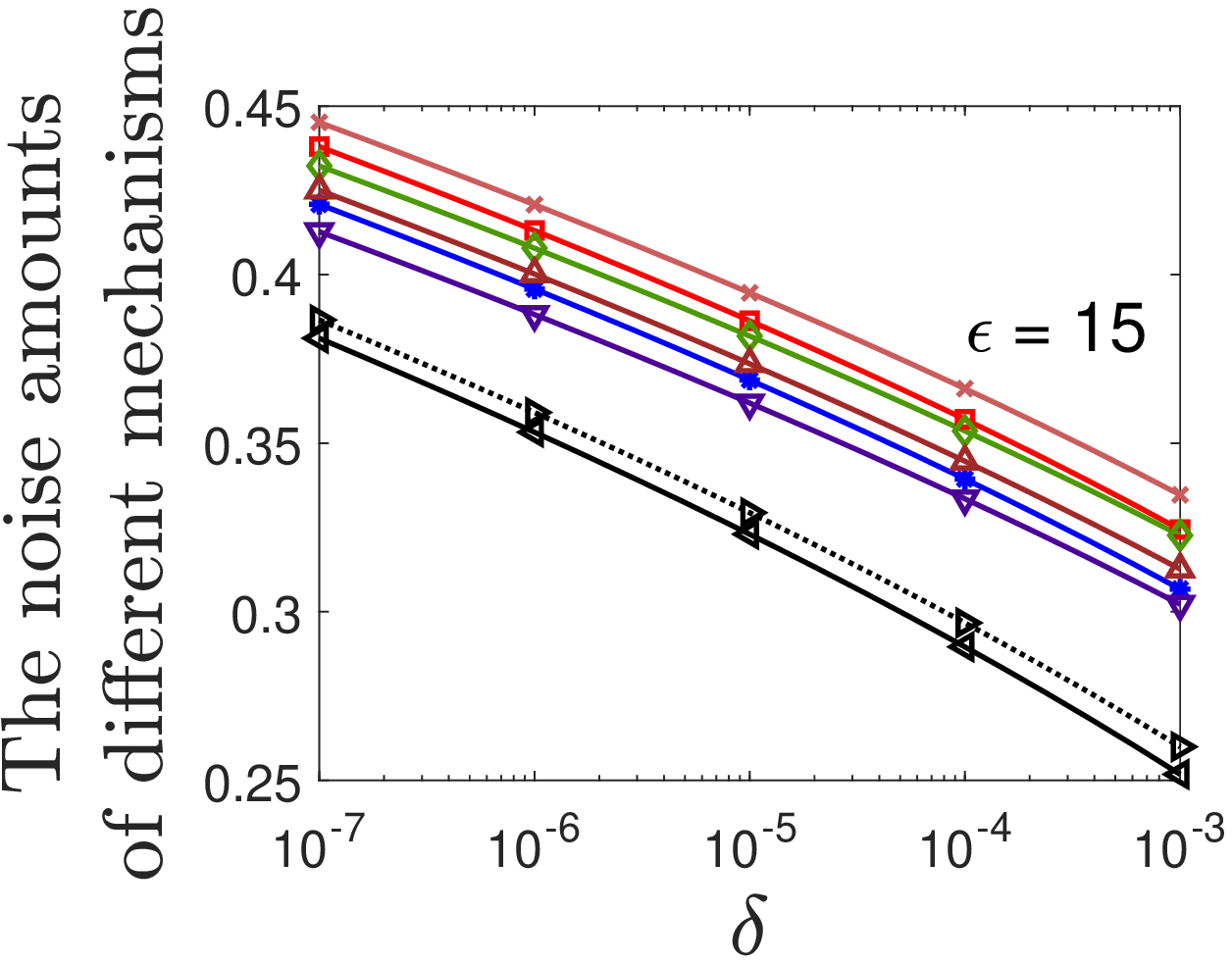} &
    \hspace{-3mm}\includegraphics[width=0.23\textwidth]{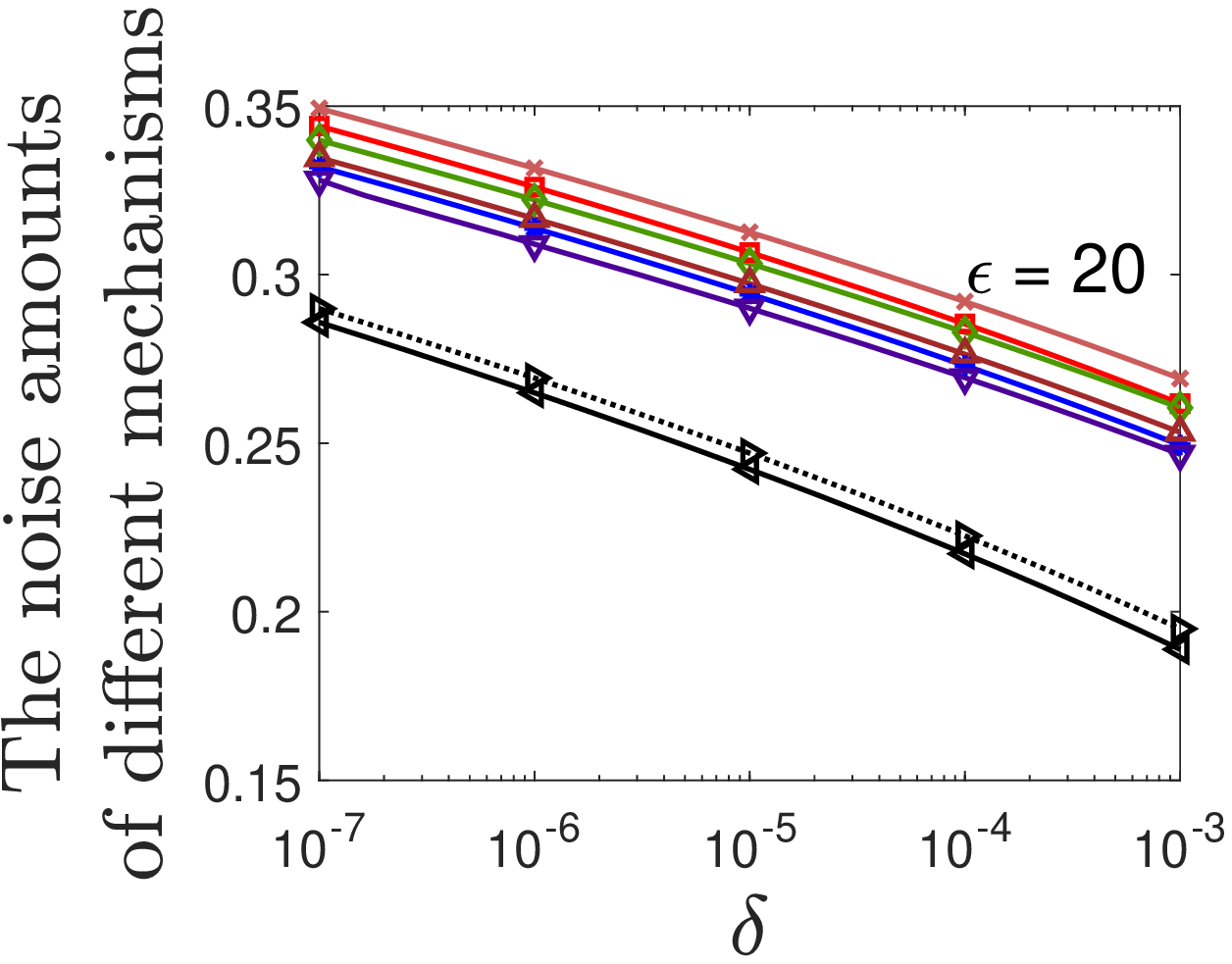} \\
  ~ & (iv)  & (v) & (vi)
   \end{tabular}
\captionsetup{singlelinecheck=off}
\caption[foo bar]{\mbox{The noise amounts of different mechanisms with respect to $\delta$, for $\epsilon$ = 0.1, 1, 5, 10, 15 and 20.} The meanings of the legends are as follows.
\begin{itemize}
\item \texttt{\upshape pDP-OPT} (resp.,~\texttt{\upshape DP-OPT}) is the optimal Gaussian mechanism to achieve $(\epsilon,\delta)$-pDP (resp.,~$(\epsilon,\delta)$-DP), where pDP is short for probabilistic differential privacy, a notion stronger than  differential privacy (DP) and to be elaborated in Section~\ref{sec-pDP-optimal-ours}.
\item \texttt{\upshape Dwork-2006} (resp.,~\texttt{\upshape Dwork-2014}) is the Gaussian mechanism proposed by Dwork~\textit{et~al.}~\shortcite{dwork2006our} in 2006 (resp.,~Dwork and Roth~\shortcite{dwork2014algorithmic} in 2014) to achieve $(\epsilon,\delta)$-DP.
\item \texttt{\upshape Mechanism~1} and \texttt{\upshape Mechanism~2}, which are our proposals to achieve $(\epsilon,\delta)$-DP and discussed in Section~\ref{sec-DP-ours}, are simpler and more computationally efficient than~\texttt{\upshape DP-OPT}.
\item \texttt{\upshape Mechanism~3} and \texttt{\upshape Mechanism~4}, which are our proposals to achieve $(\epsilon,\delta)$-pDP and will be discussed in Section~\ref{sec-pDP-opt-ours}, are simpler and more computationally efficient than~\texttt{\upshape pDP-OPT}.
\end{itemize}
}
  \label{fig:noiseamounts}
  \end{figure*}

\section{Our Proposed Gaussian Mechanisms for $(\epsilon,\delta)$-Differential Privacy}\label{sec-DP-ours} %\label{sec-main-our-proposed mechansim}

Table~\ref{table:DP-mechanisms} summarizes different mechanisms to achieve~$(\epsilon,\delta)$-differential privacy (DP), including \texttt{\upshape DP-OPT}   in Theorem~\ref{thm-DP-OPT} of the previous section as well as our \texttt{\upshape Mechanism~1} and \texttt{\upshape Mechanism~2} to be presented below.

We now detail our  Gaussian mechanisms for $(\epsilon,\delta)$-differential privacy, where the noise amounts have \mbox{closed-form}\footnote{Closed-form expressions in this paper can include functions ${\erf}()$, ${\erfc}()$, ${\inverf}()$, and ${\inverfc}()$.
%, whose meanings are given in the notation paragraph on Page~\pageref{paraNotation}.
} expressions and are more computationally efficient than the above Theorem~\ref{thm-DP-OPT}'s \texttt{\upshape DP-OPT}  which has no \mbox{closed-form} expression.
%  , as discussed in Section~\ref{sec-DP-optimal-our} above. Recall that Theorem~\ref{thm-DP-OPT} presents the optimal Gaussian noise amount $\sigma_{\texttt{\upshape DP-OPT}}$  for $(\epsilon,\delta)$-differential privacy.
 Our idea is to present computationally efficient upper bounds of $\sigma_{\texttt{\upshape DP-OPT}}$. To this end, we first present Lemma~\ref{lem-a-vs-b}, which upper bounds $a$ in Eq.~(\ref{eqn-sigma-DP-OPT}) of Theorem~\ref{thm-DP-OPT}.

\begin{lem} \label{lem-a-vs-b}
$a$ in Eq.~(\ref{eqn-sigma-DP-OPT}) is less than $b$ in Eq.~(\ref{eqn-u-Mechanism-1}).
\end{lem}

We prove Lemma~\ref{lem-a-vs-b} in Appendix~\ref{sec-prf-thm-Mechanism-1-based-on-thm-DP-OPT} of this supplementary file.
Theorem~\ref{thm-DP-OPT} and Lemma~\ref{lem-a-vs-b} imply
\begin{align} \label{eq-res1}
\text{$\sigma_{\texttt{\upshape DP-OPT}}$ in Eq.~(\ref{eqn-sigma-DP-OPT})} <
\text{$\sigma_{\texttt{\upshape Mechanism-1}}$ in Eq.~(\ref{eqn-sigma-Mechanism-1}) of Theorem~\ref{thm-Mechanism-1}},
\end{align}
% an upper bound of $\sigma_{\texttt{\upshape DP-OPT}}$ as $\sigma_{\texttt{\upshape Mechanism-1}}$ in Theorem~\ref{thm-Mechanism-1} below,
 where Theorem~\ref{thm-Mechanism-1} below presents \texttt{\upshape Mechanism~1}  to achieve $(\epsilon,\delta)$-differential privacy.

\begin{thm}[\textbf{Gaussian \texttt{\upshape Mechanism~1} for $(\epsilon,\delta)$-differential privacy}] \label{thm-Mechanism-1}
$(\epsilon,\delta)$-Differential privacy can be achieved by \texttt{\upshape Mechanism~1}, which adds Gaussian noise with standard deviation $\sigma_{\texttt{\upshape Mechanism-1}}$ to each dimension of a query with $\ell_2$-sensitivity $\Delta$, for $\sigma_{\texttt{\upshape Mechanism-1}}$ given by
\begin{subnumcases}{\hspace{-15pt}}
\hspace{-5pt}   b \hspace{-1.7pt}: =\hspace{-1.7pt} \begin{cases} \hspace{-1.7pt}  { \inverfc\hspace{-1pt}\left(\hspace{-2pt}\frac{2 \delta}{ 1\hspace{-1pt} -\hspace{-1pt} e^{\epsilon} \hspace{-1pt}\cdot \hspace{-1pt} \frac{\erfc\left(\hspace{-2pt} \sqrt{\big[\inverfc \big( 2 \delta + e^{\epsilon} \erfc\left( \sqrt{ \epsilon} \right) \big)\big]^2 + \epsilon} \right) }{ 2 \delta + e^{\epsilon} \erfc\left( \sqrt{ \epsilon} \right)}}\hspace{-2pt} \right)} \\ \text{~if $2- e^{\epsilon} \erfc\left( \sqrt{ \epsilon} \right) > 2 \delta $,} \label{eqn-u-Mechanism-1}  \\  \hspace{-1.7pt} 0 \text{~otherwise}; \end{cases}
\\ \hspace{-5pt} \textstyle{\sigma_{\texttt{\upshape Mechanism-1}}    := \frac{\left( b+\sqrt{b^2+\epsilon}\hspace{1.5pt} \right) \cdot  \Delta }{\epsilon\sqrt{2}}}   .  \label{eqn-sigma-Mechanism-1}
\end{subnumcases}
\end{thm}

The expression of $\sigma_{\texttt{\upshape Mechanism-1}}$ involves the complementary error function~$\erfc()$ and its inverse $\inverfc()$. Hence, we further present Lemma~\ref{lem-b-vs-c} below, which will enable us to propose \texttt{\upshape Mechanism 2}. Its noise amount is given by the closed-form expression of $\sigma_{\texttt{\upshape Mechanism-2}}$ and has only elementary functions.

Lemma~\ref{lem-b-vs-c} upper bounds $b$ in Eq.~(\ref{eqn-u-Mechanism-1}) of Theorem~\ref{thm-Mechanism-1}.

\begin{lem} \label{lem-b-vs-c}
$b$ in Eq.~(\ref{eqn-u-Mechanism-1}) is less than $c$ in Eq.~(\ref{eqn-u-Mechanism-2}).
\end{lem}

We prove Lemma~\ref{lem-b-vs-c} in Appendix~\ref{appenlem-lem-b-vs-c} of this supplementary file.
Theorem~\ref{thm-Mechanism-1} and Lemma~\ref{lem-b-vs-c} imply
\begin{align} \label{eq-res2}
\text{$\sigma_{\texttt{\upshape Mechanism-1}}$ in Eq.~(\ref{eqn-sigma-Mechanism-1})} <
 \text{$\sigma_{\texttt{\upshape Mechanism-2}}$ in Eq.~(\ref{eqn-u-Mechanism-2})} ,
\end{align}
  where the presented \texttt{\upshape Mechanism~2} in Theorem~\ref{thm-Mechanism-2} below is further simpler   than   \texttt{\upshape Mechanism~1} as noted above.

\begin{thm}[\textbf{Gaussian \texttt{\upshape Mechanism~2} for $(\epsilon,\delta)$-differential privacy}] \label{thm-Mechanism-2}
For $0<\delta<0.5$, $(\epsilon,\delta)$-differential privacy can be achieved by \texttt{\upshape Mechanism~2}, which adds Gaussian noise with standard deviation $\sigma_{\texttt{\upshape Mechanism-2}}$ to each dimension of a query with $\ell_2$-sensitivity $\Delta$, for $\sigma_{\texttt{\upshape Mechanism-2}}$ given by
\begin{align}
\textstyle{c : = \sqrt{\ln \frac{2}{\sqrt{16\delta+1}-1}}}   ; \label{eqn-u-Mechanism-2}
 ~~\textstyle{\sigma_{\texttt{\upshape Mechanism-2}}    := \frac{\left( c+\sqrt{c^2+\epsilon}\hspace{1.5pt} \right) \cdot  \Delta }{\epsilon\sqrt{2}} } .
\end{align}
\end{thm}

\noindent
\textbf{Superiority of our mechanisms.} The following discussions show the superiority of our proposed  mechanisms.
\begin{itemize}
\item[i)] From Inequalities~(\ref{eq-res1}) and~(\ref{eq-res2}), we have $\text{$\sigma_{\texttt{\upshape DP-OPT}}$ in Eq.~(\ref{eqn-sigma-DP-OPT})} <
\text{$\sigma_{\texttt{\upshape Mechanism-1}}$ in Eq.~(\ref{eqn-sigma-Mechanism-1})} <
 \text{$\sigma_{\texttt{\upshape Mechanism-2}}$ in Eq.~(\ref{eqn-u-Mechanism-2})} $. Among these noise amounts, $\sigma_{\texttt{\upshape Mechanism-1}}$ and $\sigma_{\texttt{\upshape Mechanism-2}}$ are straightforward to compute, whereas $\sigma_{\texttt{\upshape DP-OPT}}$ require higher computational complexity (a simple approach is the bisection method in~\cite[Page 3]{rao2010mathematical}.  Also, our plots in Figure~\ref{fig:noiseamounts} show that the noise amounts added by the optimal Gaussian mechanism \texttt{\upshape DP-OPT} and our more computationally efficient \texttt{\upshape Mechanism~1} are close.
 \item[ii)] For $0<\epsilon \leq 1$ where the proofs of \texttt{\upshape Dwork-2006} of~\cite{dwork2006our} and \texttt{\upshape Dwork-2014} of~\cite{dwork2014algorithmic}  require, we prove in Appendix~\ref{sec-Mechanism-2-Dwork-2014} that
 \begin{align}   \sigma_{\texttt{\upshape Mechanism-1}} <\sigma_{\texttt{\upshape Mechanism-2}} < \sigma_{\texttt{\upshape Dwork-2014}}< \sigma_{\texttt{\upshape Dwork-2006}}.\label{compare-eq-all}
\end{align}
%  \\ \textbullet~$$,\\ which along with Inequalities~(\ref{eq-res0}) and~(\ref{eq-res2}) further implies\\ \textbullet~$$,\\ \textbullet~$$, and\\ \textbullet~$$.
  \item[iii)] From Theorem~\ref{thm-Dwork-2014-not-work},
  %for \mbox{any $0<\delta<1$,}
  there exists a   function $G(\delta)$ such that
%   $G(\delta)$ is the maximal possible $\epsilon$ for {\texttt{\upshape Dwork-2014}} to achieve $(\epsilon,\delta)$-differential privacy and
    {\texttt{\upshape Dwork-2014}} {does~\emph{not}} achieve $(\epsilon,\delta)$-differential privacy for   $\epsilon >  G(\delta)$. Figure~\ref{fig:region-dp} shows $G(10^{-3}) = 7.47$, $G(10^{-4}) = 8.00$, $G(10^{-5}) = 8.43$, and $G(10^{-6}) = 8.79$. Result ii) above considers $0<\epsilon \leq 1$. For $1 < \epsilon \leq G(\delta)$ which the proof of {\texttt{\upshape Dwork-2014}} does not cover but {\texttt{\upshape Dwork-2014}} happens to achieve $(\epsilon,\delta)$-differential privacy, $\sigma_{\texttt{\upshape Mechanism-1}} < \sigma_{\texttt{\upshape Dwork-2014}}$ still holds as given by Figure~\ref{fig:noiseamounts}. Moreover, our \texttt{\upshape Mechanism~1} and \texttt{\upshape Mechanism~2} apply to any $\epsilon $. A similar discussion holds for $\texttt{\upshape Dwork-2006}$.
       % our plots in Figure~\ref{fig:noiseamounts} show    higher utilities of our mechanisms via \textbullet~$\sigma_{\texttt{\upshape Mechanism-2}} < \sigma_{\texttt{\upshape Dwork-2014}}$, which further implies\\ \textbullet~$\sigma_{\texttt{\upshape Mechanism-1}} < \sigma_{\texttt{\upshape Dwork-2014}}$, \textbullet~$\sigma_{\texttt{\upshape Mechanism-2}} < \sigma_{\texttt{\upshape Dwork-2006}}$, and \textbullet~$\sigma_{\texttt{\upshape Mechanism-1}} < \sigma_{\texttt{\upshape Dwork-2006}}$.
%   \item[iv)] Similar to the above result iii), there exists a positive function $C(\delta)$ such that for $1 \leq \epsilon \leq C(\delta)$ which the proof of {\texttt{\upshape Dwork-2006}} does not cover but {\texttt{\upshape Dwork-2006}} happens to achieve $(\epsilon,\delta)$-differential privacy, our plots in Figure~\ref{fig:noiseamounts} show    higher utilities of our mechanisms via \textbullet~$\sigma_{\texttt{\upshape Mechanism-2}} < \sigma_{\texttt{\upshape Dwork-2006}}$, which further implies  \textbullet~$\sigma_{\texttt{\upshape Mechanism-1}} < \sigma_{\texttt{\upshape Dwork-2006}}$.
%   \item[v)] For $G(\delta)$ in result iii), {\texttt{\upshape Dwork-2014}} {does~\emph{not}} achieve $(\epsilon,\delta)$-differential privacy for any $\epsilon >  G(\delta)$. For $C(\delta)$ in result iv), {\texttt{\upshape Dwork-2014}} {does~\emph{not}} achieve $(\epsilon,\delta)$-differential privacy for any $\epsilon >  C(\delta)$. In contrast, our \texttt{\upshape Mechanism-1} and \texttt{\upshape Mechanism-2} apply to any $\epsilon $.
\end{itemize}

\noindent
\textbf{Applications of our mechanisms.} Our proposed  mechanisms has the following applications. First, the   noise amounts of our mechanisms can be set as initial values to quickly search for  the optimal value or its tighter upper bound (as the optimal value has no closed-form expression). We use such approach in Algorithm~\ref{Alg-opt} of Appendix~\ref{appendix-sec-Alg-opt} of this supplementary file. In addition, our upper bounds may provide an \textit{intuitive} understanding about how a sufficient Gaussian noise amount changes according to $\epsilon$ and $\delta$: given $\delta$, a noise amount of $\Theta\big(\frac{1}{\epsilon}\big) + \Theta\big(\frac{1}{\sqrt{\epsilon}}\big)$ suffices; i.e., $\Theta\big(\frac{1}{\epsilon}\big) $ suffices for small $\epsilon$ and $\Theta\big(\frac{1}{\sqrt{\epsilon}}\big)$ suffices for large $\epsilon$. Finally, our mechanisms can be useful for Internet of Things (IoT) devices with little power or computational capabilities, since our mechanisms are  more computationally efficient than the optimal  Gaussian mechanism.

\section{$(\epsilon,\delta)$-Probabilistic Differential Privacy: Connection to $(\epsilon,\delta)$-Differential Privacy and Gaussian Mechanisms}\label{sec-pDP-optimal-ours}

In this section, for $(\epsilon,\delta)$-probabilistic differential privacy, we discuss its connection to $(\epsilon,\delta)$-differential privacy and its Gaussian mechanisms.

\begin{figure}
  \centering
  \footnotesize
  \begin{tabular}{cc}
    \hspace{0mm}\includegraphics[width=0.23\textwidth]{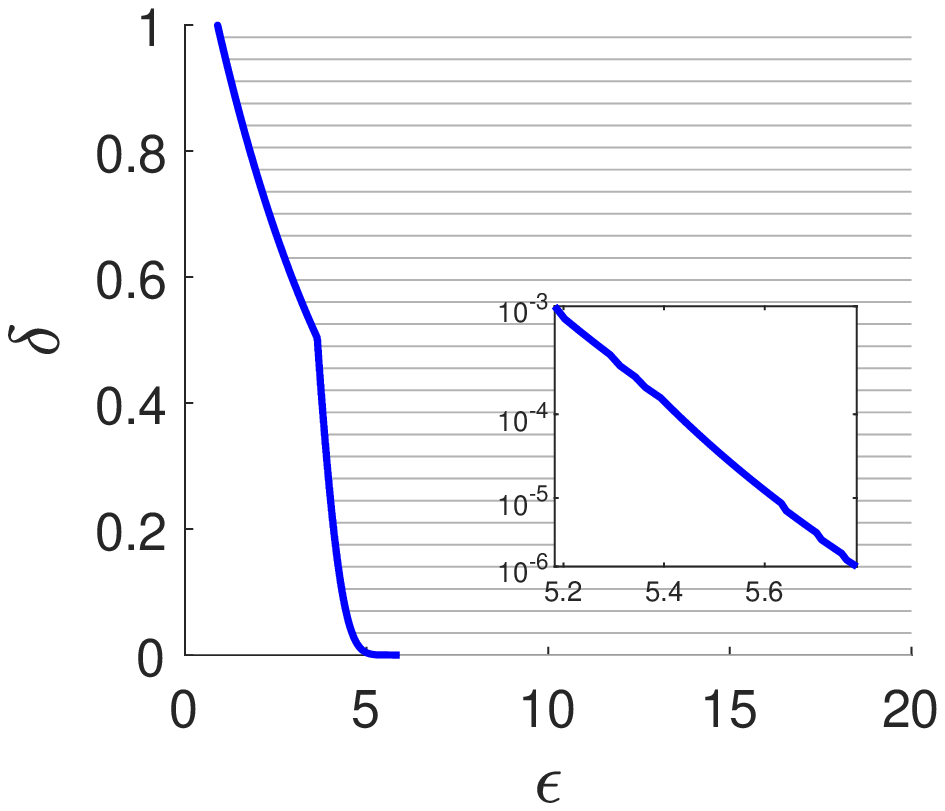} &
    \hspace{0mm}\includegraphics[width=0.23\textwidth]{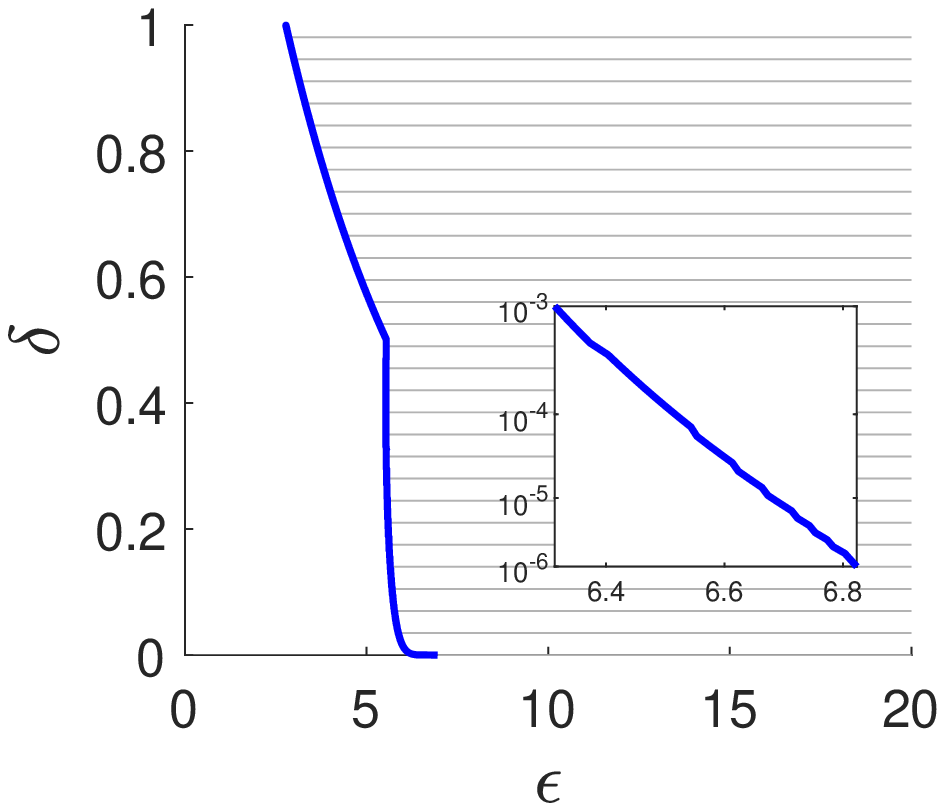}\\
(i) Mechanism \texttt{\upshape Dwork-2014} & (ii) Mechanism \texttt{\upshape Dwork-2006}
   \end{tabular}
 \vspace{-1pt} \caption{The shaded area in each subfigure represents the set of $(\epsilon,\delta)$ where Mechanism \texttt{\upshape Dwork-2014} (resp.,~\texttt{\upshape Dwork-2006}) does \textbf{not} achieve  \mbox{$(\epsilon,\delta)$-probabilistic differential privacy}. \vspace{-4mm}  }
  \label{fig:region-pdp}
  \end{figure}

\subsection{$(\epsilon,\delta)$-Probabilistic differential privacy} \label{appsecpdp}

To achieve $(\epsilon,\delta)$-differential privacy (formally given in Definition~\ref{defn-DP} on Page~\pageref{defn-DP}), most mechanisms ensure a condition on the privacy loss random variable defined below. Such condition is termed   $(\epsilon,\delta)$-probabilistic differential privacy~\cite{machanavajjhala2008privacy} and elaborated below. We will explain that $(\epsilon,\delta)$-probabilistic differential privacy is sufficient but not necessary for $(\epsilon,\delta)$-differential privacy.

For neighboring datasets $D$ and $D'$, the \textbf{privacy loss} $L_{Y,D,D'}(y)$ represents the multiplicative difference between the probabilities that the same output $y$ is observed when the randomized algorithm $Y$ is applied to $D$ and $D'$, respectively. Specifically, we define
\begin{align}
L_{Y,D,D'}(y) := \ln \frac{\fr{Y(D)=y}}{\fr{Y(D')=y}}, \label{eqn-L-Y-D-Dprime}
\end{align}
where $\fr{\cdot}$ denotes the probability density function.

For simplicity, we use probability density function $\fr{\cdot}$ in Eq.~(\ref{eqn-L-Y-D-Dprime}) above by assuming that the randomized algorithm $Y$ has continuous output. If $Y$ has discrete output, we replace $\fr{\cdot}$ by probability notation $\bp{\cdot}$.

When $y$ follows the probability distribution of random variable $Y(D)$, $L_{Y,D,D'}(y)$ follows the probability distribution of $L_{Y,D,D'}(Y(D))$, which is the \textbf{privacy loss random variable}. As a sufficient condition to enforce $(\epsilon,\delta)$-differential privacy, $(\epsilon,\delta)$-probabilistic differential privacy of~\cite{machanavajjhala2008privacy} is defined such that the privacy loss random variable $L_{Y,D,D'}(Y(D))$ falls in the interval $[-\epsilon, \epsilon]$ with probability at least $1- \delta$; i.e., $\mathbb{P} \left[ -\epsilon  \leq L_{Y,D,D'}(Y(D))  \leq \epsilon \right] \geq 1- \delta$. This is equivalent to the following definition.

\begin{defn}[\textbf{$(\epsilon,\delta)$-Probabilistic differential privacy}~\cite{machanavajjhala2008privacy}] \label{defn-Prob-DP}
A randomized algorithm $Y$ satisfies $(\epsilon,\delta)$-probabilistic differential privacy, if for any two neighboring datasets $D$ and $D'$ (elaborated in Remark~\ref{rem-neighboring-datasets} on Page~\pageref{rem-neighboring-datasets}), we have that for $y$ following the probabilistic distribution of the output $Y(D)$ (notated as $y \sim Y(D)$),
\begin{align} \label{eqn:defn-Prob-DP}
\mathbb{P}_{y \sim Y(D)}\left[ e^{-\epsilon}  \leq  \frac{\fr{Y(D)=y}}{\fr{Y(D')=y}}  \leq e^{\epsilon} \right] \geq 1- \delta,
\end{align}
where $\fr{\cdot}$ denotes the probability density function.
\end{defn}

\subsection{Relationships between differential privacy and probabilistic differential privacy}

Lemmas~\ref{lem-PDP-to-DP} and~\ref{lem-DP-to-PDP} below present the relationships between differential privacy and probabilistic differential privacy.

\begin{lem} \label{lem-PDP-to-DP}
$(\epsilon,\delta)$-Probabilistic differential privacy implies $(\epsilon,\delta)$-differential privacy.
\end{lem}

\begin{lem} \label{lem-DP-to-PDP}
$(\epsilon,\delta)$-Differential privacy implies  $(\epsilon_*,  \frac{\delta \cdot (1+e^{-\epsilon_*})}{1-e^{\epsilon-\epsilon_*}} )$-probabilistic differential privacy for any $\epsilon_* > \epsilon$.
\end{lem}

While the straightforward
Lemma~\ref{lem-PDP-to-DP} is shown in~\cite{dwork2014algorithmic}, the proof of Lemma~\ref{lem-DP-to-PDP} is not~trivial. Although~\shortcite{dwork2016concentrated} of Dwork and Rothblum, and~\shortcite{bun2016concentrated} of Bun and Steinke mention that differential privacy is equivalent, up to a
small loss in parameters, to probabilistic differential privacy, \cite{dwork2016concentrated,bun2016concentrated} do not present Lemma~\ref{lem-DP-to-PDP}. For completeness, we present the proofs of Lemmas~\ref{lem-PDP-to-DP} and~\ref{lem-DP-to-PDP}  in Appendices~\ref{secprf-lem-PDP-to-DP} and~\ref{secprf-lem-DP-to-PDP} of this supplementary file.

% to prove.
%
%Dwork and Rothblum~\cite{dwork2016concentrated} mention that with all but $\delta$
%probability the magnitude of privacy loss is bounded by $\epsilon$.?
%
%Bun and Steinke~\cite{bun2016concentrated} mention that differential privacy is equivalent, up to a
%small loss in parameters, to probabilistic differential privacy.
%Although the proofs of Lemmas~\ref{lem-PDP-to-DP} and~\ref{lem-DP-to-PDP} are not complex, we present them in Sections~\ref{secprf-lem-PDP-to-DP} and~\ref{secprf-lem-DP-to-PDP}  for completeness.

%
%\subsection{Failures of the classical Gaussian mechanisms to achieve $(\epsilon,\delta)$-probabilistic differential privacy for large $\epsilon$}

Similar to Theorem~\ref{thm-Dwork-2014-not-work} on Page~\pageref{thm-Dwork-2014-not-work}, we   show in Figure~\ref{fig:region-pdp} the failures of the classical Gaussian mechanisms of Dwork and Roth~\shortcite{dwork2014algorithmic} in 2014 and of Dwork~\textit{et~al.}~\shortcite{dwork2006our} in 2006 to achieve  $(\epsilon,\delta)$-probabilistic differential privacy for large $\epsilon$.

%We provide the details in the   full version~\cite{fullver}.

 We now present the optimal Gaussian mechanism for $(\epsilon,\delta)$-probabilistic differential privacy.

\begin{table*}[]
\caption{Different mechanisms to achieve~$(\epsilon,\delta)$-probabilistic differential privacy (pDP).}
\label{table:pDP-mechanisms}
\centering
\setlength\tabcolsep{3pt}\begin{tabular}{|l|l|}
\hline
pDP Mechanisms & Comparison                                                                               \\ \hline
\begin{tabular}[c]{@{}l@{}}Our \texttt{\upshape pDP-OPT}\\of Theorem~\ref{thm-pDP-OPT}\end{tabular} & \begin{tabular}[c]{@{}l@{}}\textbullet~the optimal Gaussian mechanism   to achieve~$(\epsilon,\delta)$-pDP,\\ \textbullet~no closed-form expression,\\ \textbullet~computed using the bisection method with the number of iterations \\ ~~being logarithmic in the given error (i.e., tolerance).\end{tabular}   \\ \cline{1-2}
\begin{tabular}[c]{@{}l@{}}Our \texttt{\upshape Mechanism~3}\\of Theorem~\ref{thm-Mechanism-3}\end{tabular} & \begin{tabular}[c]{@{}l@{}}\textbullet~closed-form expression involving \\ ~~the complementary error function's inverse $\inverfc()$,\\ \textbullet~computational complexity: dependent on\\ ~~$\inverfc()$ implementations and often very efficient, \\ \textbullet~$\sigma_{\texttt{\upshape Mechanism-3}}$ is slightly greater than $\sigma_{\texttt{\upshape pDP-OPT}}$. \end{tabular}                                                                   \\ \cline{1-2}
\begin{tabular}[c]{@{}l@{}}Our \texttt{\upshape Mechanism~4}\\of Theorem~\ref{thm-Mechanism-4}\end{tabular} & \begin{tabular}[c]{@{}l@{}}\textbullet~closed-form expression involving only elementary functions,\\ \textbullet~computed in constant amount of time, \\ \textbullet~$\sigma_{\texttt{\upshape Mechanism-4}}$ is slightly greater than $\sigma_{\texttt{\upshape Mechanism-3}}$. \end{tabular}                                                                \\ \hline
\end{tabular}
\end{table*}

\subsection{An analytical but not closed-form expression for the optimal Gaussian mechanism of $(\epsilon,\delta)$-probabilistic differential privacy} \label{sec-pDP-opt}

The optimal Gaussian mechanism of $(\epsilon,\delta)$-probabilistic differential privacy (pDP) is given in Theorem~\ref{thm-pDP-OPT} below.

\begin{thm}[\textbf{Optimal Gaussian mechanism for  $(\epsilon,\delta)$-probabilistic differential privacy}] \label{thm-pDP-OPT}
The optimal Gaussian mechanism for $(\epsilon,\delta)$-probabilistic differential privacy, denoted by Mechanism \texttt{\upshape pDP-OPT}, adds Gaussian noise with standard deviation $\sigma_{\texttt{\upshape pDP-OPT}}$ to each dimension of a query with  $\ell_2$-sensitivity $\Delta$, for $\sigma_{\texttt{\upshape pDP-OPT}}$ given by
\begin{subnumcases}{\hspace{-15pt}}
\text{Solve $d$ such that }\erfc\left(d\right) +  \erfc\left( \sqrt{d^2 + \epsilon} \right) = 2 \delta ; \label{eqn-u-pDP-OPT}
\\ \sigma_{\texttt{\upshape pDP-OPT}}    := \frac{\left( d+\sqrt{d^2+\epsilon}\hspace{1.5pt} \right)  \cdot \Delta }{\epsilon\sqrt{2}} .  \label{eqn-sigma-pDP-OPT}
\end{subnumcases}
\end{thm}

\begin{rem} \label{rem-Mechanism-pDP-OPT}
Mechanism \texttt{\upshape pDP-OPT} is just the optimal \underline{Gaussian} mechanism for $(\epsilon,\delta)$-probabilistic differential privacy in the sense that it gives the minimal required amount of noise when the noise follows a Gaussian distribution. However, it may not be the optimal mechanism for $(\epsilon,\delta)$-probabilistic differential privacy, since there may exist other perturbation methods (e.g., adding non-Gaussian noise) which may outperform a Gaussian mechanism under certain utility measure~\cite{geng2016optimal}.
% after defining a utility function to quantify the optimality.
\end{rem}

% The Optimal (Yet more computationally expensive) Gaussian Mechanism and Computationally Efficient Mechanisms

%We establish Lemma~\ref{lem-pDP-OPT-bounds} in Appendix~\ref{secprf-lem-pDP-OPT-bounds}.
%Lemma~\ref{lem-pDP-OPT-bounds} implies

We prove Theorem~\ref{thm-pDP-OPT} in Appendix~\ref{sec:proof3} of this supplementary file.
We present the asympotics of $\sigma_{\texttt{\upshape pDP-OPT}}$ as
 Theorem~\ref{thm-pDP-OPT-asymptotics} below.

\begin{thm}[\textbf{The asympotics of the optimal Gaussian noise amount for $(\epsilon,\delta)$-probabilistic differential privacy}] \label{thm-pDP-OPT-asymptotics}  ~
% We have the following asympotic results for the optimal Gaussian noise amount $\sigma_{\texttt{\upshape DP-OPT}}$ in Theorem~\ref{thm-DP-OPT}   for $(\epsilon,\delta)$-differential
% privacy.
\begin{itemize}
    \item[\ding{172}] Given a fixed $0<\delta<1$, $\sigma_{\texttt{\upshape pDP-OPT}}$ is $\Theta\left( \frac{1}{\epsilon} \right)$ as \mbox{$\epsilon \to 0$}. Specifically, given a fixed $0<\delta<1$, $ \lim_{\epsilon \to 0}\sigma_{\texttt{\upshape pDP-OPT}} \Big/ \left(\frac{\inverfc(\delta) \cdot \Delta }{\epsilon\sqrt{2} } \right) = 1$.
    % More specifically, given a fixed $0<\delta<1$, we have for $ \lim_{\epsilon \to \infty}\sigma_{\texttt{\upshape DP-OPT}} = \frac{\Delta}{2\sqrt{2} \cdot \inverf(\delta)} $.
    \item[\ding{173}] Given a fixed $0<\delta<1$, $\sigma_{\texttt{\upshape pDP-OPT}}$ is $\Theta\left( \frac{1}{\sqrt{\epsilon}} \right)$ as \mbox{$\epsilon \to \infty$}. Specifically, given a fixed $0<\delta<1$, $ \lim_{\epsilon \to \infty}\sigma_{\texttt{\upshape pDP-OPT}} \Big/ \left( \frac{\Delta}{\sqrt{2\epsilon\hspace{1.5pt}}} \right) = 1 $.
%  .\\ Given a fixed $0<\delta<1$, $\sigma_{\texttt{\upshape DP-OPT}}$ is $\Theta\left( \frac{1}{\epsilon} \right)$ as $\epsilon \to 0$. \\ Given a fixed $0<\delta<1$, $\sigma_{\texttt{\upshape DP-OPT}}$ is $\Theta\left( \frac{1}{\sqrt{\epsilon}} \right)$ as $\epsilon \to 0$. \\
%     More specifically, we have:\\
%   Given a fixed $0<\delta<1$, $ \lim_{\epsilon \to \infty}\sigma_{\texttt{\upshape DP-OPT}} \bigg/ \left( \frac{1}{\sqrt{\epsilon}} \right) = \frac{\Delta }{\sqrt{2}} $.\\
%     Given a fixed $\delta < 0.5 $, $ \lim_{\epsilon \to 0}\sigma_{\texttt{\upshape DP-OPT}} \bigg/ \left( \frac{1}{\epsilon} \right) = \sqrt{2} \cdot \Delta  \cdot \inverfc(2 \delta)  $.\\
%   Given a fixed $\delta \geq 0.5 $, $ \lim_{\epsilon \to 0}\sigma_{\texttt{\upshape DP-OPT}} \bigg/ \left( \frac{1}{\sqrt{\epsilon}} \right) = \frac{\Delta }{\sqrt{2}} $.
  \item[\ding{174}] Given a fixed $\epsilon >0 $, $\sigma_{\texttt{\upshape pDP-OPT}}$ is $\Theta\left( \sqrt{ \ln \frac{1}{\delta}  } \right)$ as $\delta \to 0$. Specifically, given a fixed $\epsilon>0  $, $ \lim_{\delta \to 0}\sigma_{\texttt{\upshape pDP-OPT}} \Big/ \left( \frac{\Delta}{\epsilon} \sqrt{2  \ln \frac{1}{\delta}  }\hspace{1.5pt}\right) = 1 $.
    \end{itemize}
\end{thm}

 Theorem~\ref{thm-pDP-OPT-asymptotics}  is proved in Appendix~\ref{secprf-thm-pDP-OPT-asymptotics}  of this supplementary file.

\begin{rem}  \label{rem-thm-pDP-OPT-asymptotics}
From Result~\ding{172} of Theorem~\ref{thm-pDP-OPT-asymptotics}, given a fixed $0<\delta<1$, $\sigma_{\texttt{\upshape pDP-OPT}} = \Theta\left( \frac{1}{\epsilon} \right) \to \infty$ as \mbox{$\epsilon \to 0$}. In contrast, from Result~\ding{172} of Theorem~\ref{thm-DP-OPT-asymptotics}, given a fixed $0<\delta<1$, $\sigma_{\texttt{\upshape DP-OPT}} \to \frac{\Delta}{2\sqrt{2} \cdot \inverf(\delta)} $ as \mbox{$\epsilon \to 0$}. This shows a fundamental difference between $(\epsilon,\delta)$-differential privacy and $(\epsilon,\delta)$-probabilistic differential privacy.

% In ? on Page ?, we will present the relationships between these two privacy notion.
\end{rem}

\begin{rem}
In Lemmas~\ref{lem-PDP-to-DP} and~\ref{lem-DP-to-PDP} above, we show the relationship between differential privacy and probabilistic differential privacy that the latter implies the former and the former implies the latter  up to possible loss in privacy parameters. Given this, one may wonder if this relationship contradicts   their difference discussed in Remark~\ref{rem-thm-pDP-OPT-asymptotics} above as \mbox{$\epsilon \to 0$}. Below we explain there is no contradiction, by showing that the Gaussian noise amount for probabilistic differential privacy obtained by first achieving differential privacy is at the same order as  the optimal Gaussian noise amount for probabilistic differential privacy when \mbox{$\epsilon \to 0$}.

From Lemma~\ref{lem-DP-to-PDP},
$(0, \frac{\delta \cdot (1-e^{-\epsilon})}{1+e^{-\epsilon}} )$-differential privacy implies $(\epsilon,\delta)$-probabilistic differential privacy. From Result~\ding{172} of Theorem~\ref{thm-DP-OPT-asymptotics},  $(0, \frac{\delta \cdot (1-e^{-\epsilon})}{1+e^{-\epsilon}} )$-differential privacy can be achieved by the Gaussian mechanism with noise amount $   \frac{\Delta}{2\sqrt{2} \cdot \inverf\big(\frac{\delta \cdot (1-e^{-\epsilon})}{1+e^{-\epsilon}}\big)} $.  Hence,  $(\epsilon,\delta)$-probabilistic differential privacy can also be achieved by the Gaussian mechanism with noise amount $  \frac{\Delta}{2\sqrt{2} \cdot \inverf\big(\frac{\delta \cdot (1-e^{-\epsilon})}{1+e^{-\epsilon}}\big)} $, which given $\delta$ is $\Theta\left( \frac{1}{\epsilon} \right)$ as \mbox{$\epsilon \to 0$} due to $\lim_{\epsilon \to 0}\frac{(1-e^{-\epsilon})}{1+e^{-\epsilon}} \big/ \epsilon = \frac{1}{2} $ and $ \lim_{x \to 0}\frac{\inverf(x)}{x} = \frac{\sqrt{\pi}}{2} $ from~\cite{carlitz1963inverse}. From Result~\ding{172} of Theorem~\ref{thm-pDP-OPT-asymptotics},  the optimal Gaussian noise amount for $(\epsilon,\delta)$-probabilistic differential privacy given $\delta$ is also $\Theta\left( \frac{1}{\epsilon} \right)$ as \mbox{$\epsilon \to 0$}. Hence, the combination of Lemma~\ref{lem-DP-to-PDP} and Result~\ding{172} of Theorem~\ref{thm-DP-OPT-asymptotics} does not contradict   Result~\ding{172} of Theorem~\ref{thm-pDP-OPT-asymptotics}.
\end{rem}

% In \texttt{\upshape Mechanism~2}, the Gaussian noise amount
% $\sigma_{\texttt{\upshape Mechanism-2}}$ is given by

% In Mechanism \texttt{\upshape pDP-OPT}, the Gaussian noise amount
% $\sigma_{\texttt{\upshape pDP-OPT}}$ is given by

% Lemma~\ref{lem-d-vs-f} upper bounds $d$ in Eq.~(\ref{eqn-u-pDP-OPT}) of Theorem~\ref{thm-pDP-OPT}.

% \begin{lem} \label{lem-d-vs-f}
% $d$ in Eq.~(\ref{eqn-u-pDP-OPT}) is less than $f$ in Eq.~(\ref{eqn-u-Mechanism-3}).
% \end{lem}

From Theorem~\ref{thm-pDP-OPT}, the optimal Gaussian mechanism \texttt{\upshape pDP-OPT} does not have a \mbox{closed-form} expression. In the next subsection, we detail our  Gaussian mechanisms for $(\epsilon,\delta)$-pDP, where the noise amounts have \mbox{closed-form} expressions and are more computationally efficient than \texttt{\upshape pDP-OPT}.

\subsection{Our Gaussian mechanisms for $(\epsilon,\delta)$-probabilistic differential privacy with \mbox{closed-form} expressions of noise amounts}\label{sec-pDP-opt-ours}

 The idea of our Gaussian mechanisms is to present computationally efficient upper bounds of $\sigma_{\texttt{\upshape pDP-OPT}}$. To this end, we first present Lemma~\ref{lem-pDP-OPT-bounds}, which upper bounds $d$ in Eq.~(\ref{eqn-u-pDP-OPT}) of Theorem~\ref{thm-pDP-OPT}.

\begin{lem} \label{lem-pDP-OPT-bounds}
$d$ in Eq.~(\ref{eqn-u-pDP-OPT}) is greater  than $\inverfc(2\delta) $ and less than $\inverfc(\delta) $.
\end{lem}

We prove Lemma~\ref{lem-pDP-OPT-bounds} in Appendix~\ref{secprf-lem-pDP-OPT-bounds} of this supplementary file. Theorem~\ref{thm-pDP-OPT} and Lemma~\ref{lem-pDP-OPT-bounds} imply an upper bound of $\sigma_{\texttt{\upshape pDP-OPT}}$ as $\sigma_{\texttt{\upshape Mechanism-3}}$ in Theorem~\ref{thm-Mechanism-3} below, where we present \texttt{\upshape Mechanism~3}  to achieve $(\epsilon,\delta)$-probabilistic differential privacy.

\begin{thm}[\textbf{Gaussian \texttt{\upshape Mechanism~3} for $(\epsilon,\delta)$-Probabilistic differential privacy}] \label{thm-Mechanism-3}
$(\epsilon,\delta)$-Probabilistic differential privacy can be achieved by \texttt{\upshape Mechanism~3}, which adds Gaussian noise with standard deviation $\sigma_{\texttt{\upshape Mechanism-3}}$ to each dimension of a query with $\ell_2$-sensitivity $\Delta$, for $\sigma_{\texttt{\upshape Mechanism-3}}$ given by
\begin{subnumcases}{}
f := \inverfc(\delta) ; \label{eqn-u-Mechanism-3}
\\ \sigma_{\texttt{\upshape Mechanism-3}}    := \frac{\left( f+\sqrt{f^2+\epsilon}\hspace{1.5pt} \right) \cdot \Delta }{\epsilon\sqrt{2}} .  \label{eqn-sigma-Mechanism-3}
\end{subnumcases}
\end{thm}

% \begin{subnumcases}{}
% g := \sqrt{\ln\frac{1}{\delta}} ; \label{eqn-u-Mechanism-3}
% \\ \sigma   := \frac{\left( g+\sqrt{g^2+\epsilon}\hspace{1.5pt} \right) \cdot \Delta }{\epsilon\sqrt{2}} .  \label{eqn-sigma-Mechanism-3}
% \end{subnumcases}

% In \texttt{\upshape Mechanism~3}, the Gaussian noise amount
% $\sigma_{\texttt{\upshape Mechanism-3}}$ is given by

The expression of $\sigma_{\texttt{\upshape Mechanism-3}}$ involves the complementary error function's inverse $\inverfc()$. Hence, we further present Lemma~\ref{lem-f-vs-g} below, which will enable us to propose \texttt{\upshape Mechanism 4}. Its noise amount is given by the closed-form expression of $\sigma_{\texttt{\upshape Mechanism-4}}$ and has only elementary functions.

Lemma~\ref{lem-f-vs-g} upper bounds $f$ in Eq.~(\ref{eqn-u-Mechanism-3}).

\begin{lem} \label{lem-f-vs-g}
$f$ in Eq.~(\ref{eqn-u-Mechanism-3}) is less than $g$ in Eq.~(\ref{eqn-u-Mechanism-4}).
\end{lem}

We prove Lemma~\ref{lem-f-vs-g} in Appendix~\ref{appenlem-erf-2} of this supplementary file.

Theorem~\ref{thm-Mechanism-3}  and Lemma~\ref{lem-f-vs-g} imply an upper bound of $\sigma_{\texttt{\upshape Mechanism-3}}$ as $\sigma_{\texttt{\upshape Mechanism-4}}$ in Theorem~\ref{thm-Mechanism-4} below,  where the presented \texttt{\upshape Mechanism~4} is further simpler   than   \texttt{\upshape Mechanism~3} as noted above.
% The closed-form expression of $\sigma_{\texttt{\upshape Mechanism-4}}$ has only elementary functions, while the expression of $\sigma_{\texttt{\upshape Mechanism-3}}$ involves the complementary error function's inverse $\inverfc()$.

\begin{thm}[\textbf{Gaussian \texttt{\upshape Mechanism~4} for $(\epsilon,\delta)$-Probabilistic differential privacy}] \label{thm-Mechanism-4}
$(\epsilon,\delta)$-Probabilistic differential privacy can be achieved by \texttt{\upshape Mechanism~4}, which adds Gaussian noise with standard deviation $\sigma_{\texttt{\upshape Mechanism-4}}$ to each dimension of a query with $\ell_2$-sensitivity $\Delta$, for $\sigma_{\texttt{\upshape Mechanism-4}}$ given by
\begin{subnumcases}{}
g : = \sqrt{\ln \frac{2}{\sqrt{8\delta+1}-1}}; \label{eqn-u-Mechanism-4}
\\ \sigma_{\texttt{\upshape Mechanism-4}}    := \frac{\left( g+\sqrt{g^2+\epsilon}\hspace{1.5pt} \right) \cdot \Delta  }{\epsilon\sqrt{2}} .  \label{eqn-sigma-Mechanism-4}
\end{subnumcases}
\end{thm}

Table~\ref{table:pDP-mechanisms} summarizes different mechanisms to achieve~$(\epsilon,\delta)$-probabilistic differential privacy discussed above.

\section{Concentrated Differential Privacy and Related Notions} \label{sec-DP-CDP-Related}

Several variants of differential privacy (DP), including mean-concentrated differential privacy (mCDP) \cite{dwork2016concentrated}, zero-concentrated differential privacy (zCDP) \cite{bun2016concentrated}, R\'{e}nyi differential privacy \cite{mironov2017renyi} (RDP), and truncated concentrated differential privacy (tCDP) \cite{bun2018composable} have been recently proposed as alternatives to $(\epsilon, \delta)$-DP. Below we show that achieving $(\epsilon,\delta)$-DP by first ensuring one of these privacy definitions (mCDP, zCDP, RDP, and tCDP) cannot give Gaussian mechanisms better than ours, based on existing results on the relationships between mCDP, zCDP, RDP,  tCDP and DP.

% If the results on the relationships can be improved, such approaches may yield better mechanisms than ours.

%\textbf{Relationship between mCDP and DP.}

\begin{lem}[\textbf{Relationship between $(\epsilon,\delta)$-DP and $(\mu, \tau )$-mCDP}] \label{relationship-mCDP-DP}
For $\epsilon>\mu$, $(\mu, \tau )$-mCDP implies $(\epsilon,\delta)$-probabilistic differential privacy (pDP) for \\ $\delta = \exp\left( - \frac{(\epsilon-\mu)^2}{2\tau^2} \right) + \exp\left( - \frac{(\epsilon+\mu)^2}{2\tau^2} \right)$, which further implies $(\epsilon,\delta)$-DP.
\end{lem}

Despite not being presented in~\cite{dwork2016concentrated} which proposes mCDP, the first part of Lemma~\ref{relationship-mCDP-DP} clearly follows from the definitions of mCDP and pDP by using the tail bounds on the privacy loss random variable of mCDP, while the second part of Lemma~\ref{relationship-mCDP-DP} is from Lemma~\ref{lem-PDP-to-DP}.

For a query with $\ell_2$-sensitivity $1$, Theorem 3.2 in~\cite{dwork2016concentrated} shows that the Gaussian mechanism with standard deviation $\sigma$ achieves $ (\frac{1}{2 \sigma^2}, \frac{1}{ \sigma})$-mCDP, which based on Lemma~\ref{relationship-mCDP-DP}   implies $(\epsilon,\delta)$-pDP for $\delta = \exp\left( - \frac{(\epsilon-\frac{1}{2 \sigma^2})^2}{2(\frac{1}{ \sigma})^2} \right) + \exp\left( - \frac{(\epsilon+\frac{1}{2 \sigma^2})^2}{2(\frac{1}{ \sigma})^2} \right)$. Expressing $\sigma$ in terms of $\epsilon$ and $\delta$ gives $\sigma$ as  $\sigma_{\texttt{\upshape pDP-OPT}}$ of Theorem~\ref{thm-pDP-OPT}. Hence, using the relationship between mCDP and (p)DP does not give a new mechanism which we have not presented.

\begin{figure*}
\centering
    \begin{tabular}{cc}
    \multicolumn{2}{c}{\includegraphics[width=1\textwidth]{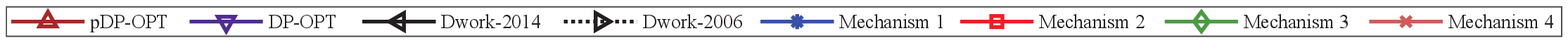}}\\
    \end{tabular}
\begin{minipage}{.62\textwidth}
    \begin{tabular}{ccc}
        \hspace{-16mm}\includegraphics[height=3.4cm]{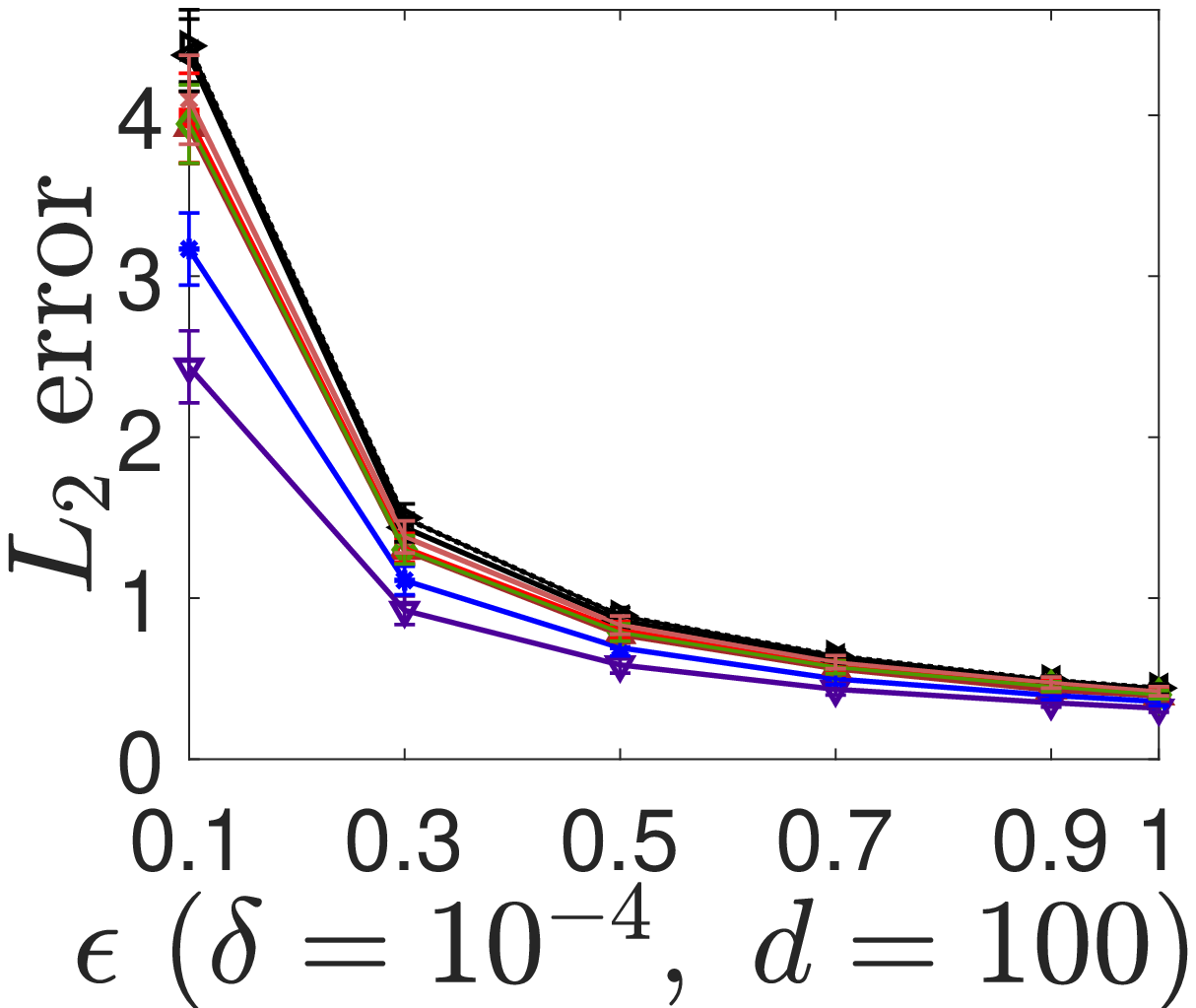} &
        \hspace{-12mm}\includegraphics[height=3.4cm]{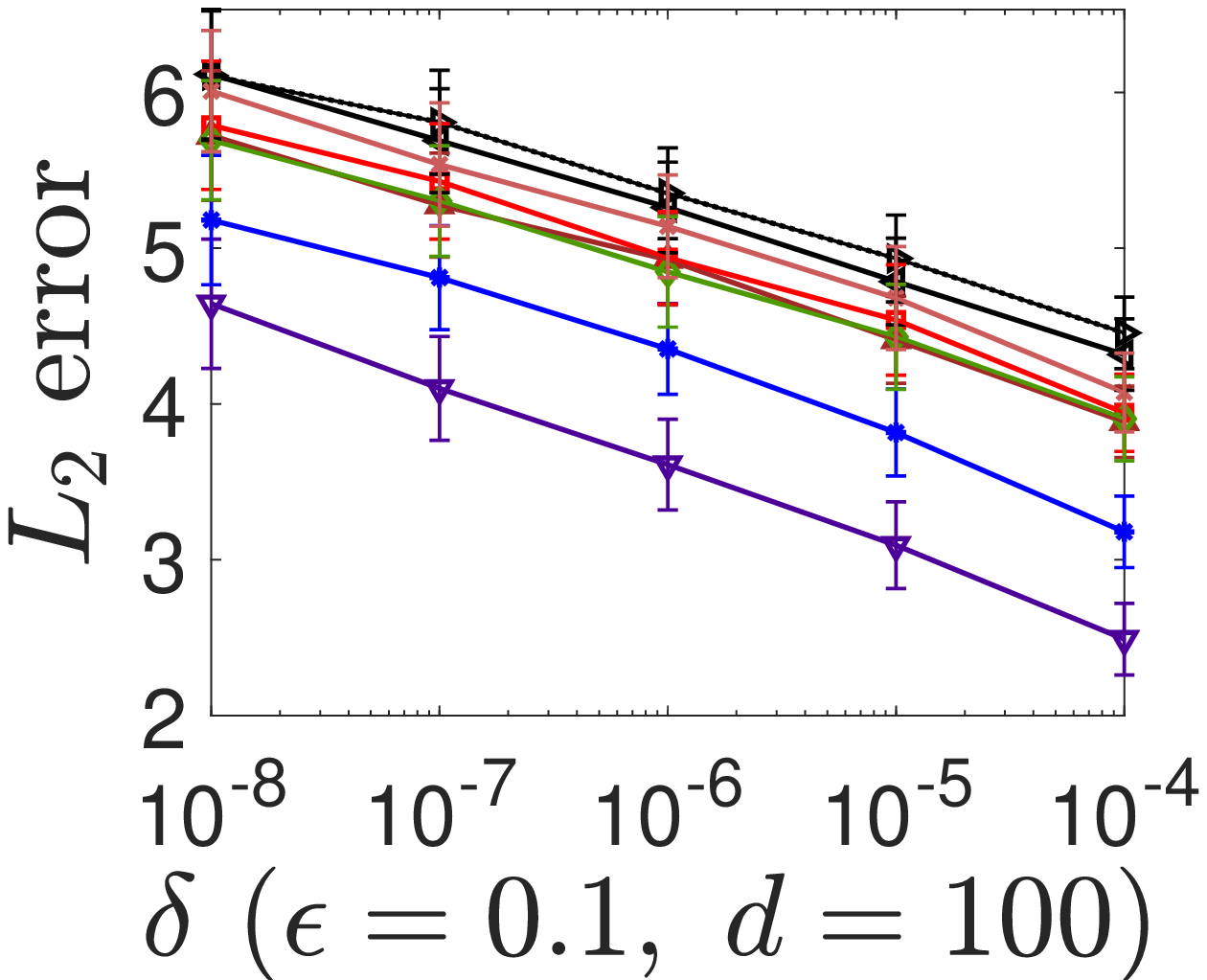} &
        \hspace{-10mm}\includegraphics[height=3.4cm]{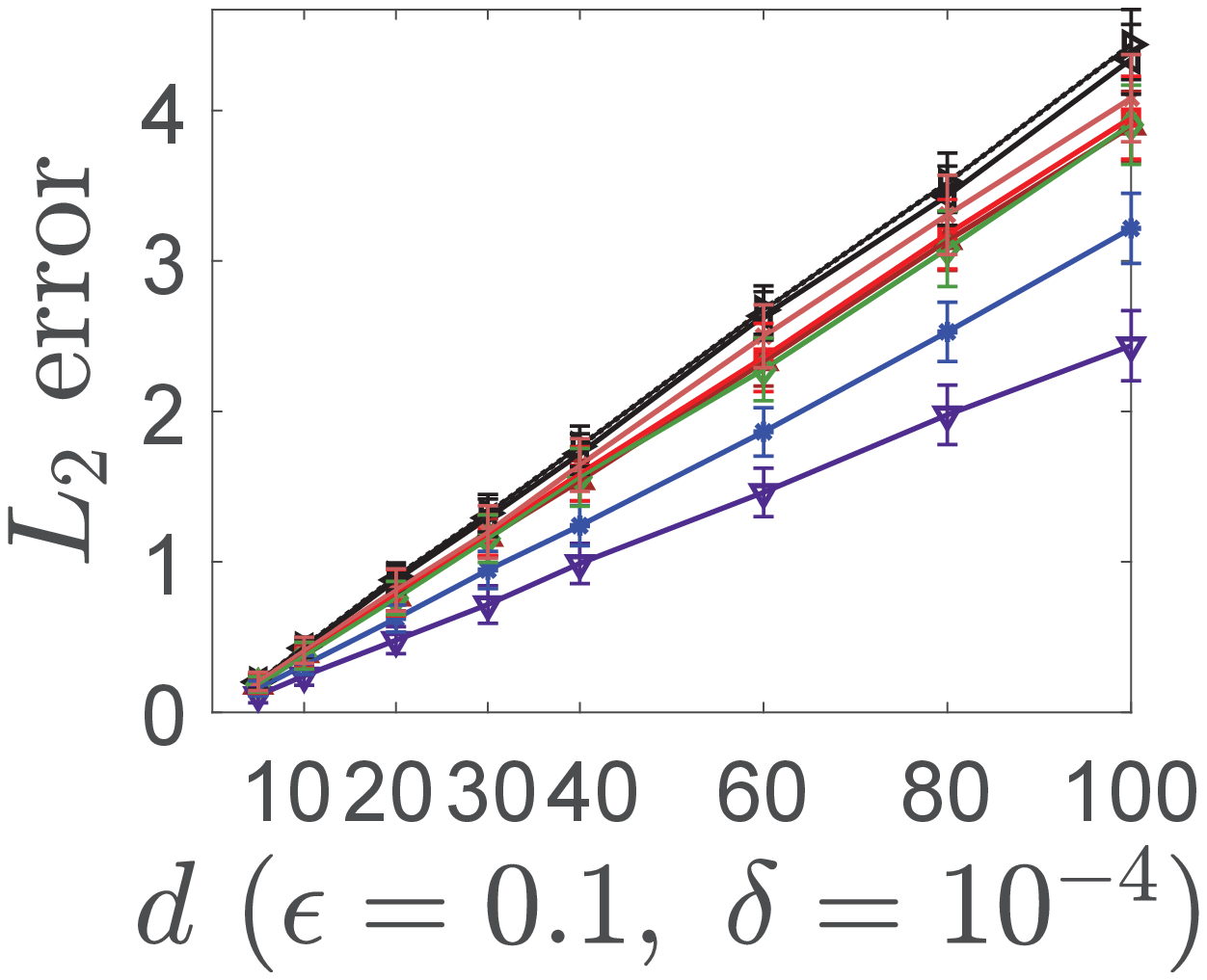} \\[-1pt]
    {\hspace{-16mm}(a) Error w.r.t. $\epsilon$} & {\hspace{-12mm}(b) Error w.r.t. $\delta$}  & {\hspace{-10mm}(c) Error w.r.t. dimension $d$}
    \end{tabular}\vspace{-5pt}
    \caption{Mean estimation. \vspace{-6pt}}\label{meanEstimation}
\end{minipage}
\begin{minipage}[c]{.36\textwidth}
    \begin{tabular}{cc}
        \hspace{-9mm}\includegraphics[height=3.4cm]{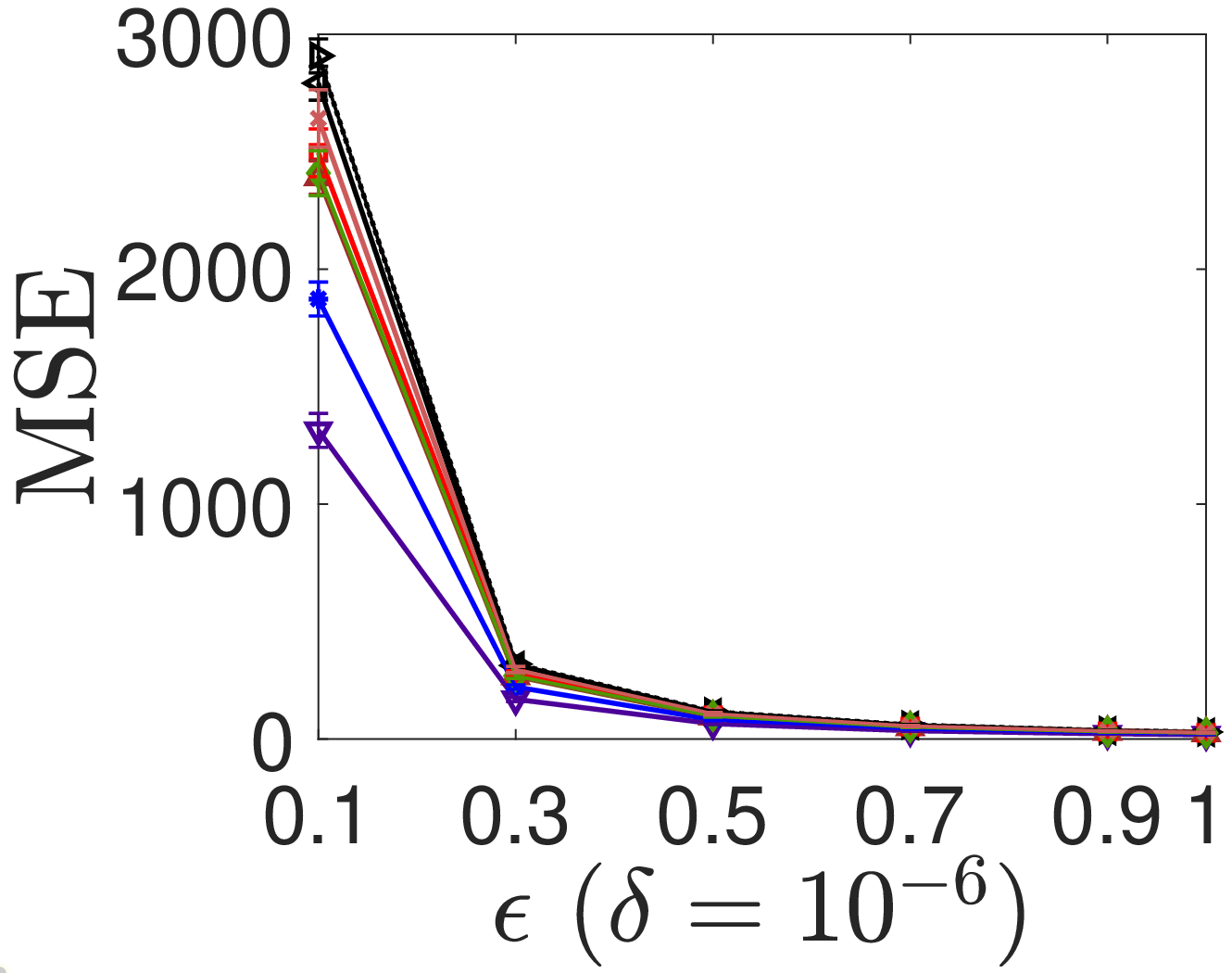} &
        \hspace{-7mm}\includegraphics[height=3.4cm]{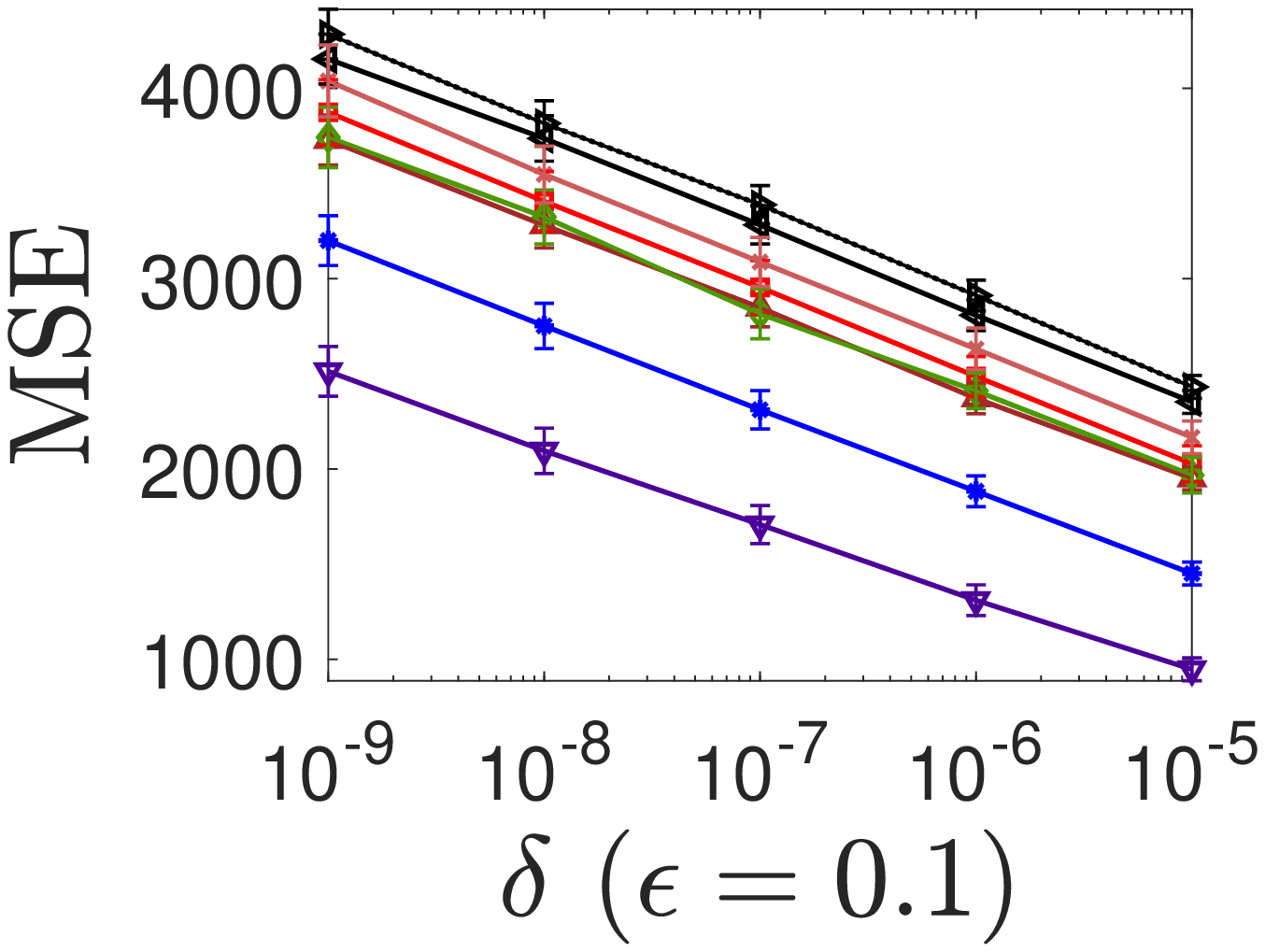}\\[-1pt]
    {\hspace{-6mm}(a) Error w.r.t. $\epsilon$} & {\hspace{-6mm}(b) Error w.r.t. $\delta$}
    \end{tabular}\vspace{-5pt}
    \caption{Histogram estimation. \vspace{-6pt}}\label{histEstimation}
\end{minipage}
\end{figure*}

\textbf{Relationship between zCDP and DP.} From   Proposition 1.3 and Proposition 1.6 in \cite{bun2016concentrated}, $\rho$-zCDP implies ($\epsilon$, $\delta$)-DP for $\epsilon = \rho + 2\sqrt{\rho \ln(\frac{1}{\delta})}$. Moreover, the Gaussian mechanism with standard deviation $\sigma$ achieves $\rho$-zCDP by~\cite{bun2016concentrated}, where $\rho= \frac{\Delta^2}{2\sigma^2}$. Combining these results, we can derive that the Gaussian mechanism with standard deviation $  \frac{\Delta\cdot \left(\sqrt{\ln\frac{1}{\delta}}+\sqrt{\ln\frac{1}{\delta}+\epsilon}\right)}{\sqrt{2}\epsilon} $    achieves \mbox{($\epsilon$, $\delta$)-DP.}
This expression   is obtained by solving $\sigma$ which satisfy $\rho= \frac{\Delta^2}{2\sigma^2}$ and $\epsilon = \rho + 2\sqrt{\rho \ln(\frac{1}{\delta})}$.
Such noise amount is even worse (i.e., higher) than our weakest  \texttt{\upshape Mechanism~4} in Theorem~\ref{thm-Mechanism-4} on  Page~\pageref{thm-Mechanism-4} in view of $\sqrt{8\delta+1}-1 > 2 \delta$ given $0<\delta<1$. Hence, achieving $(\epsilon,\delta)$-DP by first ensuring zCDP cannot give Gaussian mechanisms better than ours.

%We now explain the above formula. Suppose the Gaussian mechanism with standard deviation $\sigma$ achieves $\rho$-zCDP which further implies ($\epsilon$, $\delta$)-DP for $\rho= \frac{\Delta^2}{2\sigma^2}$ $\epsilon = \rho + 2\sqrt{\rho \ln(\frac{1}{\delta})}$

%
%
%
%Besides, we also know that if the Gaussian mechanism $\mathcal{N}(0, \sigma ^2)$ achieves $\rho$-zCDP, then we have $\rho= \frac{\Delta^2}{2\sigma^2}$.

\textbf{Relationship between RDP and DP.} Mironov~\shortcite{mironov2017renyi} shows that ($\alpha$, $\rho\alpha$)-RDP implies ($\epsilon$, $\delta$)-DP for $\epsilon =\rho\alpha+\frac{\ln(1/ \delta )}{\alpha-1}$, and the Gaussian mechanism with standard deviation $\sigma$ achieves ($\alpha, \rho\alpha$)-RDP for $\rho=\frac{\Delta^2}{2\sigma^2}$. Combining these results, we can also prove that  the Gaussian mechanism with standard deviation $  \frac{\Delta\cdot \left(\sqrt{\ln\frac{1}{\delta}}+\sqrt{\ln\frac{1}{\delta}+\epsilon}\right)}{\sqrt{2}\,\epsilon} $  achieves  \mbox{($\epsilon$, $\delta$)-DP}. This expression   is obtained by finding the smallest $\sigma$   such that there exists $\alpha > 1$ such that  $\rho= \frac{\Delta^2}{2\sigma^2}$ and $\epsilon =\rho\alpha+\frac{\ln(1/ \delta )}{\alpha-1}$ (we just express $\sigma$ and take its minimum with respect to $\alpha$). As noted above, this noise amount is even worse (i.e., higher) than our weakest  \texttt{\upshape Mechanism~4} in Theorem~\ref{thm-Mechanism-4}. Thus, achieving $(\epsilon,\delta)$-DP by first ensuring  RDP cannot give Gaussian mechanisms better than ours. We emphasize that the comparison may be different if the RDP paper~\shortcite{mironov2017renyi}'s Proposition~3 that ($\alpha$, $\rho\alpha$)-RDP implies ($\rho\alpha+\frac{\ln(1/ \delta )}{\alpha-1}$, $\delta$)-DP can be improved. Yet, we have not been able to  find such improvement after checking prior papers related to RDP.

% on in Page~\pageref{thm-Mechanism-4} in view of $\sqrt{8\delta+1}-1} > 2 \delta$ given $0<\delta<1$.

\textbf{Relationship between tCDP and DP.} Bun~\emph{et~al.}~\cite{bun2018composable} show that ($\rho,~\omega$)-tCDP implies ($\epsilon$, $\delta$)-DP for $\epsilon=\begin{cases}
\rho+2\sqrt{\rho\ln{\frac{1}{\delta}}}~\textit{if}~\ln\frac{1}{\delta} \leq (\omega-1)^2 \rho, \\
\rho\omega+\frac{\ln(\frac{1}{\delta})}{\omega-1}~\textit{if}~\ln\frac{1}{\delta} \geq (\omega-1)^2 \rho,
\end{cases}$  and the Gaussian mechanism with standard deviation $\sigma$ achieves ($\rho,~\omega$)-tCDP for $\rho=\frac{\Delta^2}{2\sigma^2}$. We can see that these results are already covered by the above discussions for the relationship between zCDP and DP, and for the relationship between RDP and DP. Therefore, achieving $(\epsilon,\delta)$-DP by first ensuring tCDP cannot give Gaussian mechanisms better than ours.

\section{Experiments} \label{sec-main-experiments}

This section presents experiments to evaluate different Gaussian mechanisms for mean estimation and histogram estimation under differential privacy.

% First, we evaluate the performances of a series of mechanisms for the task of private mean estimation using synthetic data. Then, we run the experiments on the Adult dataset to evaluate the histogram estimation using different mechanisms.

%
%Further, we also compare the performances of all mechanisms for the machine learning tasks on Adult dataset. The specific experiments setups and experimental results are presented as follows.

 \subsection{Mean Estimation}

We evaluate the utility of all mechanisms for the task of private mean estimation using synthetic data. The input dataset  $x=(x_{1},\ldots,x_{n})$ contains $n$ vectors $x_{i}\in \mathbb{R}^d$ for a given $d$, and the query for mean computation is $Q(x)=(1/n)\sum_{i=1}^{n}x_i$.  We set $n=1000$ and sample each dataset $x$ in two steps~\cite{liu2018generalized}. The first step is to sample an initial data center $x_0\in \mathbb{R}^d$, with each dimension of  $x_0$ independently following a standard Gaussian distribution with zero mean and variance being $1$.  The second step is to construct $x=(x_{1},\ldots,x_{n})$ with $x_i=x_0+\xi_i $, where each $\xi_i\in \mathbb{R}^d$ is \mbox{independently} and identically distributed (i.i.d.) with independent coordinates sampled uniformly from the interval $[-1/2,1/2]$. We consider bounded differential privacy, where two neighboring datasets have the same size $n$, and have different records at only one of the $n$ positions. Since the points $x_i$ in each dataset all lie in an $\ell_\infty $-ball of radius 1, the $\ell_2$-sensitivity  of mean estimation is $ \sqrt{d}/n$, where $d$ is a   record's dimension.
 %We sample each dataset $x$ the same as in~\cite{liu2018generalized}.
%  Thus, the points $x_i$ in each dataset all lie in an $\ell_\infty $-ball of radius 1, leading to an  $\ell_2$-sensitivity  of $ \sqrt{d}/n$, where $d$ is a data record's dimension.

 For the above query $Q$ on the dataset $x$, we consider different Gaussian mechanisms to achieve $(\epsilon,\delta)$-differential privacy. Let $\widetilde{Q}$ be such a Gaussian mechanism. We report the $\ell_2$ error $\left \| \widetilde{Q}(x)-Q(x) \right \|_2$. The results for  different Gaussian mechanisms are presented in Figure~\ref{meanEstimation}. The plots consider $\epsilon \leq 1$ since this is required by the proofs of \texttt{\upshape Dwork-2006} of~\cite{dwork2006our} and \texttt{\upshape Dwork-2014} of~\cite{dwork2014algorithmic}. Figure~\ref{meanEstimation}-(a) fixes $\delta=10^{-4}$ and varies $\epsilon$; Figure~\ref{meanEstimation}-(b) fixes $\epsilon=0.1$ and varies $\delta$; and Figure~\ref{meanEstimation}-(c) with $\epsilon=0.1$ and $\delta=10^{-4}$ evaluates the impact of a data record's dimension $d$. All subfigures of Figure~\ref{meanEstimation} show that our proposed Gaussian mechanisms achieve better utilities than the classical Gaussian mechanisms~\cite{dwork2006our,dwork2014algorithmic} \texttt{\upshape Dwork-2014} and \texttt{\upshape Dwork-2006}; In fact, \texttt{\upshape Dwork-2014} and \texttt{\upshape Dwork-2006} have the largest   $\ell_2$-errors. Moreover, the utilities of our proposed   mechanisms are  close to that of the optimal yet more computationally expensive Gaussian mechanism  \texttt{\upshape DP-OPT}.

\subsection{Histogram Estimation}

We now run experiments on the Adult dataset from the UCI machine learning repository\footnote{http://archive.ics.uci.edu/ml\label{adult}}, to evaluate different Gaussian mechanisms for histogram estimation with differential privacy. The Adult dataset contains census information with 45222 records and 15 attributes. The attributes include both categorical ones such as race, gender, and education level, as well as numerical ones such as capital gain,  capital loss, and weight. We consider the combination of all categorical attributes and let the histogram query be a vector of the counts. Here we tackle unbounded differential privacy, where a neighboring dataset is obtained by deleting or adding one record, so the sensitivity of the histogram query is $1$.  For different Gaussian mechanisms satisfying  $(\epsilon,\delta)$-differential privacy, we compare their Mean Squared Error (MSE) and plot the results in Figure~\ref{histEstimation}.

% We ignore those bins with $\text{count}=0$ when preprocessing the dataset, thus leading to a histogram with 36800 bins. We aim to publish the private count of each bin with differential privacy via the Gaussian mechanism. Clearly, the  sensitivity $\Delta$ of the query equals $1$. We test different parameters and report the Mean Squared Error (MSE) of each mechanism.

  In Figure~\ref{histEstimation}-(a), we vary $\epsilon$ from $0.1$ to $1.0$ while fixing $\delta=10^{-6}$. In Figure~\ref{histEstimation}-(b), we vary $\delta$ from $10^{-9}$ to $10^{-5}$ while fixing $\epsilon=0.1$. Both subfigures  show that the utilities of our proposed Gaussian mechanisms are higher than those of the classical ones~\cite{dwork2006our,dwork2014algorithmic} and close to that of the optimal yet more computationally expensive \texttt{\upshape DP-OPT} mechanism.

\section{Conclusion}
\label{sec-main-conclusion}
Differential privacy (DP) has received considerable interest recently since it provides a rigorous framework to quantify data privacy.
%  Roughly speaking, a randomized mechanism satisfying $(\epsilon,\delta)$-differential privacy means that except with a small probability~$\delta$,
% altering a record in a dataset cannot change the probability
% that an output is seen by more than a
%   multiplicative factor $e^{\epsilon} $.
Well-known solutions to   $(\epsilon,\delta)$-DP are the Gaussian mechanisms by Dwork~\textit{et~al.}~\cite{dwork2006our} in 2006 and by Dwork and Roth~\cite{dwork2014algorithmic} in 2014, where a certain amount of Gaussian noise is added independently to each dimension of the query result. Although the two classical Gaussian mechanisms~\cite{dwork2006our,dwork2014algorithmic} explicitly state their usage for $\epsilon \leq 1$ only,   many studies applying them neglect the constraint on $\epsilon$, rendering the obtained results inaccurate. In this paper, for $(\epsilon,\delta)$-DP, we present  Gaussian mechanisms which    work for {every} $\epsilon$. Another improvement is that our     mechanisms   achieve   {higher utilities} than those of the   classical ones~\cite{dwork2006our,dwork2014algorithmic}. Since most mechanisms proposed in the literature for \mbox{$(\epsilon,\delta)$-DP} are obtained by ensuring a condition called $(\epsilon,\delta)$-probabilistic differential privacy (pDP), we also present  the difference/relationship between $(\epsilon,\delta)$-DP and $(\epsilon,\delta)$-pDP, and Gaussian mechanisms for $(\epsilon,\delta)$-pDP.
%  We further apply our Gaussian mechanisms to existing differentially private algorithms for deep learning and hypothesis testing, respectively, and show that the incorporation of our Gaussian mechanisms improves performance.
Our research on reviewing and improving the Gaussian mechanisms will benefit differential privacy applications built based on the primitive.

\small
%\small

%\bibliographystyle{unsrt}
%
%\bibliographystyle{IEEEtran}
%\bibliography{related2}


\normalsize

~\vspace{-20pt}

\appendix
%\begin{appendices}

%  \section*{Appendices}

The appendices   are organized as follows. Appendices~\ref{sec-Mechanism-2-Dwork-2014} and~\ref{secprf-thm-Dwork-2014-not-work} are also provided in the submission, while other appendices are given in this submitted supplementary file (the same as~\cite{zhao2019DP}).
\begin{itemize}
\item Appendix~\ref{sec-Mechanism-2-Dwork-2014} presents the proof of\\ \mbox{$\sigma_{\texttt{\upshape Mechanism-1}} <\sigma_{\texttt{\upshape Mechanism-2}} < \sigma_{\texttt{\upshape Dwork-2014}}< \sigma_{\texttt{\upshape Dwork-2006}}$.}
\item Appendix~\ref{secprf-thm-Dwork-2014-not-work} presents the proof of Theorem~\ref{thm-Dwork-2014-not-work}.
\item Appendix~\ref{appsecprf-DP-OPT}  presents the proof of Theorem~\ref{thm-DP-OPT}.
\item Appendix~\ref{subsection-rem-Mechanism-DP-OPT-ru-detailed}  presents more discussions about Remark~\ref{rem-Mechanism-DP-OPT-ru} of Page~\pageref{rem-Mechanism-DP-OPT-ru}.
\item Appendix~\ref{secprf-thm-DP-OPT-asymptotics}  presents the proof of Theorem~\ref{thm-DP-OPT-asymptotics}.
\item Appendix~\ref{Appendix-lem-DP-OPT-bounds}  presents the proof of Lemma~\ref{lem-DP-OPT-bounds}.
\item Appendix~\ref{sec-prf-thm-Mechanism-1-based-on-thm-DP-OPT}  proves Lemma~\ref{lem-a-vs-b}, which along with Theorem~\ref{thm-DP-OPT} implies Theorem~\ref{thm-Mechanism-1}.
% \item Appendix~\ref{sec-prf-rem-Mechanism-1}  presents the proof of Remark~\ref{rem-Mechanism-1} based on Theorem~\ref{thm-Mechanism-1}.
\item Appendix~\ref{appenlem-lem-b-vs-c}  proves Lemma~\ref{lem-b-vs-c}, which along with Theorem~\ref{thm-Mechanism-1} implies Theorem~\ref{thm-Mechanism-2}.
\item Appendix~\ref{secprf-lem-PDP-to-DP}  presents the proof of Lemma~\ref{lem-PDP-to-DP}.
\item Appendix~\ref{secprf-lem-DP-to-PDP}  presents the proof of Lemma~\ref{lem-DP-to-PDP}.
\item Appendix~\ref{sec:proof3}  presents the proof of Theorem~\ref{thm-pDP-OPT}.
\item Appendix~\ref{secprf-thm-pDP-OPT-asymptotics}  presents the proof of Theorem~\ref{thm-pDP-OPT-asymptotics}.
\item Appendix~\ref{secprf-lem-pDP-OPT-bounds}  proves Lemma~\ref{lem-pDP-OPT-bounds}, which along with Theorem~\ref{thm-pDP-OPT} implies Theorem~\ref{thm-Mechanism-3}.
\item Appendix~\ref{appenlem-erf-2}  proves Lemma~\ref{lem-f-vs-g}, which along with Theorem~\ref{thm-Mechanism-3} implies Theorem~\ref{thm-Mechanism-4}.
\item Appendix~\ref{appendix-sec-Alg-opt}  presents Algorithm~\ref{Alg-opt} to compute $\sigma_{\texttt{\upshape DP-OPT}}$ of Theorem~\ref{thm-DP-OPT}.
    \item   Appendix~\ref{sec-main-composition}   provides  analyses for the composition of Gaussian mechanisms to achieve $(\epsilon,\delta)$-DP or $(\epsilon,\delta)$-pDP.
        \item Appendix~\ref{sec-lem-inverfc}  shows Lemma~\ref{lem-inverfc}, which is used in  the proofs of Lemmas~\ref{lem-b-vs-c} and~\ref{lem-f-vs-g}.
\end{itemize}

% \section{Proofs of Our Theorems and Lemmas} \label{sec-main-proof}

% In this section?, we present the proofs  for theorems and lemmas presented in earlier parts of this paper.

\subsection{Proving $\sigma_{\texttt{\upshape Mechanism-1}} <\sigma_{\texttt{\upshape Mechanism-2}} < \sigma_{\texttt{\upshape Dwork-2014}}< \sigma_{\texttt{\upshape Dwork-2006}}$} \label{sec-Mechanism-2-Dwork-2014}

To prove $\sigma_{\texttt{\upshape Mechanism-1}} <\sigma_{\texttt{\upshape Mechanism-2}} < \sigma_{\texttt{\upshape Dwork-2014}}< \sigma_{\texttt{\upshape Dwork-2006}}$, from Inequalities~(\ref{eq-res0}) and~(\ref{eq-res2}), we just need to establish $\sigma_{\texttt{\upshape Mechanism-2}} < \sigma_{\texttt{\upshape Dwork-2014}}$. Recalling Eq.~(\ref{dwork-2014}) and~(\ref{eqn-u-Mechanism-2}),
 we will prove
% \begin{align}
% & \sqrt{2\ln \frac{1.25}{\delta}} \nonumber \\ &   \stackrel{\text{(a)}}{>} \Big({\sqrt{\ln \frac{2}{\sqrt{16\delta+1}-1} +\epsilon} + \sqrt{\ln \frac{2}{\sqrt{16\delta+1}-1}}}\hspace{1pt}\Big)\hspace{-1pt}\Big/\hspace{-1pt}{\sqrt{2}} \nonumber \\ &   \stackrel{\text{(b)}}{>} \Big({\sqrt{\left[ \inverfc(\delta)\right]^2 +\epsilon\hspace{2pt}} \hspace{3pt}+\hspace{3pt} \inverfc(\delta)}\Big)\Big/{\sqrt{2}} \nonumber \\ &   \stackrel{\text{(c)}}{>} ({\sqrt{\lambda^2 +\epsilon\hspace{2pt}} \hspace{3pt}+\hspace{3pt} \lambda})/{\sqrt{2}} \text{ for $\lambda$ in (\ref{eqsec-lambdaexpr})}.
% \end{align}
\begin{align}
& \sqrt{2\ln \frac{1.25}{\delta}}   \nonumber \\& > \bigg({\sqrt{\ln \frac{2}{\sqrt{16\delta+1}-1} +\epsilon} \hspace{1pt}+\hspace{1pt} \sqrt{\ln \frac{2}{\sqrt{16\delta+1}-1}}} \hspace{2pt}\bigg)\bigg/{\sqrt{2}} ,  \nonumber \\ & \hspace{10pt} \text{for $\epsilon \leq 1$ and $0<\delta < 0.5$}  .\label{eqsec-DPeqn0}
\end{align}

Since the term after ``$>$'' in Inequality~(\ref{eqsec-DPeqn0}) is increasing with respect to $\epsilon$, we can just let $\epsilon $ be $ 1$ in Inequality~(\ref{eqsec-DPeqn0}). Hence, we will obtain Inequality~(\ref{eqsec-DPeqn0})  once proving
\begin{align}
& \sqrt{2\ln \frac{1.25}{\delta}}    \nonumber \\& > \bigg({\sqrt{\ln \frac{2}{\sqrt{16\delta+1}-1} +1} \hspace{1pt}+\hspace{1pt} \sqrt{\ln \frac{2}{\sqrt{16\delta+1}-1}}}\hspace{1pt} \hspace{2pt}\bigg)\bigg/{\sqrt{2}} ,  \nonumber \\ & \hspace{10pt} \text{for $0<\delta < 0.5$}  .\label{eqsec-DPeqn1}
\end{align}
With $a$ denoting $\sqrt{\ln\frac{1.25}{\delta}}$ and $b$ denoting $\sqrt{\ln \frac{2}{\sqrt{16\delta+1}-1}}$, then Inequality~(\ref{eqsec-DPeqn1}) means $\sqrt{2}\,a > (\sqrt{b^2+1}+\sqrt{b}\,)/\sqrt{2}$, which is equivalent to $b < a - \frac{0.25}{a}$ since setting $b$ as $a - \frac{0.25}{a}$ will let $ (\sqrt{b^2+1}+\sqrt{b}\,)/\sqrt{2}$ be $\sqrt{2}\,a$ exactly (note that $a - \frac{0.25}{a}>0$ clearly holds for $0<\delta < 0.5$). Hence,
the desired result Inequality~(\ref{eqsec-DPeqn1}) is equivalent to
\begin{align}
&\sqrt{\ln \frac{2}{\sqrt{16\delta+1}-1}}  < \sqrt{\ln\frac{1.25}{\delta}}-\frac{0.25}{\sqrt{\ln\frac{1.25}{\delta}}},   \hspace{10pt}  \nonumber \\& \text{for $0<\delta < 0.5$}, \label{eqsec-DPeqn2equiv}
\end{align}
which clearly is implied by the following after taking the square on both sides:
\begin{align}
 & {\ln \frac{2}{\sqrt{16\delta+1}-1}}< \ln\frac{1.25}{\delta} - 0.5,   \text{ for $0<\delta < 0.5$}.\label{eqsec-DPeqn2}
\end{align}
Due to {$ 1.25 \times \exp(-0.5)  \approx 0.7582 > 0.75$}, Inequality~(\ref{eqsec-DPeqn2}) is implied by
\begin{align}
&   {\frac{2}{\sqrt{16\delta+1}-1}} < \frac{0.75}{\delta}   \text{ for $0<\delta < 0.5$}.\label{eqsec-DPeqn7}
\end{align}
We define $f(\delta):={\frac{2}{\sqrt{16\delta+1}-1}} - \frac{0.75}{\delta}  $. Taking the derivative of $f(\delta)$ with respect to $\delta$, we obtain
\begin{align}
f'(\delta)&= -\frac{16}{(\sqrt{16\delta+1}-1)^2\sqrt{16\delta+1}} + \frac{3}{4\delta^2} \nonumber \\& = {\frac{5\sqrt{16\delta+1}-(8\delta+1)}{8\delta^2\sqrt{16\delta+1}}}\nonumber \\& = {\frac{25(16\delta+1)-(8\delta+1)^2}{8[5\sqrt{16\delta+1}+(8\delta+1)]\delta^2\sqrt{16\delta+1}}} \nonumber \\& = {\frac{64\delta(1-\delta)+320\delta+24}{8[5\sqrt{16\delta+1}+(8\delta+1)]\delta^2\sqrt{16\delta+1}}}  \nonumber \\& >0, \text{~for $0<\delta < 0.5$}. \label{eqsec-derivativef}
\end{align}
Hence, $f(\delta) $ is strictly increasing for $0<\delta < 0.5$, resulting in $f(\delta)<f(0.5)={\frac{2}{\sqrt{16\delta+1}-1}} - \frac{0.75}{\delta} =-0.5<0  $, so that Inequality~(\ref{eqsec-DPeqn7}) is proved. Then following the explanation above, we complete establishing $\sigma_{\texttt{\upshape Mechanism-2}} < \sigma_{\texttt{\upshape Dwork-2014}}$. \qeda

%  \subsection{Proof of Remark~\ref{thm-rem-Mechanism-6}} \label{secprf-thm-rem-Mechanism-6}

%
%  \begin{figure}
%  \includegraphics[width=\linewidth]{figures/remark6.eps}
%  \caption{The noise amounts of UpperBound and Mechanism---3.}
%  \label{fig:remark4}
%\end{figure}

%

\subsection{Proof of Theorem~\ref{thm-Dwork-2014-not-work}} \label{secprf-thm-Dwork-2014-not-work}

From Theorem~\ref{thm-DP-OPT}, $\sigma_{\texttt{\upshape DP-OPT}}$ is the minimal required amount of Gaussian noise to achieve $(\epsilon,\delta)$-differential privacy. Hence, to show that the Gaussian noise amount $F(\delta) \times {\Delta}/{\epsilon}$ is not sufficient for $(\epsilon,\delta)$-differential privacy, we will prove that for any $0<\delta<1$,  there exists a positive function $G(\delta)$ such that for any $\epsilon > G(\delta)$, we have
% $F(\delta) \times \frac{\Delta}{\epsilon} < \sigma_{\texttt{\upshape DP-OPT}}$.
\begin{align}
& F(\delta) \times {\Delta}/{\epsilon} < \sigma_{\texttt{\upshape DP-OPT}} \label{general-vs-DP-OPT}.
\end{align}

 We can show that the function $x+\sqrt{x^2+\epsilon}$   strictly increases as $x$ increases for $x \in (-\infty, \infty)$ by noting its derivative $1+ \frac{x}{\sqrt{x^2+\epsilon}} $ is positive. Also, $\lim_{x \to -\infty} (x+\sqrt{x^2+\epsilon}) = \lim_{x \to -\infty} \frac{\epsilon}{-x+\sqrt{x^2+\epsilon}} = 0 $ and $\lim_{x \to \infty} (x+\sqrt{x^2+\epsilon}) = \infty$. Hence, the values that $x+\sqrt{x^2+\epsilon}$ for $x \in (-\infty, \infty)$ can take constitutes the open interval $(0,\infty)$. Then due to
 $F(\delta)>0$, we can define $h$ such that
 \begin{align}
F(\delta) = \frac{h+\sqrt{h^2+\epsilon}  }{\sqrt{2}}. \label{defhAdelta}
\end{align}
From Eq.~(\ref{defhAdelta}) and $\sigma_{\texttt{\upshape DP-OPT}}  = \frac{\left(a+\sqrt{a^2+\epsilon} \hspace{1.5pt}\right)  \cdot \Delta }{\epsilon\sqrt{2}}$ of (\ref{eqn-sigma-DP-OPT}), clearly Inequality~(\ref{general-vs-DP-OPT}) is equivalent to $  \frac{h+\sqrt{h^2+\epsilon}  }{\sqrt{2}} <   \frac{a+\sqrt{a^2+\epsilon}  }{\sqrt{2}}$ and further equivalent to $h<a$.

 As shown in Appendix~\ref{subsection-rem-Mechanism-DP-OPT-ru-detailed}, \mbox{$r(u) : = \erfc\left(u \right)   -  e^{\epsilon} \erfc\left( \sqrt{u^2 + \epsilon} \right) $}    strictly decreases as $u$ increases for $u \in (-\infty, \infty)$. Then $h<a$ is equivalent to $  r(h) >   r(a)$. We will prove $\lim_{\epsilon \to \infty} r(h)  = 2 $, which along with $r(a) = 2 \delta$ in~Eq.~(\ref{eqn-sigma-DP-OPT}) implies that for any $0<\delta<1$, there exists a positive function $G(\delta)$ such that for any $\epsilon  > G(\delta)$, we have $  r(h) >   r(a)$ and thus $h<a$.

 From the above discussion, the desired result~Eq.~(\ref{general-vs-DP-OPT}) follows once we show $\lim_{\epsilon \to \infty} r(h)  = 2 $. From~Eq.~(\ref{defhAdelta}), it holds that  $ h =  {\frac{F(\delta)}{\sqrt{2}}} - \frac{\epsilon}{F(\delta)\cdot 2 \sqrt{2} }  $. Hence, for any $\epsilon  \geq 4 \times [F(\delta)]^2 $, we have $ h \leq - \frac{\epsilon}{4 \sqrt{2} \cdot F(\delta)  } $, which implies
\begin{align}
& e^{\epsilon} \erfc\left( \sqrt{h^2 + \epsilon} \right)  \nonumber
\\ &\leq e^{\epsilon} \erfc\left( |h| \right)  \nonumber
\\ &\leq e^{\epsilon} \erfc\left( \frac{\epsilon}{4 \sqrt{2} \cdot F(\delta)  }  \right) \nonumber
\\ & \leq e^{\epsilon} \times  \exp\left( -\left( \frac{\epsilon}{4 \sqrt{2} \cdot F(\delta)  }  \right)^2 \right)  \nonumber
\\ & \to 0, \text{~as~} \epsilon \to \infty ,  \label{eq-espdilon-h}
\end{align}
where the last ``$\leq$'' uses $\erfc\left(x \right) \leq \exp\left( -x^2 \right)  $ for $x > 0$. The above result~Eq.~(\ref{eq-espdilon-h}) implies $\lim_{\epsilon \to \infty} [e^{\epsilon} \erfc\left( \sqrt{h^2 + \epsilon} \right)] = 0 $. Combining this and $\lim_{\epsilon \to \infty} \erfc\left(h \right)  = 2 $, we derive $\lim_{\epsilon \to \infty} r(h)  = 2 $. Then as already explained, the desired result is proved. \qeda

\newpage

\subsection{Proof of Theorem~\ref{thm-DP-OPT}} \label{appsecprf-DP-OPT}

\noindent\textbf{Proving Theorem~\ref{thm-DP-OPT}'s Property (i):}

 The optimal Gaussian mechanism for $(\epsilon,\delta)$-differential privacy, denoted by Mechanism \texttt{\upshape DP-OPT}, adds Gaussian noise with standard deviation $\sigma_{\texttt{\upshape DP-OPT}}$ to each dimension of a query with $\ell_2$-sensitivity $\Delta$, for $\sigma_{\texttt{\upshape DP-OPT}}$ obtained by Theorem 8 of Balle and Wang~\shortcite{balle2018improving} to satisfy
\begin{align}
& \Phi\left( \frac{\Delta}{2 \sigma_{\texttt{\upshape DP-OPT}} } - \frac{\epsilon \sigma_{\texttt{\upshape DP-OPT}} }{\Delta} \right)  \nonumber \\ & \quad  - e^{\epsilon}  \Phi\left( -\frac{\Delta}{2 \sigma_{\texttt{\upshape DP-OPT}} } - \frac{\epsilon \sigma_{\texttt{\upshape DP-OPT}} }{\Delta} \right)  = \delta, \label{eqn-Phi-DP-OPT}
\end{align}
where $\Phi\left( \cdot \right)$ denotes the cumulative distribution function of the standard
univariate Gaussian probability distribution with mean $0$ and variance $1$.

We define
\begin{align}   \label{DP-eq}
a:=\frac{1}{\sqrt{2}} \big( \frac{\epsilon \sigma_{\texttt{\upshape DP-OPT}} }{\Delta} - \frac{\Delta}{2 \sigma_{\texttt{\upshape DP-OPT}} } \big)  .
\end{align}
Then $\sigma_{\texttt{\upshape DP-OPT}}$ equals $ \frac{\left(a+\sqrt{a^2+\epsilon} \hspace{1.5pt}\right) \cdot \Delta }{\epsilon\sqrt{2}}$, as given by Eq.~(\ref{eqn-sigma-DP-OPT}). Also, \mbox{$\frac{1}{\sqrt{2}}\big( -\frac{\Delta}{2 \sigma_{\texttt{\upshape DP-OPT}} } - \frac{\epsilon \sigma_{\texttt{\upshape DP-OPT}} }{\Delta} \big)    $}  in Eq.~(\ref{eqn-Phi-DP-OPT}) equals $- \sqrt{a^2 + \epsilon}$, since \mbox{$ \frac{1}{2}\left(-\frac{\Delta}{2 \sigma_{\texttt{\upshape DP-OPT}} } -\frac{\epsilon \sigma_{\texttt{\upshape DP-OPT}} }{\Delta}\right)^2-  \frac{1}{2}\left(\frac{\epsilon \sigma_{\texttt{\upshape DP-OPT}} }{\Delta} - \frac{\Delta}{2 \sigma_{\texttt{\upshape DP-OPT}} }\right)^2=\epsilon$.} Thus, Eq.~(\ref{eqn-Phi-DP-OPT}) becomes
\begin{align}   \label{DP-eq}
\Phi\left( -a\sqrt{2} \right)
  - e^{\epsilon}  \Phi\left(- \sqrt{2(a^2 + \epsilon)} \right) =\delta.
\end{align}
 Given
  \begin{align}   \label{DP-eq}
\Phi\left( -a\sqrt{2} \right)  &= \frac{1}{2} +  \frac{1}{2} \erf\left(-a \right)  \nonumber \\ &= \frac{1}{2} -  \frac{1}{2} \erf\left(a \right) \nonumber \\ & =  \frac{1}{2} \erfc\left(a \right)
\end{align}
 and
  \begin{align}   \label{DP-eq}
 &\Phi\left(- \sqrt{2(a^2 + \epsilon)} \right)  \nonumber \\ &= \frac{1}{2} +  \frac{1}{2} \erf\left(- \sqrt{a^2 + \epsilon}\right)  \nonumber \\ & = \frac{1}{2} -  \frac{1}{2} \erf\left( \sqrt{a^2 + \epsilon}\right)   \nonumber \\ &=  \frac{1}{2}\erfc\left( \sqrt{a^2 + \epsilon}\right).
\end{align}   Then we write Eq.~(\ref{eqn-Phi-DP-OPT}) as   \mbox{$\frac{1}{2} \erfc\left(a \right)
  - e^{\epsilon} \cdot  \frac{1}{2} \erfc\left( \sqrt{a^2 + \epsilon}\right)   = \delta$,} so $a$ is given by Eq.~(\ref{eqn-sigma-DP-OPT}).

\noindent\textbf{Proving Theorem~\ref{thm-DP-OPT}'s Property (ii):}

For $\epsilon \geq 0.01$ and $0<\delta \leq 0.05$, we know from Appendix~\ref{subsection-rem-Mechanism-DP-OPT-ru-detailed} to be presented soon that $  1 - e^{\epsilon} \erfc\left( \sqrt{ \epsilon} \right) > 2 \delta  $ and   $a > 0$. Using this in $\sigma_{\texttt{\upshape DP-OPT}}  := \frac{\left(a+\sqrt{a^2+\epsilon} \hspace{1.5pt}\right)  \cdot \Delta }{\epsilon\sqrt{2}}$, we clearly have $\sigma_{\texttt{\upshape DP-OPT}} >  \frac{\Delta}{\sqrt{2\epsilon\,}} $.

\noindent\textbf{Proving Theorem~\ref{thm-DP-OPT}'s Property (iii):}

From Theorem~\ref{thm-DP-OPT}'s Property (ii) proved above, $a   > 0$. Given \mbox{$\erfc\left(a \right)   -  e^{\epsilon} \erfc\left( \sqrt{a^2 + \epsilon} \right)  =  2 \delta$,}  we use $\erfc\left(a \right) > 2 \delta$ to derive $a< \inverfc(2\delta)   < \sqrt{\ln \frac{1}{2\delta}}$, where the last step uses $0<\delta \leq 0.05$ and Proposition~\ref{prop-erfc-vs-ln} below.
%  Moreover, from Appendix~\ref{subsection-rem-Mechanism-DP-OPT-ru-detailed}, it holds that $ 2 \delta \leq 0.01 < 1 - e \cdot \erfc\left(1 \right) \leq 1 - e^{\epsilon} \erfc\left( \sqrt{ \epsilon} \right) $, which implies $a > 0$.
 Then we have  \begin{align}
 \sigma_{\texttt{\upshape DP-OPT}}  : & = \frac{\left(a+\sqrt{a^2+\epsilon} \hspace{1.5pt}\right)  \cdot \Delta }{\epsilon\sqrt{2}} \nonumber  \\ &  <  \frac{\left(a+\sqrt{(a+\sqrt{\epsilon})^2} \hspace{1.5pt}\right)  \cdot \Delta }{\epsilon\sqrt{2}} \nonumber  \\ & = \frac{\left(2a+\sqrt{\epsilon} \hspace{1.5pt}\right)  \cdot \Delta }{\epsilon\sqrt{2}} \nonumber  \\ & <  \sqrt{2\ln \frac{1}{2\delta} }\cdot\frac{\Delta}{\epsilon} + \frac{\Delta}{\sqrt{2\epsilon\,}}
\end{align}
% , where ?; i.e., $\sqrt{\ln \frac{1}{2\delta}}$ is an upper bound for $a$.
 \qeda

\begin{prop} \label{prop-erfc-vs-ln}
$\inverfc(x) < \sqrt{\ln \frac{1}{x}}$ for $0<x <1$.
\end{prop}

\noindent \textbf{Proof of Proposition~\ref{prop-erfc-vs-ln}.}~The desired result $\inverfc(x) < \sqrt{\ln \frac{1}{x}}$  follows from
 %our submitted supplementary file's
  Lemma~\ref{lem-inverfc} (i.e., $\inverfc(x) < \sqrt{\ln \frac{2}{\sqrt{8x+1}-1}}  $ for $0<x <1$) and the obvious inequality $ \sqrt{8x+1}-1 > 2x $ for $0<x <1$.\qeda

\begin{lem} \label{lem-inverfc}
For $0<y<1$, it holds that $\inverfc(y)<\sqrt{\ln \frac{2}{\sqrt{8y+1}-1}} $.
\end{lem}

We defer the proof of Lemma~\ref{lem-inverfc} to the end (i.e.,
Appendix~\ref{sec-lem-inverfc}).

% \begin{lem} The following bounds hold for $a$ in Eq.~(\ref{eqn-sigma-DP-OPT}) and the optimal Gaussian noise amount $\sigma_{\texttt{\upshape DP-OPT}}$ in Eq.~(\ref{eqn-sigma-DP-OPT}).
% \end{lem}

%For simplicity, we consider one-dimensional real-valued queries in the proofs. Due to space limitation, the extension to higher-dimensional real-valued queries is presented in the full version~\cite{fullver}.

%In Section \ref{secompareutility}, we prove Eq.~(\ref{comparenoise0}); i.e., Utility of the mechanism by~Dwork and Roth~\cite{dwork2014algorithmic}    $<$ Utility of \texttt{\upshape Mechanism~1} \mbox{$<$ Utility of \texttt{\upshape Mechanism~2}} $<$  Utility of \texttt{\upshape Mechanism~3}. Sections \ref{seceps0eps1eq4}, \ref{seceps0eps2eq4}, and \ref{seceps0eps3eq4} present the proofs of Eq.~(\ref{eps0eps1eq4}), (\ref{eps0eps2eq4}), and~(\ref{eps0eps3eq4}), respectively.

\subsection{More discussions about Remark~\ref{rem-Mechanism-DP-OPT-ru} of Page~\pageref{rem-Mechanism-DP-OPT-ru}}\label{subsection-rem-Mechanism-DP-OPT-ru-detailed}

%\begin{rem}  \label{rem-Mechanism-DP-OPT-ru-detailed}
With $r(u) : = \erfc\left(u \right)   -  e^{\epsilon} \erfc\left( \sqrt{u^2 + \epsilon} \right) $, the term $a$ in Eq.~(\ref{eqn-sigma-DP-OPT}) satisfies $r(a) =   2 \delta  $.
We know that $r(u)$ strictly decreases as $u$ increases for $u \in (-\infty, \infty)$ in view of the derivative $r'(u)  =  \frac{2}{\sqrt{\pi}} \exp(-u^2) \cdot \frac{u - \sqrt{u^2 + \epsilon}  }{\sqrt{u^2 + \epsilon}} < 0 $. Moreover, we now show $r(0) = 1 - e^{\epsilon} \erfc\left( \sqrt{ \epsilon} \right) > 0 $. With $s(\epsilon) : =  e^{\epsilon} \erfc\left( \sqrt{ \epsilon} \right) $, we know from Lemma~\ref{lemma-subsection-rem-Mechanism-DP-OPT-ru-detailed} below that $s(\epsilon)$ strictly \mbox{decreases} as $\epsilon$ increases for $\epsilon > 0$.  The above analysis induces $r(0) = 1 - s(\epsilon) > 1 - s(0) = 0 $.

% Also, from $\erfc\left(\sqrt{ \epsilon}\right) < \frac{\exp(-\epsilon)}{\sqrt{\pi \epsilon\hspace{2pt}}} $ above, it is clear that $\lim_{u \to \infty}r(u) = 0$.

% $\erfc\left(x \right) < \exp\left( -x^2 \right)  $ for $x > 0$.

Summarizing the above results, $r(a) =   2 \delta  $, \mbox{$r(0) = 1 - e^{\epsilon} \erfc\left( \sqrt{ \epsilon} \right) > 0$,}
%  $\lim_{u \to \infty}r(u) = 0$, and $\lim_{u \to -\infty}r(u) = 2 $,
 we define $\epsilon_*$ as the solution to $1 - e^{\epsilon_*} \erfc\left( \sqrt{ \epsilon_*} \right)  =  2 \delta$ ($\epsilon_*$ exists for $0<\delta<0.5$ from Lemma~\ref{lemma-subsection-rem-Mechanism-DP-OPT-ru-detailed} below), and have the following results for $a$ in Eq.~(\ref{eqn-sigma-DP-OPT}), where ``iff'' is short for ``if and only if'':
\begin{enumerate}%[itemindent=2em]
    \item \mbox{$a\hspace{-1.5pt}>\hspace{-1.5pt} 0$ iff $ 1\hspace{-1.5pt}  -\hspace{-1.5pt}   e^{\epsilon} \erfc\left( \sqrt{ \epsilon} \right)  \hspace{-1.5pt} >\hspace{-1.5pt}  2 \delta $ (i.e., iff $\epsilon\hspace{-1.5pt}>\hspace{-1.5pt}\epsilon_*$ when $\epsilon_*$ exists);}
    \item \mbox{$a\hspace{-1.5pt}=\hspace{-1.5pt} 0$ iff $ 1\hspace{-1.5pt}  -\hspace{-1.5pt}   e^{\epsilon} \erfc\left( \sqrt{ \epsilon} \right)  \hspace{-1.5pt}=\hspace{-1.5pt}  2 \delta $ (i.e., iff $\epsilon\hspace{-1.5pt}=\hspace{-1.5pt}\epsilon_*$ when $\epsilon_*$ exists);}
    \item \mbox{$a\hspace{-1.5pt}<\hspace{-1.5pt} 0$ iff $ 1\hspace{-1.5pt}  -\hspace{-1.5pt}   e^{\epsilon} \erfc\left( \sqrt{ \epsilon} \right)  \hspace{-1.5pt} <\hspace{-1.5pt}  2 \delta $ (i.e., iff $\epsilon\hspace{-1.5pt}<\hspace{-1.5pt}\epsilon_*$ when $\epsilon_*$ exists).}
\end{enumerate}
In most real-world applications with $\epsilon \geq 0.01$ and $\delta \leq 0.05$, case 1) above holds since $ 1 -  e^{\epsilon} \erfc\left( \sqrt{ \epsilon} \right) =  1 -  s(\epsilon) \geq 1- s(0.01) > 0.1 \geq  2 \delta $, where we use the above result that $s(\epsilon)$ strictly decreases as $\epsilon$ increases.

\begin{lem}\label{lemma-subsection-rem-Mechanism-DP-OPT-ru-detailed}
The following results hold.
\begin{itemize}
\item[i)] With $s(\epsilon) : =  e^{\epsilon} \erfc\left( \sqrt{ \epsilon} \right) $, $s(\epsilon)$ strictly \mbox{decreases} as $\epsilon$ increases for $\epsilon > 0$.
\item[ii)] The values that $s(\epsilon)$ for $\epsilon\in(0,\infty)$ can take constitutes the open interval $(0,1)$.
\end{itemize}

\end{lem}

\noindent \textbf{Proof of Lemma~\ref{lemma-subsection-rem-Mechanism-DP-OPT-ru-detailed}:}

\noindent Proving Result i):
We obtain the desired result in view of the derivative \mbox{$s'(\epsilon) : = - \frac{1}{\sqrt{\pi \epsilon\hspace{2pt}}} + e^{\epsilon} \erfc\left( \sqrt{ \epsilon} \right)  < 0 $,}  where the last step holds from $\erfc\left(\sqrt{ \epsilon}\right) < \frac{\exp(-\epsilon)}{\sqrt{\pi \epsilon\hspace{2pt}}} $, which we obtain by replacing $x$ with $\sqrt{ \epsilon}$ in Reference~\cite{karagiannidis2007improved}'s Inequality (4): $\erfc\left(x\right) < \frac{\exp(-x^2)}{x\sqrt{\pi}} $.

\noindent Proving Result ii): From $\erfc\left(\sqrt{ \epsilon}\right) < \frac{\exp(-\epsilon)}{\sqrt{\pi \epsilon\hspace{2pt}}} $ given above, we have $s(\epsilon)   =  e^{\epsilon} \erfc\left( \sqrt{ \epsilon} \right)  < \frac{1}{\sqrt{\pi \epsilon\hspace{2pt}}} \to 0$ as $\epsilon \to \infty$. Also, $s(0) =  1 $. Since we know from Result i) that $s(\epsilon)$ strictly \mbox{decreases} as $\epsilon$ increases for $\epsilon > 0$, the values that $s(\epsilon)$ for $\epsilon\in(0,\infty)$ can take constitutes the open interval $(0,1)$.

%  Then for $\epsilon \geq 0.01$ and $\delta \leq 0.05$, it holds that $s(\epsilon) \geq s(0.01) \approx 0.1035 > 0. 1 \geq  2 \delta $; i.e., we have $ 2 \delta  < 1 - e^{\epsilon} \erfc\left( \sqrt{ \epsilon} \right)$ so case 1) above holds.
% , which implies $ 2 \delta  < 1 - e^{\epsilon} \erfc\left( \sqrt{ \epsilon} \right)$. Hence, we often have case
%We note that
%\end{rem}

%
%\begin{align} \label{eqn:privacy-loss}
%L_{Y,D,D'}(y) := \ln \frac{\fr{Y(D)=y}}{\fr{Y(D')=y}}.
%\end{align}
%
%for $y$ following the probabilistic distribution of the output $Y(D)$, $L_{Y,D,D'}(y)$ is a  Gaussian variable with mean $\frac{[\|Q(D) - Q(D')\|_{2}]^2}{2{\sigma}^2}$ and  variance $\frac{[\|Q(D) - Q(D')\|_{2}]^2}{{\sigma}^2}$.
%
%for a Gaussian variable $V$ with mean $\tau^2/2$ and  variance $\tau^2$, the term $\bp{V \geq \epsilon } - e^{\epsilon} \bp{V \leq -\epsilon }$ strictly increases as $\tau$ increases.

%\subsection{Proof of Lemma~\ref{lem-eqn-u-DP-OPT-bound-a}}

\subsection{Proof of Theorem~\ref{thm-DP-OPT-asymptotics}}\label{secprf-thm-DP-OPT-asymptotics}

We first present Lemma~\ref{lem-DP-OPT-bounds}, which is proved in Appendix~\ref{Appendix-lem-DP-OPT-bounds} below.

\begin{lem}[\textbf{Bounds of the optimal Gaussian noise amount for $(\epsilon,\delta)$-differential privacy}] \label{lem-DP-OPT-bounds}
Given a fixed $0<\delta<1$, we have:
\begin{subnumcases}{\hspace{-27pt}}
\hspace{-5pt}\mathsmaller{\textup{For $  \epsilon \hspace{-1pt}>\hspace{-1pt}0 $: $  \sigma_{\texttt{\upshape DP-OPT}} \hspace{-1pt}<\hspace{-1pt} \frac{\Delta}{2\sqrt{2} \cdot \inverfc(1-\delta)}  = \frac{\Delta}{2\sqrt{2} \cdot \inverf(\delta)} $;}}  \label{eqn-subnumcases-case1}
\\ \hspace{-5pt}\mathsmaller{\textup{For $  \epsilon \hspace{-1pt}>\hspace{-1pt}0 $: $  \sigma_{\texttt{\upshape DP-OPT}} \hspace{-1.5pt}<\hspace{-1.5pt} \frac{\left(\inverfc(2\delta) +\sqrt{[\inverfc(2\delta) ]^2+\epsilon} \hspace{1.5pt}\right)  \cdot \Delta }{\epsilon\sqrt{2}} $.}} \label{eqn-subnumcases-case2}
\end{subnumcases}
If $0<\delta<0.5$, with $\epsilon_*$ denoting the solution to $e^{\epsilon_*} \erfc\left( \sqrt{ \epsilon_*} \right)  =  1 -  2 \delta$,  we have:
\begin{subnumcases}{\hspace{-36pt}}
 \hspace{-5pt}\mathsmaller{\textup{For $0 \hspace{-3pt}< \hspace{-3pt}\epsilon \hspace{-3pt}\leq \hspace{-3pt}\epsilon_*$:\hspace{-2pt} $ \sigma_{\texttt{\upshape DP-OPT}} \hspace{-3pt}> \hspace{-3pt}\frac{\Delta}{\sqrt{2}  \{\inverfc(\frac{2-2 \delta}{e^{\epsilon}+1}) \hspace{-1pt}+\hspace{-1pt}\sqrt{[\inverfc(\frac{2-2 \delta}{e^{\epsilon}+1}) ]^2+\epsilon}\}}$\hspace{-3pt};}}   \label{eqn-subnumcases-case3}
\\ \hspace{-5pt}\mathsmaller{\textup{For $\epsilon > \epsilon_*$: $  \sigma_{\texttt{\upshape DP-OPT}} > \frac{\Delta }{\sqrt{2\epsilon\hspace{1pt}}} $}}  . \label{eqn-subnumcases-case4}
\end{subnumcases}
If $ 0.5 \leq \delta <1 $ (\underline{which does not hold in practice and is} \underline{presented here only for completeness}), with $\epsilon_{\#}$ denoting the solution to $e^{\epsilon_{\#}} \erfc\left( \sqrt{ \epsilon_{\#}} \right)  =  1 -  \delta$,  we have:
\begin{subnumcases}{\hspace{-27pt}}
 \hspace{-5pt}\mathsmaller{\textup{For $\epsilon >0 $:\hspace{-2pt} $ \sigma_{\texttt{\upshape DP-OPT}} \hspace{-3pt}> \hspace{-3pt}\frac{\Delta}{\sqrt{2}  \{\inverfc(\frac{2-2 \delta}{e^{\epsilon}+1}) \hspace{-1pt}+\hspace{-1pt}\sqrt{[\inverfc(\frac{2-2 \delta}{e^{\epsilon}+1}) ]^2+\epsilon}\}}$\hspace{-1pt};}}   \label{eqn-subnumcases-case5}
\\ \hspace{-5pt}\mathsmaller{\textup{For $\epsilon > \epsilon_{\#}$: $  \sigma_{\texttt{\upshape DP-OPT}} > \frac{  \Delta }{\left(\inverf(\delta)+\sqrt{[\inverf(\delta)]^2+\epsilon} \hspace{1.5pt}\right) \cdot \sqrt{2}} $}}  . \label{eqn-subnumcases-case6}
\end{subnumcases}

% \begin{align}
% \hspace{-5pt}\mathsmaller{\textup{For $  \epsilon >0 $:\hspace{-2pt} $ \sigma_{\texttt{\upshape DP-OPT}} \hspace{-3pt}> \hspace{-3pt}\frac{\Delta}{\sqrt{2}  \{\inverfc(\frac{2-2 \delta}{e^{\epsilon}+1}) \hspace{-1pt}+\hspace{-1pt}\sqrt{[\inverfc(\frac{2-2 \delta}{e^{\epsilon}+1}) ]^2+\epsilon}\}}$\hspace{-3pt}}}  . \label{eqn-subnumcases-case5}
% \end{align}

% \begin{itemize}
% \item[\ding{172}] For $  \epsilon >0 $, it holds that \\ $  \sigma_{\texttt{\upshape DP-OPT}} < \frac{\Delta}{2\sqrt{2} \cdot \inverfc(1-\delta)}  = \frac{\Delta}{2\sqrt{2} \cdot \inverf(\delta)} $.
% \item[\ding{173}]  If $0 < \epsilon  \leq \epsilon_*$, then \\ $ \sigma_{\texttt{\upshape DP-OPT}} > \frac{\Delta}{\sqrt{2} \cdot \{\inverfc(\frac{2-2 \delta}{e^{\epsilon}+1}) +\sqrt{[\inverfc(\frac{2-2 \delta}{e^{\epsilon}+1}) ]^2+\epsilon}\}}$.
% \item[\ding{174}]  If $\epsilon > \epsilon_*$, then \\ $ \frac{\Delta }{\sqrt{2\epsilon\hspace{1pt}}} < \sigma_{\texttt{\upshape DP-OPT}} < \frac{\left(\inverfc(2\delta) +\sqrt{[\inverfc(2\delta) ]^2+\epsilon} \hspace{1.5pt}\right)  \cdot \Delta }{\epsilon\sqrt{2}}  $.
% \end{itemize}

% \begin{align}
% \hspace{-130pt} \sigma_{\texttt{\upshape DP-OPT}} \leq \nonumber
% \end{align}
% \begin{subnumcases}{\hspace{-23pt}}
% \hspace{-5pt}\frac{\Delta }{\sqrt{2\epsilon\hspace{1pt}}}, \text{ if $\epsilon  \leq \epsilon_*$} ;
% \\ \hspace{-5pt}\frac{\left(\inverfc(2\delta) +\sqrt{[\inverfc(2\delta) ]^2+\epsilon} \hspace{1.5pt}\right)  \cdot \Delta }{\epsilon\sqrt{2}} ,  \text{ if $\epsilon > \epsilon_*$} .
% \end{subnumcases}
\end{lem}

We prove Lemma~\ref{lem-DP-OPT-bounds} in Appendix~\ref{Appendix-lem-DP-OPT-bounds}. Below we use Lemma~\ref{lem-DP-OPT-bounds} to show Theorem~\ref{thm-DP-OPT-asymptotics}.

% In particular, Eq.~(\ref{eqn-subnumcases-case2}) uses the following Lemma~\ref{lem-a-vs-erfc-2delta}.

Eq.~(\ref{eqn-subnumcases-case1}) is Result~\ding{172} of Theorem~\ref{thm-DP-OPT-asymptotics}.
Eq.~(\ref{eqn-subnumcases-case1})~(\ref{eqn-subnumcases-case3}) and~(\ref{eqn-subnumcases-case5}) imply Result~\ding{173} of Theorem~\ref{thm-DP-OPT-asymptotics}. If $0<\delta<0.5$,
Eq.~(\ref{eqn-subnumcases-case2})  and~(\ref{eqn-subnumcases-case4}) imply Result~\ding{174} of Theorem~\ref{thm-DP-OPT-asymptotics}. If $ 0.5 \leq \delta <1 $, Eq.~(\ref{eqn-subnumcases-case2})  and~(\ref{eqn-subnumcases-case6}) imply Result~\ding{174} of Theorem~\ref{thm-DP-OPT-asymptotics}.

To prove
Result~\ding{175} of Theorem~\ref{thm-DP-OPT-asymptotics} (i.e., $ \lim_{\delta \to 0}\sigma_{\texttt{\upshape DP-OPT}} \Big/ \left( \frac{\Delta}{\epsilon} \sqrt{2  \ln \frac{1}{\delta}  }\hspace{1.5pt}\right) = 1 $), below we use the sandwich method. Specifically, we find an upper bound and a lower bound for $\sigma_{\texttt{\upshape DP-OPT}}$, and show that dividing each bound by $ \frac{\Delta}{\epsilon} \sqrt{2  \ln \frac{1}{\delta}  }$ converges to $1$ as $\delta \to 0$.

For the upper bound part,  given a fixed $\epsilon >0 $, we use Theorem~\ref{thm-DP-OPT}'s Property (iii) to derive
\begin{align}
&\sigma_{\texttt{\upshape DP-OPT}} \bigg/ \left( \frac{\Delta}{\epsilon} \sqrt{2  \ln \frac{1}{\delta}  }\right) \nonumber  \\ &   < \left( \sqrt{2\ln \frac{1}{2\delta} }\cdot\frac{\Delta}{\epsilon} + \frac{\Delta}{\sqrt{2\epsilon\,}}\right)  \bigg/ \left( \frac{\Delta}{\epsilon} \sqrt{2  \ln \frac{1}{\delta}  }\right) \nonumber  \\ &  \to 1, \text{ as $\delta \to 0$} . \label{ubeq}
\end{align}

The proof for the  lower bound part is more complex and is presented below.

We define $ f(x) =  e^{x} \erfc\left( \sqrt{a^2 + x} \right) $. Then we have the first-order derivative  $f'(x)$ and second-order derivative $f''(x)$ as follows:
$$f'(x) = e^{x} \erfc\left( \sqrt{a^2 + x} \right) - \frac{\exp(-a^2)}{\sqrt{\pi(a^2 + x)}}   $$ and
$$f''(x) = e^{x} \erfc\left( \sqrt{a^2 + x} \right) - \frac{\exp(-a^2)}{\sqrt{\pi(a^2 + x)}} + \frac{\exp(-a^2)}{2\sqrt{\pi} (a^2 + x)^{3/2}} .$$ We have the following two propositions. After stating their proofs, we continue proving Theorem~\ref{thm-DP-OPT}.

%, whose proofs are provided in the full version~\cite{fullver}.

\begin{prop} \label{prop-f-prime-x}
$f'(x)  < 0 $ for $x \geq 0$.
\end{prop}

\begin{prop} \label{prop-f-prime-prime-x}
$f''(x)  >
0 $ for $x \geq 0$.
\end{prop}

%d/dx(e^x erfc(sqrt(1 + x))) = e^x erfc(sqrt(1 + x)) - 1/(e sqrt(π) sqrt(1 + x))

%d/dx(e^x erfc(sqrt(1 + x)) - 1/(e sqrt(π) sqrt(1 + x))) = e^x erfc(sqrt(1 + x)) - 1/(e sqrt(π) sqrt(1 + x)) + 1/(2 e sqrt(π) (1 + x)^(3/2))

%plot e^x erfc(sqrt(1 + x)) - 1/(e sqrt(π) sqrt(1 + x)) + 1/(2 e sqrt(π) (1 + x)^(3/2))

\noindent
\textbf{Proof of Proposition~\ref{prop-f-prime-x}:} From Proposition~\ref{prop-f-prime-prime-x}, we have $f'(x) \leq f'(0) $ for $x \geq 0$, which along with $ \textstyle{f'(0) = \erfc\left(|a|\right) - \frac{\exp(-a^2)}{|a|\sqrt{\pi}} < 0}  $ from Reference [52]'s Inequality (4) implies $f'(x)  < 0 $ for $x \geq 0$.

\noindent
\textbf{Proof of Proposition~\ref{prop-f-prime-prime-x}:} We can write $\textstyle{f''(x) = e^{x} u(\sqrt{a^2 + x})  }$ for function $u(y)$ defined by \mbox{$u(y):= \erfc\left(y \right) - \frac{\exp(-y^2)}{y\sqrt{\pi}} (1-\frac{1}{2y^2}) $.}  We have $u(y)>0$ from the asymptotic expansion (i.e., Inequality 7.12.1 in~\cite{abramowitz1964handbook}) of the  complementary error function $\erfc\left(\cdot\right)$. Hence, the desired result is proved.

Now we continue the proof of Theorem~\ref{thm-DP-OPT}.

% From ?, the Taylor series of the scaled complementary error function $\erfc\left(t\right) \cdot \exp(t^2)$ is given by
% \begin{align}
% &\erfc\left(t\right) \cdot \exp(t^2) \nonumber \\ & = \frac{1}{t\sqrt{\pi}}  + \frac{1}{\sqrt{\pi}} \sum_{i=1}^n \left[ (-1)^i \cdot \frac{1\cdot 3 \ldots \cdot (2i-1)}{2^i t^{2i+1}} \right] . \label{scaled-erfc-Taylor-series}
% \end{align}

% $a>0$

Propositions~\ref{prop-f-prime-x} and ~\ref{prop-f-prime-prime-x} induce $f'(x) \leq f'(\epsilon) < 0 $ for $x \geq 0$. Then we have
$f(0) - f(\epsilon) = - \int_{x=0}^{\epsilon} f'(x) \, \text{d} x \geq - \int_{x=0}^{\epsilon} f'(\epsilon) \, \text{d} x = - f'(\epsilon) \epsilon  $, which implies
\begin{align}
- f'(\epsilon) = \frac{\exp(-a^2)}{\sqrt{\pi(a^2 + \epsilon)}}  -  e^{\epsilon} \erfc\left( \sqrt{a^2 + \epsilon} \right)   \leq \frac{ 2 \delta}{\epsilon } .
\end{align}
 For notation convenience, we define
\begin{align}
A:= \sqrt{a^2 + \epsilon} .
\end{align}
Then
\begin{align}
\frac{\exp(-A^2)}{A\sqrt{\pi}}  -  \erfc\left(A\right)   \leq \frac{ 2 \delta}{ \epsilon \cdot \exp(\epsilon) }  . \label{eq-A-epsilon}
\end{align}
% $ \frac{\exp(-a^2-\epsilon)}{\sqrt{\pi(a^2 + \epsilon)}}  -  \erfc\left( \sqrt{a^2 + \epsilon} \right)   \leq \frac{ 2 \delta}{ \epsilon \cdot \exp(\epsilon) }   $.

% \begin{align} \erfc\left(t\right) \cdot \exp(t^2) \leq \frac{1}{\sqrt{\pi}}  + \frac{1}{t\sqrt{\pi}} \sum_{i=1}^n \left[ -\frac{1}{2t^{3}} + \frac{3}{4t^{5}} \right] .
% \end{align}

From the inverse factorial series of the  complementary error function $\erfc\left(\cdot\right)$~\cite{abramowitz1964handbook}, it holds that
\begin{align}
& \erfc\left( A \right) \nonumber  \\ & \leq \frac{\exp(-A^2)}{A\sqrt{\pi}} \left[1 - \frac{1}{2(A^2+1)} + \frac{1}{4(A^2+1)(A^2+2)}\right] \nonumber  \\ &    =  \frac{\exp(-A^2)}{A\sqrt{\pi}} \left[1 -  \frac{2A^2+3}{4(A^2+1)(A^2+2)}\right],
\end{align}
  which we use in the above  Inequality~(\ref{eq-A-epsilon}) to obtain \begin{align}
\frac{\exp(-A^2)}{A\sqrt{\pi}} \cdot \frac{2A^2+3}{4(A^2+1)(A^2+2)} \leq \frac{ 2 \delta}{ \epsilon \cdot \exp(\epsilon) } .
\end{align}   This further induces
  \begin{align}
& \exp(-A^2) \nonumber  \\&  \leq \frac{ 2 \delta}{ \epsilon \cdot \exp(\epsilon) }  \cdot \frac{4A\cdot (A^2+1)(A^2+2)\sqrt{\pi}}{2A^2+3}   \nonumber  \\ &\leq \frac{ 2 \delta}{ \epsilon \cdot \exp(\epsilon) }  \cdot 4A\cdot (0.5A^2+0.75)\sqrt{\pi} , \end{align}
where the last step uses \begin{align}
  &  (A^2+1)(A^2+2)-(2A^2+3)(0.5A^2+0.75)  \nonumber  \\ & = - 0.25 < 0.
\end{align}

Then
  \begin{align}
& \exp(-A^2) \nonumber  \\&  \leq \frac{ 2 \delta}{ \epsilon \cdot \exp(\epsilon) }  \cdot 4A\cdot (0.5A^2+0.75)\sqrt{\pi}  \nonumber  \\ & \leq \frac{ 2 \delta}{ \epsilon \cdot \exp(\epsilon) }  \cdot 4\sqrt{\ln \frac{1}{\delta}}\Big[0.5\Big(\sqrt{\ln \frac{1}{\delta}}\,\Big)^2+0.75\Big]\sqrt{\pi}. \label{eq-exp-A1}
\end{align}

% From Lemma~\ref{lem-eqn-u-DP-OPT-bound-a} on Page~\pageref{lem-eqn-u-DP-OPT-bound-a}, we have $A< \inverfc(2\delta) < \inverfc(\delta) < \sqrt{\ln \frac{1}{\delta}}$

We consider $0< \delta \leq 0.005$. We can assume $\delta \leq e^{-1.5}$. For such $\delta \leq e^{-1.5}$, it holds that
\begin{align}
0.5\Big(\sqrt{\ln \frac{1}{\delta}}\,\Big)^2+0.75 \leq \Big(\sqrt{\ln \frac{1}{\delta}}\,\Big)^2 ,
\end{align}
 which we use in Inequality~(\ref{eq-exp-A1}) to derive
\begin{align}
 \exp(-A^2) \leq  8 \delta  \Big(\sqrt{\ln \frac{1}{\delta}}\,\Big)^3    \frac{\sqrt{\pi}}{{ \epsilon \cdot \exp(\epsilon) }}  .
\end{align}
 Then
  for $\delta \leq e^{-1.5}$ and $   8 \delta  \Big(\sqrt{\ln \frac{1}{\delta}}\,\Big)^3\sqrt{\pi} /\epsilon  \leq  1 $, which clearly holds given a fixed $\epsilon >0 $ and $\delta \to 0$,
 we have
 \begin{align}
A &\geq \sqrt{\ln \frac{\epsilon \cdot \exp(\epsilon) }{ 8 \delta  \Big(\sqrt{\ln \frac{1}{\delta}}\,\Big)^3\sqrt{\pi}  }} \nonumber  \\ & =  \sqrt{ \epsilon + \ln \frac{1}{\delta} +  \ln \frac{\epsilon}{ 8 \sqrt{\pi} } - \frac{3}{2} \ln \ln \frac{1}{\delta} }.
\end{align}
 For $\epsilon \geq 1$ and $0< \delta \leq 0.005$, we can verify that \textbullet~$\delta \leq e^{-1.5}$, \textbullet~$   8 \delta  \Big(\sqrt{\ln \frac{1}{\delta}}\,\Big)^3\sqrt{\pi} /\epsilon  \leq  1 $, and $\ln \frac{\epsilon}{ 8 \sqrt{\pi} } - \frac{3}{2} \ln \ln \frac{1}{\delta} \geq \ln \frac{1}{ 8 \sqrt{\pi} } - \frac{3}{2} \ln \ln \frac{1}{0.005}>0.0057$. Hence,
\begin{align} \label{align-bound-eq-a}
A \geq \sqrt{ \ln \frac{1}{\delta} +  \ln \frac{\epsilon}{ 8 \sqrt{\pi} } - \frac{3}{2} \ln \ln \frac{1}{\delta} } >\sqrt{ \ln \frac{0.0057}{\delta} },
\end{align}
which implies
\begin{align}
&\sigma_{\texttt{\upshape DP-OPT}}  \nonumber  \\ & = \frac{\left(a+\sqrt{a^2+\epsilon} \hspace{1.5pt}\right)  \cdot \Delta }{\epsilon\sqrt{2}}\nonumber  \\ &  =\frac{\left(\sqrt{A^2-\epsilon}\,+\,A \hspace{1.5pt}\right)  \cdot \Delta }{\epsilon\sqrt{2}}   \nonumber  \\ &    >   \frac{\Delta}{\epsilon\sqrt{2}} \left( \sqrt{  \ln \frac{1}{\delta} +\ln 0.0057-\epsilon}  + \sqrt{  \ln \frac{1}{\delta} +\ln 0.0057} \right) . \label{lbeq}
\end{align}
Clearly, dividing the above lower bound of~(\ref{lbeq}) by $ \frac{\Delta}{\epsilon} \sqrt{2  \ln \frac{1}{\delta}  }$ converges to $1$ as $\delta \to 0$. Combining this with~(\ref{ubeq}), we complete proving Result~\ding{175} of Theorem~\ref{thm-DP-OPT-asymptotics}.\qeda

% Below we use Theorem~\ref{thm-DP-OPT} to prove
% Result~\ding{175} of Theorem~\ref{thm-DP-OPT-asymptotics}.

%  Moreover,  we obtain $\sigma_{\texttt{\upshape DP-OPT}}  := \frac{\left(a+\sqrt{a^2+\epsilon} \hspace{1.5pt}\right)  \cdot \Delta }{\epsilon\sqrt{2}} >   \frac{2a  \cdot \Delta }{\epsilon\sqrt{2}}   > \sqrt{ 2 \ln \frac{0.0057}{\delta} } \cdot\frac{\Delta }{\epsilon} $, so that $\sigma_{\texttt{\upshape DP-OPT}} \Big/ \left( \frac{\Delta}{\epsilon} \sqrt{2  \ln \frac{1}{\delta}  }\right)  > \left(\sqrt{2\ln \frac{0.0057}{\delta} } \cdot\frac{\Delta }{\epsilon}  \right) \Big/ \left( \frac{\Delta}{\epsilon} \sqrt{2  \ln \frac{1}{\delta}  }\right) \to 1 $ as $\delta \to 0$.  Hence,  $ \lim_{\delta \to 0}\sigma_{\texttt{\upshape DP-OPT}} \Big/ \left( \frac{\Delta}{\epsilon} \sqrt{2  \ln \frac{1}{\delta}  }\right) = 1 $.?
% % From Theorem~\ref{thm-DP-OPT}'s Property (ii), we have
% % Now, we find a lower bound for $a$.

%
%\textbf{Proof of Proposition~\ref{prop-f-prime-x}.} \qeda
%
%\textbf{Proof of Proposition~\ref{prop-f-prime-prime-x}.} \qeda

\subsection{Proof of Lemma~\ref{lem-DP-OPT-bounds}} \label{Appendix-lem-DP-OPT-bounds}

% \subsection{Proof of Lemma~\ref{lem-DP-OPT-bounds}}\label{secprf-thm-DP-OPT-bounds}

\noindent \textbf{Proof of Lemma~\ref{lem-DP-OPT-bounds}'s Eq.~(\ref{eqn-subnumcases-case1}) for $  \epsilon >0 $:}

We write $\sigma_{\texttt{\upshape DP-OPT}}$ of Theorem~\ref{thm-DP-OPT} as a function $\sigma_{\texttt{\upshape DP-OPT}}(\epsilon,\delta)$. Given a fixed $0<\delta<0.5$, clearly $\sigma_{\texttt{\upshape DP-OPT}}(\epsilon,\delta)$ strictly decreases as $\epsilon$ increases, which implies for $\epsilon>0$ that $\sigma_{\texttt{\upshape DP-OPT}}(\epsilon,\delta) $ is less than $ \lim_{\epsilon \to 0}\sigma_{\texttt{\upshape DP-OPT}}(\epsilon,\delta)$ (if such limit exists). When $\epsilon \to 0$, $a$ in Eq.~(\ref{eqn-sigma-DP-OPT}) is negative and satisfies $\erfc\left(a \right)   -   \erfc\left( -a\right)  \to  2 \delta$ so that $a \to -\inverfc(1-\delta)$ due to $\erfc\left( -a\right)=2-\erfc\left(a \right) $. This further implies for $\epsilon \to 0$ that
\begin{align}
\sigma_{\texttt{\upshape DP-OPT}}(\epsilon,\delta)   & =  \frac{\left(a+\sqrt{a^2+\epsilon} \hspace{1.5pt}\right)  \cdot \Delta }{\epsilon\sqrt{2}} \nonumber  \\ &= \frac{  \Delta }{\left(-a+\sqrt{a^2+\epsilon} \hspace{1.5pt}\right) \cdot \sqrt{2}} \nonumber  \\ &\to \frac{\Delta}{2\sqrt{2} \cdot \inverfc(1-\delta)}.
\end{align}
  Hence, for $\epsilon>0$, we have\begin{align}
 \sigma_{\texttt{\upshape DP-OPT}}(\epsilon,\delta)  &  <\lim_{\epsilon \to 0} \sigma_{\texttt{\upshape DP-OPT}}(\epsilon,\delta)\nonumber  \\ &  = \frac{\Delta}{2\sqrt{2} \cdot \inverfc(1-\delta)} \nonumber  \\ & = \frac{\Delta}{2\sqrt{2} \cdot \inverf(\delta)}  .
\end{align}

%\begin{lem} \label{lem-sigma-DP-OPT-increase}
%Given a fixed $0<\delta<1$, $\sigma_{\texttt{\upshape DP-OPT}}$ of Theorem~\ref{thm-DP-OPT} strictly decreases as $\epsilon$ increases.
%\end{lem}

\noindent \textbf{Proof of Lemma~\ref{lem-DP-OPT-bounds}'s Eq.~(\ref{eqn-subnumcases-case2}) for $  \epsilon >0 $:} Eq.~(\ref{eqn-subnumcases-case2}) follows from Eq.~(\ref{eqn-sigma-DP-OPT}) and Lemma~\ref{lem-a-vs-erfc-2delta} presented at the end of this subsection.

\noindent \textbf{Proof of Lemma~\ref{lem-DP-OPT-bounds}'s Eq.~(\ref{eqn-subnumcases-case3}) for $0<\delta<0.5$ and $0 < \epsilon  \leq \epsilon_*$:} \label{proof-lem-DP-OPT-bounds}

We   consider $0<\delta<0.5$ and $0 < \epsilon  \leq \epsilon_*$ here. In this case, from Appendix~\ref{subsection-rem-Mechanism-DP-OPT-ru-detailed},   $a$ in Eq.~(\ref{eqn-sigma-DP-OPT}) is negative or zero. Then we have $\erfc\left( \sqrt{a^2 + \epsilon} \right) < \erfc\left( |a|\right) = \erfc\left( - a\right) $, which along with $\erfc\left(a \right)   -  e^{\epsilon} \erfc\left( \sqrt{a^2 + \epsilon} \right)  =  2 \delta$ and $\erfc\left(  a\right)= 2- \erfc\left(-  a\right)$ implies $2- \erfc\left(-  a\right)  -  e^{\epsilon} \erfc\left(-  a\right) < 2 \delta$. Then we have $\erfc\left(-  a\right) > \frac{2-2 \delta}{e^{\epsilon}+1} $, which along with the aforementioned result $a \leq 0$ implies
\begin{align}
-\inverfc\bigg(\frac{2-2 \delta}{e^{\epsilon}+1}\bigg)<a \leq 0. \label{eqanega0}
\end{align}
% $-a<\inverfc(\frac{2-2 \delta}{e^{\epsilon}+1})$.
Thus,
\begin{align}
& \sigma_{\texttt{\upshape DP-OPT}}(\epsilon,\delta) \nonumber  \\ & = \frac{\left(a+\sqrt{a^2+\epsilon} \hspace{1.5pt}\right)  \cdot \Delta }{\epsilon\sqrt{2}} \nonumber  \\ &= \frac{  \Delta }{\left(-a+\sqrt{a^2+\epsilon} \hspace{1.5pt}\right) \cdot \sqrt{2}} \nonumber  \\ & > \frac{\Delta}{\sqrt{2} \cdot \{\inverfc(\frac{2-2 \delta}{e^{\epsilon}+1}) +\sqrt{[\inverfc(\frac{2-2 \delta}{e^{\epsilon}+1}) ]^2+\epsilon}\}}.\label{eqanega0v0}
\end{align}

\noindent \textbf{Proof of Lemma~\ref{lem-DP-OPT-bounds}'s Eq.~(\ref{eqn-subnumcases-case4}) for $0<\delta<0.5$ and $\epsilon  > \epsilon_*$:}

We   consider $0<\delta<0.5$ and $\epsilon  > \epsilon_*$ here. In this case, from Appendix~\ref{subsection-rem-Mechanism-DP-OPT-ru-detailed},   $a$ in Eq.~(\ref{eqn-sigma-DP-OPT}) is positive. Then $ \sigma_{\texttt{\upshape DP-OPT}}(\epsilon,\delta) = \frac{\left(a+\sqrt{a^2+\epsilon} \hspace{1.5pt}\right)  \cdot \Delta }{\epsilon\sqrt{2}} > \frac{\sqrt{\epsilon}  \cdot \Delta }{\epsilon\sqrt{2}} = \frac{\Delta }{\sqrt{2\epsilon\hspace{1pt}}} $.

\noindent \textbf{Proof of Lemma~\ref{lem-DP-OPT-bounds}'s Eq.~(\ref{eqn-subnumcases-case5}) for $ 0.5 \leq \delta <1 $ and  $\epsilon  > 0$:}

% We   consider $0<\delta<0.5$ and $0 < \epsilon  \leq \epsilon_*$ here.
  The proof is similar to that for Eq.~(\ref{eqn-subnumcases-case3}) above. First, with $ 0.5 \leq \delta <1 $ and  $\epsilon  > 0$, from Appendix~\ref{subsection-rem-Mechanism-DP-OPT-ru-detailed},   $a$ in Eq.~(\ref{eqn-sigma-DP-OPT}) is negative.  Then similar to the proof of Eq.~(\ref{eqanega0}), we  have
\begin{align}
-\inverfc\bigg(\frac{2-2 \delta}{e^{\epsilon}+1}\bigg)<a < 0. \label{eqanega0v2}
\end{align}
Then we also obtain Eq.~(\ref{eqn-subnumcases-case5}) in a way similar to the proof of Eq.~(\ref{eqanega0v0}).

\noindent \textbf{Proof of Lemma~\ref{lem-DP-OPT-bounds}'s Eq.~(\ref{eqn-subnumcases-case6}) for $ 0.5 \leq \delta <1 $ and  $\epsilon  > \epsilon_{\#}$:}

Since
$e^{\epsilon} \erfc\left( \sqrt{ \epsilon} \right)$  strictly decreases as $\epsilon$ increases from Lemma~\ref{lemma-subsection-rem-Mechanism-DP-OPT-ru-detailed} on Page \pageref{lemma-subsection-rem-Mechanism-DP-OPT-ru-detailed}, for $\epsilon_{\#}$ denoting the solution to $e^{\epsilon_{\#}} \erfc\left( \sqrt{ \epsilon_{\#}} \right)  =  1 -  \delta$, we have for $\epsilon > \epsilon_{\#}$ that $  e^{\epsilon} \erfc\left( \sqrt{ \epsilon} \right) < e^{\epsilon_{\#}} \erfc\left( \sqrt{ \epsilon_{\#}} \right)  =  1 -  \delta$, which gives a lower bound on $a$ of Eq.~(\ref{eqn-sigma-DP-OPT}): \\
\begin{align}
  a   &>  \inverfc \big( 2 \delta + e^{\epsilon} \erfc\left( \sqrt{ \epsilon} \right) \big)   \nonumber  \\ &  > \inverfc (1 + \delta)   \nonumber  \\ &  = -\inverf ( \delta).
\end{align}
 Then \begin{align}
  &  \sigma_{\texttt{\upshape DP-OPT}}(\epsilon,\delta)  \nonumber  \\ & = \frac{\left(a+\sqrt{a^2+\epsilon} \hspace{1.5pt}\right)  \cdot \Delta }{\epsilon\sqrt{2}}  \nonumber  \\ & = \frac{  \Delta }{\left(-a+\sqrt{a^2+\epsilon} \hspace{1.5pt}\right) \cdot \sqrt{2}}   \nonumber  \\ &  >\frac{  \Delta }{\left(\inverf(\delta)+\sqrt{[\inverf(\delta)]^2+\epsilon} \hspace{1.5pt}\right) \cdot \sqrt{2}} .
\end{align}
\qeda

\begin{lem} \label{lem-a-vs-erfc-2delta}
$a$ in Eq.~(\ref{eqn-sigma-DP-OPT}) is less than $\inverfc(2\delta) $.
\end{lem}

\noindent \textbf{Proof of Lemma~\ref{lem-a-vs-erfc-2delta}:} The result follows since Eq.~(\ref{eqn-sigma-DP-OPT}) induces $\erfc\left(a \right)    =  2 \delta + e^{\epsilon} \erfc\left( \sqrt{a^2 + \epsilon} \right) >  2 \delta$ and $\erfc\left( \cdot \right)$ is a strictly decreasing function.

% Belowe we use

% we have $\erfc\left( \sqrt{a^2 + \epsilon} \right) < \erfc\left( |a|\right) = \erfc\left( - a\right) $,

% $a+\sqrt{a^2+\epsilon} = \frac{\epsilon}{-a+\sqrt{a^2+\epsilon}} \leq  \frac{\epsilon}{\sqrt{\epsilon}} = \sqrt{\epsilon}$, which implies that $\sigma_{\texttt{\upshape DP-OPT}}$ in Eq.~(\ref{eqn-sigma-DP-OPT}) is no greater than $\frac{\Delta }{\sqrt{2\epsilon\hspace{1pt}}} $.

% $2 - \frac{2 e^{\epsilon}+2 \delta}{e^{\epsilon}+1} = \frac{2-2 \delta}{e^{\epsilon}+1}$

% $1 -  \frac{2-2 \delta}{e^{\epsilon}+1}$

% In Theorem~\ref{thm-DP-OPT-asymptotics}, we further have asympotic results for $\sigma_{\texttt{\upshape DP-OPT}}$.

\subsection{Establishing Lemma~\ref{lem-a-vs-b}, which along with Theorem~\ref{thm-DP-OPT} implies Theorem~\ref{thm-Mechanism-1}}\label{sec-prf-thm-Mechanism-1-based-on-thm-DP-OPT}

 When $ e^{\epsilon} \erfc\left( \sqrt{ \epsilon} \right) + 2 \delta \geq 2$, we have $b=0$ from Eq.~(\ref{eqn-u-Mechanism-1}) on Page~\pageref{eqn-u-Mechanism-1} and $a \leq 0$ from Appendix~\ref{subsection-rem-Mechanism-DP-OPT-ru-detailed} on Page~\pageref{subsection-rem-Mechanism-DP-OPT-ru-detailed}, so the desired result $a \leq b $ follows. Below we focus on the case of $ e^{\epsilon} \erfc\left( \sqrt{ \epsilon} \right) + 2 \delta < 2$.

%\textbf{Case 1:
%$ e^{\epsilon} \erfc\left( \sqrt{ \epsilon} \right) + 2 \delta < 1$.}
%
%\textbf{Case 2:
%$   e^{\epsilon} \erfc\left( \sqrt{ \epsilon} \right) + 2 \delta = 1$.}
%
%\textbf{Case 3:
%$ 1 < e^{\epsilon} \erfc\left( \sqrt{ \epsilon} \right) + 2 \delta < 2$.}
%
%\textbf{Case 4:
%$ e^{\epsilon} \erfc\left( \sqrt{ \epsilon} \right) + 2 \delta \geq 2$.}

% \subsection{Proof of Theorem~\ref{thm-Mechanism-1} based on Theorem~\ref{thm-DP-OPT}}

We use Theorem~\ref{thm-DP-OPT} to  prove Theorem~\ref{thm-Mechanism-1}. In particular, we will show that $a$ specified in~Eq.~(\ref{eqn-sigma-DP-OPT}) is less than $b$ defined in~Eq.~(\ref{eqn-u-Mechanism-1}).

Recall that Eq.~(\ref{eqn-sigma-DP-OPT}) presents
\begin{align}
\erfc\left(a \right)   -  e^{\epsilon} \erfc\left( \sqrt{a^2 + \epsilon} \right)  =  2 \delta. \label{eqn-erfc-a}
\end{align}
We will find an upper bound for $a$ and this upper bound will be $b$. To this end, we will show $\erfc\left(a \right)$ is at least some fraction of $\erfc\left( \sqrt{a^2 + \epsilon} \right)$. This will be done by i) proving a lower bound for $a$, and ii) showing that $\frac{\erfc\left( \sqrt{u^2 + \epsilon} \right) }{\erfc\left(u \right)} $ strictly increases as $u$ increases for $u \in (-\infty, \infty)$.

We first give a lower bound for $a$. From~Eq.~(\ref{eqn-erfc-a}), we have $\erfc\left(a \right) =    2 \delta + e^{\epsilon} \erfc\left( \sqrt{a^2 + \epsilon} \right) <  2 \delta + e^{\epsilon} \erfc\left( \sqrt{ \epsilon} \right)$, which implies that if $2 \delta + e^{\epsilon} \erfc\left( \sqrt{ \epsilon} \right)<2$,
\begin{align}
 &a > \inverfc \big( 2 \delta + e^{\epsilon} \erfc\left( \sqrt{ \epsilon} \right) \big), \label{eqn-a-lower-bound}
\end{align}
where we note that the image domain of $\erfc\left(\cdot\right)$ is $(0,2)$ since the image domain of $\erf\left(\cdot\right)$ is $(-1,1)$ and $\erfc\left(\cdot\right)=1-\erf\left(\cdot\right)$.

We now prove $h(u) := \frac{\erfc\left( \sqrt{u^2 + \epsilon} \right) }{\erfc\left(u \right)} $ strictly increases as $u$ increases for $u \in (-\infty, \infty)$. Taking the derivative of  $h(u)$ with respect to $u$, we obtain
\begin{align}
 h '(u)   &  = \frac{\erfc'\left( \sqrt{u^2 + \epsilon} \right) \times \erfc\left(u \right) - \erfc\left( \sqrt{u^2 + \epsilon} \right) \times \erfc'\left(u \right) }{\erfc^2\left(u \right)} \nonumber \\  & = \frac{2}{\sqrt{\pi}} \times \exp(-u^2) \times  \frac{\kappa(u)}{\erfc^2\left(u \right)} , \label{hprimeu}
\end{align}
for $\kappa(u)$ defined by
\begin{align}
\kappa(u) := - \exp\left( - \epsilon \right) \times \frac{u}{\sqrt{u^2 + \epsilon}} \times \erfc\left(u \right) + \erfc\left( \sqrt{u^2 + \epsilon} \right).
\end{align}

We will prove $\kappa(u)>0$. To this end, we first investigate the monotonicity of $\kappa(u)$ for $u \in (-\infty, \infty)$. Taking the derivative of  $\kappa(u)$ with respect to $u$, we get
\begin{align}
\kappa'(u) & =   \exp\left( - \epsilon \right) \times \frac{u}{\sqrt{u^2 + \epsilon}} \times \frac{2}{\sqrt{\pi}} \exp\left( - u^2 \right) \nonumber \\  & \quad -  \exp\left( - \epsilon \right) \times \frac{\sqrt{u^2 + \epsilon} - \frac{u^2}{\sqrt{u^2 + \epsilon}}}{u^2 + \epsilon} \times \erfc\left(u \right)\nonumber \\  & \quad  - \frac{2}{\sqrt{\pi}}   \exp\left( - u^2 - \epsilon \right) \times \frac{u}{\sqrt{u^2 + \epsilon}}  \nonumber \\  & = -  \exp\left( - \epsilon \right) \times \frac{\epsilon}{(u^2 + \epsilon)^{3/2}} \times \erfc\left(u \right) \nonumber \\  & < 0.
\end{align}
Hence, $\kappa(u)$ strictly decreases as $u$ increases for $u \in (-\infty, \infty)$. Combining this and $\lim_{u \to \infty} \frac{\kappa(u)}{\erfc\left(u \right)} := 1 - \exp\left( - \epsilon \right) > 0$,
% \begin{align}
% \lim_{u \to \infty} \frac{\kappa(u)}{\erfc\left(u \right)} := 1 - \exp\left( - \epsilon \right) > 0.
% \end{align}
we conclude for $u \in (-\infty, \infty)$ that $\frac{\kappa(u)}{\erfc\left(u \right)} > 0$ and hence $\kappa(u)> 0$. Thus, $h '(u)$ in~Eq.~(\ref{hprimeu}) is positive, so that $h(u)$ is increasing for $u \in (-\infty, \infty)$. This along with~Eq.~(\ref{eqn-a-lower-bound}) implies
\begin{align}
h(a)  &\geq h \left(\inverfc \big( 2 \delta + e^{\epsilon} \erfc\left( \sqrt{ \epsilon} \right) \big)\right). \label{habound}
\end{align}

From Eq.~(\ref{eqn-erfc-a}) and~(\ref{habound}), and $h(a) := \frac{\erfc\left( \sqrt{a^2 + \epsilon} \right) }{\erfc\left(a \right)} $, we derive
\begin{align}
\erfc\left(a \right) &= \frac{2\delta}{1 - e^{\epsilon}\cdot h(a)} \nonumber \\  & \geq \frac{2\delta}{1 - e^{\epsilon}\cdot h \left(\inverfc \big( 2 \delta + e^{\epsilon} \erfc\left( \sqrt{ \epsilon} \right) \big)\right)}\nonumber \\  &  = \erfc\left(b \right) , \label{eqn-erfca}
\end{align}
where the last step uses the expression of $b$ in~Eq.~(\ref{eqn-u-Mechanism-1}). Hence, it holds that $a \leq b$. Then we obtain the desired result of
Theorem~\ref{thm-DP-OPT} implying Theorem~\ref{thm-Mechanism-1}.\qeda

\subsection{Establishing Lemma~\ref{lem-b-vs-c}, which along with Theorem~\ref{thm-Mechanism-1} implies Theorem~\ref{thm-Mechanism-2}} \label{appenlem-lem-b-vs-c}

% \subsection{Proof of Theorem~\ref{thm-Mechanism-2} based on Theorem~\ref{thm-Mechanism-1}}

 From~Eq.~(\ref{eqn-erfca}), it holds that
 $\erfc\left(b \right) > 2\delta$, which implies $b < \inverfc ( 2 \delta )$. For $0<\delta<0.5$, we  replace $y$ in Lemma~\ref{lem-inverfc} on Page~\pageref{lem-inverfc} with $2\delta$  to obtain  $\inverfc ( 2 \delta ) <  c$ for $c : = \sqrt{\ln \frac{2}{\sqrt{16\delta+1}-1}} $. Then we have $b <  c$. Thus,  Theorem~\ref{thm-Mechanism-1} implies Theorem~\ref{thm-Mechanism-2}. \qeda

  \subsection{Proof of Lemma~\ref{lem-PDP-to-DP}} \label{secprf-lem-PDP-to-DP}

The  sktech of the following proof is given in~\cite{dwork2014algorithmic}. We present the full details for completeness.

Recall that a mechanism $Y$ achieves $(\epsilon, \delta)$-differential privacy if
\begin{align}
& \pr{Y(x) \in \mathcal{Y}}\leq e^{\epsilon} \pr{Y(x')\in \mathcal{Y}} + \delta,\nonumber \\  &  \text{for any output set } \mathcal{Y}, \text{ neighboring datasets } x \text{ and } x', \label{DPeq1}
\end{align}
where the probability space is over the coin flips of the randomized mechanism $Y$, $D$ and $D'$ iterate through all pairs of neighboring datasets, and $\mathcal{Y}$ iterates through all subsets of the output range.

To achieve $(\epsilon, \delta)$-differential privacy, we first show that it suffices to ensure
\begin{align}
 & \pr{ \frac{\fr{Y(x)=y}}{\fr{Y(x')=y}}  \leq e^{\epsilon} } \geq 1- \delta,\nonumber \\  &  \text{for any output } y, \text{ neighboring datasets } x \text{ and } x',  \label{DPeq2}
\end{align}
where the probability space is over the coin flips of the randomized mechanism $Y$, $D$ and $D'$ iterate through all pairs of neighboring datasets, and $y$ iterates through the output range $\mathcal{O}$.
Specifically, we will prove that Eq.~(\ref{DPeq2}) implies Eq.~(\ref{DPeq1}).

We define set $\mathcal{S}$ by
\begin{align}
\mathcal{S} & := \left\{ y ~\bigg|~ \frac{\fr{Y(x)=y}}{\fr{Y(x')=y}}  \leq e^{\epsilon} \right\} . \label{DPeq4}
\end{align}
Then if Eq.~(\ref{DPeq2}) holds, we have
\begin{align}
  \pr{Y(x) \in \mathcal{S}}  & \geq 1- \delta. \label{DPeq5x}
\end{align}
With  $\mathcal{O}$ being the output range, $\mathcal{O} \setminus \mathcal{S} $ is the   complement set  of $\mathcal{S}$. Then Eq.~(\ref{DPeq5x}) implies
\begin{align}
  \pr{Y(x) \in \mathcal{O} \setminus \mathcal{S} }  = 1 -   \pr{Y(x) \in \mathcal{S}} & \leq   \delta. \label{DPeq5}
\end{align}
To show that Eq.~(\ref{DPeq2}) implies Eq.~(\ref{DPeq1}), we have
\begin{align}
& \pr{Y(x) \in \mathcal{Y}}  \nonumber \\ & = \pr{Y(x) \in \mathcal{Y} \cap \mathcal{S}} +  \pr{Y(x) \in \mathcal{Y} \setminus \mathcal{S}}
 \nonumber \\ & = \int_{y\in \mathcal{Y} \cap \mathcal{S}} \fr{Y(x)=y} \dr y +  \pr{Y(x) \in \mathcal{Y} \setminus \mathcal{S}}
  \nonumber \\ & \stackrel{\text{(*)}}{\leq} \int_{y\in \mathcal{Y} \cap \mathcal{S}} e^{\epsilon} \fr{Y(x')=y} \dr y +    \pr{Y(x) \in \mathcal{O} \setminus \mathcal{S}  }
   \nonumber \\ &  \stackrel{\text{(\#)}}{\leq} e^{\epsilon}\pr{Y(x') \in \mathcal{Y} \cap \mathcal{S}}  +  \delta
     \nonumber \\ & \leq e^{\epsilon} \fr{Y(x')\in \mathcal{Y}} + \delta, \label{DPeq3}
\end{align}
where the above step (*) uses Eq.~(\ref{DPeq4}) and $\mathcal{Y} \setminus \mathcal{S} \subseteq \mathcal{O} \setminus \mathcal{S}$, and step (\#) uses Eq.~(\ref{DPeq5}). \qeda

  \subsection{Proof of Lemma~\ref{lem-DP-to-PDP}}\label{secprf-lem-DP-to-PDP}

%Recall from Eq.~(\ref{eqn-L-Y-D-Dprime}) that the privacy loss is defined as
%\begin{align}
%L_{Y,D,D'}(y) := \ln \frac{\fr{Y(D)=y}}{\fr{Y(D')=y}} .
%\end{align}

For neighboring datasets $D$ and $D'$, the privacy loss $L_{Y,D,D'}(y)$ represents the multiplicative difference between the probabilities that the same output $y$ is observed when the randomized algorithm $Y$ is applied to $D$ and $D'$, respectively. Specifically, we define
\begin{align}
L_{Y,D,D'}(y) := \ln \frac{\fr{Y(D)=y}}{\fr{Y(D')=y}},
\end{align}
where $\fr{\cdot}$ denotes the probability density function.

For simplicity, we use probability density function $\fr{\cdot}$ in Eq.~(\ref{eqn-L-Y-D-Dprime}) above by assuming that the randomized algorithm $Y$ has continuous output. If $Y$ has discrete output, we replace $\fr{\cdot}$ by probability notation $\bp{\cdot}$.

When $y$ follows the probability distribution of random variable $Y(D)$, $L_{Y,D,D'}(y)$ follows the probability distribution of random variable $L_{Y,D,D'}(Y(D))$.

We have Lemmas~\ref{lem-eps-delta-DP-LD-vs-LDprime} and~\ref{lem-LD-vs-LDprime} below, which will be proved soon.

\begin{lem} \label{lem-eps-delta-DP-LD-vs-LDprime}
Given datasets $D$, $D'$, and an $(\epsilon,\delta)$-differentially private randomized algorithm $Y$, for any real number $t$, it holds that
\begin{align}
\bp{L_{Y,D,D'}(Y(D)) \geq t} &  \leq \frac{\delta}{1-e^{\epsilon-t} }.
\end{align}
\end{lem}

\begin{lem} \label{lem-LD-vs-LDprime}
The relationships between privacy loss random variables $L_{Y,D,D'}(Y(D))$ and $L_{Y,D',D}(Y(D'))$ are as follows. Given datasets $D$, $D'$, and a randomized algorithm $Y$, for any real number $t$, it holds that
\begin{align}
&\bp{L_{Y,D,D'}(Y(D)) \leq - t}  \leq  e^{-t} \bp{L_{Y,D',D}(Y(D')) \geq t}.
\end{align}
\end{lem}

\noindent \textbf{Proof of Lemma~\ref{lem-DP-to-PDP}:} The result follows from Lemmas~\ref{lem-eps-delta-DP-LD-vs-LDprime} and~\ref{lem-LD-vs-LDprime}. \qeda

\noindent \textbf{Proof of Lemma~\ref{lem-eps-delta-DP-LD-vs-LDprime}:} Since $L_{Y,D,D'}(Y(\cdot)) $ can be seen as post-processing on $Y(\cdot)$ and hence also satisfies $(\epsilon,\delta)$-differential privacy, we have
\begin{align}
&\bp{L_{Y,D,D'}(Y(D)) \geq t }  \nonumber
\\ & \leq \delta +  e^{\epsilon} \bp{L_{Y,D,D'}(Y(D')) \geq t }   \nonumber
\\ & = \delta + e^{\epsilon} \int_{\mathcal{Y}} \bfu{Y(D') = y} \bp{L_{Y,D,D'}(y) \geq t } \, \textup{d}y   \nonumber
\\ & = \delta + e^{\epsilon} \int_{\mathcal{Y}} \bfu{Y(D') = y} \bp{ \begin{array}{l}  \fr{Y(D)=y} \\ \geq e^{t} \fr{Y(D')=y}\end{array}} \, \textup{d}y   \nonumber
\\ & \leq \delta + e^{\epsilon} \int_{\mathcal{Y}} e^{-t}\bfu{Y(D) = y} \bp{ \begin{array}{l}  \fr{Y(D)=y} \\ \geq e^{t} \fr{Y(D')=y}\end{array}} \, \textup{d}y   \nonumber
\\ & = \delta + e^{\epsilon-t} \bp{L_{Y,D,D'}(Y(D)) \geq t }
\end{align}\qeda

\noindent \textbf{Proof of Lemma~\ref{lem-LD-vs-LDprime}:} We have
\begin{align}
&\bp{L_{Y,D,D'}(Y(D)) \leq - t}  \nonumber
% \\ & =  \int_{\mathcal{Y}} \bfu{Y(D) = y} \bp{ \begin{array}{l}  \fr{Y(D)=y} \\ < e^{-t} \fr{Y(D')=y}\end{array}} \, \textup{d}y  \nonumber
\\ & =  \int_{\mathcal{Y}} \bfu{Y(D) = y} \bp{ \begin{array}{l}  \fr{Y(D)=y} \\  \leq  e^{-t} \fr{Y(D')=y}\end{array}} \, \textup{d}y \nonumber
\\ &  \leq  \int_{\mathcal{Y}} e^{-t} \fr{Y(D')=y} \bp{ \begin{array}{l} \fr{Y(D')=y}  \\ \geq e^{t} \fr{Y(D)=y} \end{array}} \, \textup{d}y  \nonumber
\\ &  = e^{-t} \bp{L_{Y,D',D}(Y(D')) \geq t} . \nonumber
%\\ &  \leq   e^{-t} \cdot \frac{\delta}{1-e^{\epsilon-t}}
\end{align}  \qeda

\subsection{Proof of Theorem~\ref{thm-pDP-OPT}} \label{sec:proof3}

% \subsection{Proof of \texttt{\upshape Mechanism~3} achieving \texorpdfstring{$(\epsilon, \delta)$-differential privacy}:}
%

%
%
%In this Section, we will show the relationships of $(\epsilon,\delta)$-Differential Privacy, $(\epsilon,\delta)$-Probabilistic Differential Privacy, and Concentrated differential privacy.

% \subsection{The Relationships of $(\epsilon,\delta)$-Differential Privacy and $(\epsilon,\delta)$-Probabilistic Differential Privacy}

For the proposed Gaussian mechanism, we now
  prove
\begin{align}
&\mathbb{P}_{y \sim Y(D) }\left[{ e^{-\epsilon}  \leq  \frac{\fr{Y(D)=y}}{\fr{Y(D')=y}}  \leq e^{\epsilon} }\right] \geq 1- \delta, \nonumber \\  &  \text{for any output } y, \text{ neighboring datasets } D \text{ and } D'. \label{DPeq6}
\end{align}
The desired result Eq.~(\ref{DPeq6}) can also be written as
\begin{align}
&\mathbb{P}_{y \sim Y(D) }\left[{  \left| \ln \frac{\fr{Y(D)=y}}{\fr{Y(D')=y}} \right|   \leq \epsilon } \right] \geq 1- \delta, \nonumber \\  &  \text{for any output } y, \text{ neighboring datasets } D \text{ and } D'. \label{DPeq7}
\end{align}
With  $(\epsilon, \delta)$-differential privacy being translated to Eq.~(\ref{DPeq7}), we will show that the minimal noise amount  can be derived, while the classic mechanism by Dwork and Roth~\shortcite{dwork2014algorithmic} presents only a loose bound.

%(\ref{DPeq7}) is in fact $(\epsilon, \delta)$-probabilistic differential privacy of Machanavajjhala~\textit{et al.}~\cite{machanavajjhala2008privacy}.

%===========
%
%{\color{red}
%
 Let the output of the query $Q$ on the dataset $D$ be an $m$-dimensional vector. We define notation $r_1, \ldots, r_m$ such that
\begin{align}
y - Q(D) = [r_1, \ldots, r_m] .\label{YtQtmt1}
\end{align}

Since $Y(D)$ is the result of adding a zero-mean Gaussian noise with standard deviation $\sigma$ to $Q(D)$, we have
\begin{align}
&\fr{Y(D)=y}  = \prod_{j=1}^m \left( \frac{1}{\sqrt{2\pi {\sigma}^2\,}} e^{-\frac{{r_j}^2}{2{\sigma}^2}}\right) . \label{YtGauss1}
\end{align}

 We introduce notation $s_1, \ldots, s_m$ such that
\begin{align}
Q(D)  - Q(D') = [s_1, \ldots, s_m] .\label{YtQtmst1}
\end{align}
From   Eq.~(\ref{YtQtmt1}) and (\ref{YtQtmst1}), it holds that
\begin{align}
y - Q(D') = [r_1+s_1, \ldots, r_m+s_m] .\label{YtQtmt1v2}
\end{align}
Since $Y(D')$ is the result of adding a zero-mean Gaussian noise with standard deviation $\sigma$ to $Q(D')$, we have
\begin{align}
&\fr{Y(D')=y}  = \prod_{j=1}^m \left( \frac{1}{\sqrt{2\pi {\sigma}^2\,}} e^{-\frac{{(r_j+s_j)}^2}{2{\sigma}^2}}\right) . \label{YtGauss1v2}
\end{align}
The combination of Eq.~(\ref{YtGauss1}) and (\ref{YtGauss1v2}) induces
\begin{align}
\ln \frac{\fr{Y(D)=y}}{\fr{Y(D')=y}} & =\ln  \frac{~~~ \prod_{j=1}^m \left( \frac{1}{\sqrt{2\pi {\sigma}^2\,}} e^{-\frac{{r_j}^2}{2{\sigma}^2}}\right) ~~~}{~~~ \prod_{j=1}^m \left( \frac{1}{\sqrt{2\pi {\sigma}^2\,}} e^{-\frac{{(r_j+s_j)}^2}{2{\sigma}^2}}\right)~~~}    \nonumber \\ & = \sum_{j=1}^m \left[\frac{(r_j+s_j)^2}{2{\sigma}^2} - \frac{{r_j}^2}{2{\sigma}^2} \right] \nonumber \\ & = \frac{ \sum_{j=1}^m(s_j r_j)}{{\sigma}^2} +   \frac{\sum_{j=1}^m{s_j}^2}{2{\sigma}^2}
.\label{DPeq14newq}
\end{align}

We define
\begin{align}
S: = \sqrt{ \sum_{j=1}^m  {s_j}^2} ,\label{defineSt}
\end{align}
and
\begin{align}
G: = \frac{ \sum_{j=1}^m(s_j r_j)}{S} . \label{defineGS}
\end{align}
From   Eq.~(\ref{YtQtmst1}), $S$ is the $\ell_2$ distance between $Q(D) $ and $ Q(D') $; i.e.,
\begin{align}
S: =  \|Q(D) - Q(D')\|_{2} .\label{defineStl2}
\end{align}
 Note that $r_j$ for each $j\in\{1,2,\ldots,m\}$ defined in Eq.~(\ref{YtQtmt1}) is a zero-mean Gaussian random variable with standard deviation $\sigma$. In addition, $r_1, \ldots, r_m$ are independent. Hence,
\begin{align}
&\text{$G$ defined as $\frac{ \sum_{j=1}^m(s_j r_j)}{S}$ is a zero-mean Gaussian} \nonumber \\ & \text{random variable with variance $\frac{\sum_{j=1}^m ({s_j}^2 {\sigma}^2)}{S^2} =   {\sigma}^2 $}, \label{defineGS2}
\end{align}
 where the last step uses   Eq.~(\ref{defineSt}).
 For notational simplicity, we write $G \sim \text{Gaussian}(0, {\sigma}^2 ) $.

From Eq.~(\ref{defineSt}) and (\ref{defineGS}), it follows that
\begin{align}
\ln \frac{\fr{Y(D)=y}}{\fr{Y(D')=y}} &   = \frac{GS}{{\sigma}^2} +   \frac{S^2}{2{\sigma}^2}
.\label{DPeq14newqv3}
\end{align}
Hence, we have
\begin{align}
&\mathbb{P}_{y \sim Y(D) }\left[{  \left| \ln \frac{\fr{Y(D)=y}}{\fr{Y(D')=y}} \right|   \leq \epsilon } \right] \nonumber \\ &=  \mathbb{P}_{G \sim \text{Gaussian}(0, {\sigma}^2 )}\left[{  \left|  \frac{GS}{{\sigma}^2} +   \frac{S^2}{2{\sigma}^2}\right|   \leq \epsilon} \right]  .\label{DPeq192mec31}
\end{align}
%Based on (\ref{DPeq192mec31}), below we discuss  $h > 0$, $h < 0$, and $h = 0$, respectively.
If $S > 0$, given the result Eq.~(\ref{defineGS}) that $G$ is a zero-mean Gaussian random variable with standard deviation $\sigma$, we obtain
\begin{align}
&\mathbb{P}_{G \sim \text{Gaussian}(0, {\sigma}^2 ) }\left[{  \left|  \frac{GS}{{\sigma}^2} +   \frac{S^2}{2{\sigma}^2}\right|   \leq \epsilon} \right] \nonumber \\ &=  \pr{ -\frac{\epsilon{\sigma}^2}{S}-\frac{S}{2} \leq  G  \leq \frac{\epsilon{\sigma}^2}{S}-\frac{S}{2} }  \nonumber \\ &= \frac{1}{2} \left[1+\erf\left(\frac{\frac{\epsilon{\sigma}^2}{S}-\frac{S}{2}}{\sigma\sqrt{2}}\right)\right] -\frac{1}{2} \left[1+\erf\left(\frac{-\frac{\epsilon{\sigma}^2}{S}-\frac{S}{2}}{\sigma\sqrt{2}}\right)\right]  \nonumber \\ &= \frac{1}{2}  \erf\left(\frac{\frac{\epsilon{\sigma}^2}{S}-\frac{S}{2}}{\sigma\sqrt{2}}\right) + \frac{1}{2} \erf\left(\frac{\frac{\epsilon{\sigma}^2}{S}+\frac{S}{2}}{\sigma\sqrt{2}}\right) \nonumber \\ &= f_{\epsilon,\,\sigma}(S) ,\label{DPeq192mec32}
\end{align}
for $f_{\epsilon,\,\sigma}(S)$ defined by
\begin{align}
f_{\epsilon,\,\sigma}(S):= \frac{1}{2} \left[\erf\left(\frac{\frac{\epsilon{\sigma}^2}{S}-\frac{S}{2}}{\sigma\sqrt{2}}\right) +\erf\left(\frac{\frac{\epsilon{\sigma}^2}{S}+\frac{S}{2}}{\sigma\sqrt{2}}\right)\right]  ,\label{deffh}
\end{align}
 where Eq.~(\ref{DPeq192mec32}) uses the cumulative distribution function of a zero-mean Gaussian random variable $G$ as well as the fact that $\erf(\cdot)$ is an odd function; i.e., $\erf(-x)=-\erf(x)$.

%If $h < 0$, we use the result (\ref{defineGS}) that $G$ is a zero-mean Gaussian random variable with standard deviation $\sigma$, and obtain
%\begin{align}
% &\pr{  \left|  \frac{GS}{{\sigma}^2} +   \frac{S^2}{2{\sigma}^2}\right|   \leq \epsilon} \nonumber \\ &=  \pr{\frac{\epsilon{\sigma}^2}{h}-\frac{h}{2} \leq  G  \leq  -\frac{\epsilon{\sigma}^2}{h}-\frac{h}{2} }   \nonumber \\ &=   \frac{1}{2} \left[1+\erf\left(\frac{-\frac{\epsilon{\sigma}^2}{h}-\frac{h}{2}}{\sigma\sqrt{2}}\right)\right] - \frac{1}{2} \left[1+\erf\left(\frac{\frac{\epsilon{\sigma}^2}{h}-\frac{h}{2}}{\sigma\sqrt{2}}\right)\right]  \nonumber \\ &= -f_{\epsilon,\,\sigma}(h),\label{DPeq192mec33}
%\end{align}
%for $f_{\epsilon,\,\sigma}(h)$ defined by (\ref{deffh}), where the last step of (\ref{DPeq192mec33}) uses the cumulative distribution function of a zero-mean Gaussian random variable $G$ and the fact that the  error function $\erf(\cdot)$ is an odd function.

If $S = 0$, it is clear that
\begin{align}
\mathbb{P}_{G \sim \text{Gaussian}(0, {\sigma}^2 ) }\left[{  \left|  \frac{GS}{{\sigma}^2} +   \frac{S^2}{2{\sigma}^2}\right|   \leq \epsilon} \right]  =  1  .\label{DPeq192mec35}
\end{align}

%
%
%\section{Two Alternative Gaussian Mechanisms with Simpler Expressions for the Noise Amount}
%
%
%\section{A Higher-Utility Gaussian Mechanism with More Complex Expression for the Noise Amount}
%

The $\ell_2$-sensitivity $\Delta$ of the query $Q$ is the maximal $\ell_2$ distance between the (true) query outputs for any two neighboring datasets $D$ and $D'$ that differ in one record: \mbox{$\Delta  = \max_{\textrm{neighboring $D,D'$}} \|Q(D) - Q(D')\|_{2}$}. From  Eq.~(\ref{defineStl2}), we have $0 \leq S \leq \Delta$. Then
summarizing Eq.~(\ref{DPeq192mec32})  and (\ref{DPeq192mec35}),  to guarantee Eq.~(\ref{DPeq7}), it suffices to ensure
% \begin{subnumcases}{\hspace{-3pt}}
% \hspace{-3pt}\text{For any }0<h\leq \Delta, \text{ it holds that} \nonumber \\
% \hspace{-3pt}\erf\left(\frac{\frac{\epsilon{\sigma}^2}{h}-\frac{h}{2}}{\sigma\sqrt{2}}\right) - \erf\left(\frac{-\frac{\epsilon{\sigma}^2}{h}-\frac{h}{2}}{\sigma\sqrt{2}}\right) \geq 2( 1 - \delta). \\
% \hspace{-3pt}\text{For any }-\Delta \leq h < 0, \text{ it holds that} \nonumber \\
% \hspace{-3pt}\erf\left(\frac{-\frac{\epsilon{\sigma}^2}{h}-\frac{h}{2}}{\sigma\sqrt{2}}\right) -\erf\left(\frac{\fray{\epsilon{\sigma}^2}{h}-\frac{h}{2}}{\sigma\sqrt{2}}\right)
% \geq 2( 1 - \delta).
% \end{subnumcases}
\begin{align}
 f_{\epsilon,\,\sigma}(S)  \geq  1 - \delta, \text{ for } 0<S\leq \Delta. \label{fh1delta1}
\end{align}
%\begin{subnumcases}{\hspace{-19pt}}
%\hspace{-3pt}\text{For any }0\hspace{-1pt}<\hspace{-1pt}S\hspace{-1pt}\leq\hspace{-1pt} \Delta, \text{ it holds that }  f_{\epsilon,\,\sigma}(S) \hspace{-1pt} \geq\hspace{-1pt}  1 \hspace{-1pt}-\hspace{-1pt} \delta . \label{fh1delta1} \\
%\hspace{-3pt}\text{For any }-\hspace{-2pt}\Delta \hspace{-1pt}\leq\hspace{-1pt} S \hspace{-1pt}<\hspace{-1pt} 0, \text{ it holds that } -\hspace{-2pt}f_{\epsilon,\,\sigma}(S)  \hspace{-1pt}\geq\hspace{-1pt}  1\hspace{-1pt} -\hspace{-1pt} \delta. \nonumber \\ \label{fh1delta2}
%\end{subnumcases}
%We note that $f_{\epsilon,\,\sigma}(S)$ defined by (\ref{deffh}) is an odd function (i.e., $f_{\epsilon,\,\sigma}(-S)=-f_{\epsilon,\,\sigma}(S)$), since $\erf(\cdot)$ is an odd function. Then
%(\ref{fh1delta1}) and (\ref{fh1delta2}) are equivalent, so we only need to require (\ref{fh1delta1}).

We can prove  that $ f_{\epsilon,\,\sigma}(S)$ is a decreasing function of $S$. Hence, the optimal Gaussian mechanism to achieve  $(\epsilon,\delta)$-probabilistic differential privacy satisfies $ f_{\epsilon,\,\sigma}(\Delta)  =  1 - \delta$. Defining $d$ as $\frac{\frac{\epsilon{\sigma}^2}{\Delta}-\frac{\Delta}{2}}{\sigma\sqrt{2}}$ and solving $\sigma$, we obtain the desired result. \qeda

  \subsection{Proof of Theorem~\ref{thm-pDP-OPT-asymptotics}} \label{secprf-thm-pDP-OPT-asymptotics}

 \ding{172} Given a fixed $0<\delta<1$, we have $ \lim_{\epsilon \to 0 } d = \inverfc(\delta) $, which results in $ \lim_{\epsilon \to 0}\sigma_{\texttt{\upshape pDP-OPT}} \Big/ \left(\frac{\inverfc(\delta) \cdot \Delta }{\epsilon\sqrt{2} } \right) = 1$.

 \ding{173} Given a fixed $0<\delta<1$, we have $ \lim_{\epsilon \to \infty } d = 0 $, which leads to $ \lim_{\epsilon \to \infty}\sigma_{\texttt{\upshape pDP-OPT}} \Big/ \left( \frac{\Delta}{\sqrt{2\epsilon\hspace{1.5pt}}} \right) = 1 $.

\ding{174} Given a fixed $\epsilon >0 $, we use Lemma~\ref{lem-pDP-OPT-bounds} to derive $ \lim_{\delta \to 0} d  \sqrt{  \ln \frac{1}{\delta}  }= 1 $ and thus $ \lim_{\delta \to 0}\sigma_{\texttt{\upshape pDP-OPT}} \Big/ \left( \frac{\Delta}{\epsilon} \sqrt{2  \ln \frac{1}{\delta}  }\hspace{1.5pt}\right) = 1 $. \qeda

 \subsection{Establishing  Lemma~\ref{lem-pDP-OPT-bounds}, which along with Theorem~\ref{thm-pDP-OPT} implies Theorem~\ref{thm-Mechanism-3}}\label{secprf-lem-pDP-OPT-bounds}

 From the definition of $d$ in Eq.~(\ref{eqn-u-pDP-OPT}): $\erfc\left(d\right) +  \erfc\left( \sqrt{d^2 + \epsilon} \right) = 2 \delta $, we clearly have $  \delta < \erfc\left(d\right) <   2 \delta $, which implies   $\inverfc(2\delta) <d < \inverfc(\delta) $. \qeda

\def\x{\protect\ref{lem-f-vs-g}}
\def\y{\protect\ref{thm-Mechanism-3}}
\def\z{\protect\ref{thm-Mechanism-4}}

\subsection{Proving Lemma~\protect\x, which along with Theorem~\protect\y~implies Theorem~\protect\z}~ \label{appenlem-erf-2}

To show Lemma~\ref{lem-f-vs-g}, it suffices to prove  $\sqrt{\ln \frac{2}{\sqrt{8\delta+1}-1}} > \inverfc(\delta)$ for \mbox{$0<\delta<1$}. This clearly follows from Lemma~\ref{lem-inverfc} proved in Appendix~\ref{sec-lem-inverfc} by replacing $y$ with $\delta$. \pfe
 %$\inverfc(\delta) \leq \sqrt{\ln \frac{2}{\sqrt{8\delta+1}\hspace{2pt}-\hspace{2pt}1}}$.
 %\noindent \textbf{Proof of Lemma \ref{lem-erf-2}:}
% Eq.~(\ref{lem2eqCraig5}) can also be shown using
% the tail bound for the probabilty mass function of a Gaussian variable. ?
%https://mikespivey.wordpress.com/2011/10/21/normaltails/
% The combination of~(\ref{eqCraigerf}) and~(\ref{lem2eqCraig5}) induces~(\ref{lem2erfcfx}).

\def\x{\protect\ref{Alg-opt}}
\def\y{\protect\ref{thm-DP-OPT}}

\subsection{Algorithm~\protect\x~to compute $\sigma_{\texttt{\upshape DP-OPT}}$ of Theorem~\protect\y} \label{appendix-sec-Alg-opt}

As discussed at the end of Section~\ref{sec-DP-ours}, the   noise amounts of our mechanisms can be set as initial values to quickly search for  the optimal value $\sigma_{\texttt{\upshape DP-OPT}}$. In particular, Algorithm~\protect\x~to compute $\sigma_{\texttt{\upshape DP-OPT}}$ will
use Lemma~\ref{lem-eqn-u-DP-OPT-bound-a} below.

\begin{algorithm}
\caption{Computing $\sigma_{\texttt{\upshape DP-OPT}}$ of Theorem~\ref{thm-DP-OPT} based on Lemma~\ref{lem-eqn-u-DP-OPT-bound-a}.} \label{Alg-opt}
\begin{algorithmic}[1]
\STATE $ \text{diff}  \leftarrow 1  -   e^{\epsilon} \erfc\left( \sqrt{ \epsilon} \right) - 2 \delta$;
\IF{$\text{diff} = 0$}
\STATE $a \leftarrow 0$;
\ELSIF{$\text{diff} > 0$}
\STATE $\textit{N} \leftarrow 1$;
\STATE $\textit{lower} \leftarrow 0$;
\STATE $\textit{upper} \leftarrow \text{any one of the following:}$ \label{algupper}
\item[]\mbox{$b$ of Eq.~(\ref{eqn-u-Mechanism-1}) in our Theorem~\ref{thm-Mechanism-1} for \texttt{\upshape Mechanism~1},} \item[]\mbox{$c$ of Eq.~(\ref{eqn-u-Mechanism-2}) in our Theorem~\ref{thm-Mechanism-2} for \texttt{\upshape Mechanism~2};}
% \item[]$b_*$ of Eq.~(\ref{eq-bstar}),
% \item[]$b_{\#}$ of Eq.~(\ref{eq-bpound});
\WHILE{$N\leq\textit{the allowed maximum number of iterations}$}
\IF{$\hspace{2pt}\textit{upper}-\textit{lower} < \textit{tolerance}$}
\STATE $a \leftarrow upper$;
\STATE $\textbf{break}$
\ENDIF
\STATE $\textit{mid} \leftarrow (lower+upper)/2$;
\IF{$\erfc\left(\textit{mid} \right)   -  e^{\epsilon} \erfc\left( \sqrt{\textit{mid}^2 + \epsilon} \right)  =  2 \delta \hspace{2pt} $}
\STATE $a \leftarrow \textit{mid}$;
\STATE $\textbf{break}$
\ELSIF{$\erfc\left(\textit{mid} \right)   -  e^{\epsilon} \erfc\left( \sqrt{\textit{mid}^2 + \epsilon} \right)  >  2 \delta$}
\STATE $\textit{lower} \leftarrow \textit{mid}$;
\ELSE
\STATE $\textit{upper} \leftarrow \textit{mid}$;
\ENDIF
\STATE $N \leftarrow N+1$;
\ENDWHILE
\ELSE
\STATE $\textit{N} \leftarrow 1$;
\STATE $\textit{lower} \leftarrow - \inverfc\left(\frac{2-2 \delta}{e^{\epsilon}+1}\right) $;
\STATE $\textit{upper} \leftarrow 0$;

\WHILE{$N\leq\textit{the allowed maximum number of iterations}$}
\IF{$\hspace{2pt}\textit{upper}-\textit{lower} < \textit{tolerance}$}
\STATE $a \leftarrow upper$;
\STATE $\textbf{break}$
\ENDIF
\STATE $\textit{mid} \leftarrow (lower+upper)/2$;
\IF{$\erfc\left(\textit{mid} \right)   -  e^{\epsilon} \erfc\left( \sqrt{\textit{mid}^2 + \epsilon} \right)  =  2 \delta \hspace{2pt} $}
\STATE $a \leftarrow \textit{mid}$;
\STATE $\textbf{break}$
\ELSIF{$\erfc\left(\textit{mid} \right)   -  e^{\epsilon} \erfc\left( \sqrt{\textit{mid}^2 + \epsilon} \right)  >  2 \delta$}
\STATE $\textit{lower} \leftarrow \textit{mid}$;
\ELSE
\STATE $\textit{upper} \leftarrow \textit{mid}$;
\ENDIF
\STATE $N \leftarrow N+1$;
\ENDWHILE
\ENDIF
\STATE $\sigma_{\texttt{\upshape DP-OPT}}  \leftarrow \frac{\left(a+\sqrt{a^2+\epsilon} \hspace{1.5pt}\right)  \cdot \Delta }{\epsilon\sqrt{2}} $;
\RETURN $\sigma_{\texttt{\upshape DP-OPT}}$
\end{algorithmic}
\end{algorithm}

\begin{lem} \label{lem-eqn-u-DP-OPT-bound-a}
We have the following bounds for $a$ in Eq.~(\ref{eqn-sigma-DP-OPT}) of Theorem~\ref{thm-DP-OPT}:
\begin{subnumcases}{\hspace{-25pt}}
\hspace{-5pt} 0  < a < \text{$b$ of Eq.~(\ref{eqn-u-Mechanism-1})}
 &\nonumber\\
%  ~~~~~< \text{$b_*$ of Eq.~(\ref{eq-bstar})}& \nonumber\\   ~~~~~< \text{$b_{\#}$ of Eq.~(\ref{eq-bpound})} &\nonumber\\
 ~~~~~
< \text{$c$ of Eq.~(\ref{eqn-u-Mechanism-2})}, &  ~~~\text{if $ 1  \hspace{-2pt}-\hspace{-2pt}   e^{\epsilon} \erfc\left( \sqrt{ \epsilon} \right)  \hspace{-2pt} > \hspace{-2pt} 2 \delta $;} \label{eqn-bound-on-a-case1}
\\ \hspace{-5pt}a\hspace{-2pt}=\hspace{-2pt} 0, & ~~~\text{if $ 1  \hspace{-2pt}-\hspace{-2pt}   e^{\epsilon} \erfc\left( \sqrt{ \epsilon} \right) \hspace{-2pt} = \hspace{-2pt} 2 \delta $;}  \label{eqn-bound-on-a-case2} \\ \hspace{-5pt}\mathsmaller{- \inverfc\left(\frac{2-2 \delta}{e^{\epsilon}+1}\right) \hspace{-2pt}<\hspace{-2pt} a \hspace{-2pt}<\hspace{-2pt} 0,} & ~~~\text{if $ 1 \hspace{-2pt} -\hspace{-2pt}   e^{\epsilon} \erfc\left( \sqrt{ \epsilon} \right)  \hspace{-2pt} <\hspace{-2pt} 2 \delta $.} \label{eqn-bound-on-a-case3}
\end{subnumcases}
\end{lem}

\noindent \textbf{Proof of Lemma~\ref{lem-eqn-u-DP-OPT-bound-a}:}
First, (\ref{eqn-bound-on-a-case1}) follows from
Lemmas~\ref{lem-a-vs-b} and~\ref{lem-b-vs-c}. Second, (\ref{eqn-bound-on-a-case2}) holds from Eq.~(\ref{eqn-sigma-DP-OPT}) and Remark~\ref{rem-Mechanism-DP-OPT-ru}. Next, we prove Eq.~(\ref{eqn-bound-on-a-case3}) as follows.

From Lemma~\ref{lemma-subsection-rem-Mechanism-DP-OPT-ru-detailed} on Page~\pageref{lemma-subsection-rem-Mechanism-DP-OPT-ru-detailed}, we have
\begin{subnumcases}{\hspace{-10pt}}
\textup{If $ 0.5 \leq \delta <1 $,} \nonumber \\ \textup{then any $\epsilon > 0$ satisfies } \nonumber\\ \textup{$ 1 - e^{\epsilon} \erfc\left( \sqrt{ \epsilon} \right) < 2 \delta $.} \nonumber\\[4pt]
\textup{If $0<\delta<0.5$, then  $0<\epsilon<\epsilon_*$ satisfies } \nonumber\\ \textup{$ 1 - e^{\epsilon} \erfc\left( \sqrt{ \epsilon} \right) < 2 \delta $,} \nonumber\\ \textup{where $\epsilon_*$ denotes the solution to $e^{\epsilon_*} \erfc\left( \sqrt{ \epsilon_*} \right)  =  1 -  2 \delta$.} \nonumber
\end{subnumcases}
Then we use Eq.~(\ref{eqanega0}) and~(\ref{eqanega0v2}) to obtain (\ref{eqn-bound-on-a-case3}).
\qeda

\newpage

Note that in practice, due to $\epsilon \geq 0.01$ and $\delta \leq 0.05$, we have (\ref{eqn-bound-on-a-case1}) as explained in Appendix~\ref{subsection-rem-Mechanism-DP-OPT-ru-detailed}. We present  (\ref{eqn-bound-on-a-case2}) and (\ref{eqn-bound-on-a-case3}) for completeness.
%
%\begin{lem} \label{Alg-opt-accuracy}

To ensure $\sigma$ returned by Algorithm~\ref{Alg-opt} satisfies \mbox{$0 \leq \sigma - \sigma_{\texttt{\upshape DP-OPT}} \leq \zeta$}  for some $\zeta \geq 0$, we clearly have the following results on the computational complexity of Algorithm~\ref{Alg-opt}:
\begin{itemize}
    \item If $ 1  -   e^{\epsilon} \erfc\left( \sqrt{ \epsilon} \right) > 2 \delta $,  then % ``for'' loops in Line ? of
    Algorithm~\ref{Alg-opt} takes at most $\log_2 \frac{b}{\zeta} $ iterations (resp., $\log_2 \frac{c}{\zeta} $) if Line~\ref{algupper} uses $b$ of Eq.~(\ref{eqn-u-Mechanism-1}) (resp., $c$ of Eq.~(\ref{eqn-u-Mechanism-2})), with each iteration having $O(1)$ complexity. The total complexity is $O\left(\log_2 \frac{b}{\zeta}\right)$ (resp., $O\left(\log_2 \frac{c}{\zeta}\right)$).
    \item If $ 1  -   e^{\epsilon} \erfc\left( \sqrt{ \epsilon} \right) < 2 \delta $,  then %``for'' loops in Line ? of
     Algorithm~\ref{Alg-opt} takes at most $\log_2 \frac{\inverfc\left(\frac{2-2 \delta}{\exp(\epsilon)+1}\right)}{\zeta} $ iterations, with each iteration having $O(1)$ complexity. The total complexity is $O\left(\log_2 \frac{\inverfc\left(\frac{2-2 \delta}{\exp(\epsilon)+1}\right)}{\zeta} \right)$.
\end{itemize}

\def\x{\protect\ref{Gaussiancomposition}}

\subsection{Analyses of $(\epsilon,\delta)$-Differential Privacy and $(\epsilon,\delta)$-Probabilistic Differential Privacy for the Composition of Gaussian Mechanisms} \label{sec-main-composition}

\iffalse Comment out in CCS

Meiser and Mohammadi~\shortcite{meiser2018tight}

 Comment out in CCS \fi

This section provides   analyses of $(\epsilon,\delta)$-differential privacy and $(\epsilon,\delta)$-probabilistic differential privacy for the composition of Gaussian mechanisms.

% $\sigma_{\texttt{\upshape DP-OPT}}$?

\begin{lem}\label{Gaussiancomposition}
For $m$ queries $Q_1, Q_2, \ldots, Q_m$ with
$\ell_2$-sensitivity   $\Delta_1, \Delta_2, \ldots, \Delta_m$, if the query result of $Q_i$ is added with independent Gaussian noise of standard deviation $\sigma_i$, we have the following results.
\begin{itemize}
\item[i)] The \textbf{differential privacy (DP)} level for the composition of the $m$ noisy answers is the same as that of a Gaussian mechanism with noise amount
\begin{align}
\sigma_{*}:= \left(\sum_{i=1}^m \frac{  {\Delta_i}^2 }{{\sigma_i}^2}\right)^{-1/2} \label{eqnsigmastar}
\end{align}
 for a query with $\ell_2$-sensitivity   $1$. %  the composition of the $m$ noisy answers satisfies \textbf{$(\epsilon,\delta)$-differential privacy} if
% % for
% % \begin{align}\label{lnYxYxprimechainruleY3eq1}
% %  \frac{1}{2} \left[ \erf\left( \frac{\epsilon - A/2}{\sqrt{2 A}} \right) + \erf\left( \frac{\epsilon + A/2}{\sqrt{2 A}} \right)  \right]   & =   1- \delta,  \text{~with } A:= \sum_{t=1}^T \frac{  {\Delta_t}^2 }{{\sigma_t}^2}.\label{lnYxYxprimechainruleY3eq2}
% % \end{align}
% % \begin{lemma}\label{seccomposition}
% %  \texttt{\upshape Gaussian noise based mechanism}, or  $\frac{1}{\epsilon} \sqrt{2\ln \frac{1.25}{\delta}}$ of Dwork and Roth~\cite{DR13} are just different upper bounds for $ \frac{1}{\sqrt{A}} $.
% %  \end{lemma}
% \begin{align}
% \sum_{i=1}^m \frac{  {\Delta_i}^2 }{{\sigma_i}^2} \leq \frac{1}{({\sigma_{\epsilon,\delta}^{\textup{DP}}})^2},
% \end{align}
%  for $\sigma_{\epsilon,\delta}$
% given by Eq.~(\ref{eqn-sigma-DP-OPT}) and~(\ref{eqn-sigma-DP-OPT}) with $\Delta=1$ in Theorem~\ref{thm-DP-OPT};
\item[ii)]   The \textbf{probabilistic differential privacy (pDP)} level for the composition of the $m$ noisy answers is the same as that of a Gaussian mechanism with noise amount $\sigma_{*}$ in Eq.~(\ref{eqnsigmastar})
 for a query with $\ell_2$-sensitivity   $1$. % the composition of the $m$ noisy answers satisfies \textbf{$(\epsilon,\delta)$-probabilistic differential privacy} if \begin{align}
% \sum_{i=1}^m \frac{  {\Delta_i}^2 }{{\sigma_i}^2} \leq \frac{1}{({\sigma_{\epsilon,\delta}^{\textup{pDP}}})^2},
% \end{align}
%  for $\sigma_{\texttt{\upshape pDP-OPT}}$
% given by Equations~(\ref{eqn-u-pDP-OPT}) and~(\ref{eqn-sigma-pDP-OPT}) with \mbox{$\Delta=1$}  in Theorem~\ref{thm-pDP-OPT}.
\end{itemize}

% ; i.e.,
% \begin{subnumcases}{\hspace{-22pt}}
% \text{Solve $a$ such that } \erfc\left(a \right)   -  e^{\epsilon} \erfc\left( \sqrt{a^2 + \epsilon} \right)  =  2 \delta; \label{eqn-u-DP-OPT-repeat}
% \\ \sigma_{\epsilon,\delta}  := \frac{\left(a+\sqrt{a^2+\epsilon} \hspace{1.5pt}\right)  \cdot \Delta }{\epsilon\sqrt{2}}  . \label{eqn-sigma-DP-OPT-repeat}
% \end{subnumcases}
 \end{lem}

\begin{rem}~
\begin{itemize}
\item Result i) of Lemma~\ref{Gaussiancomposition} implies the following. Let ${\sigma_{\epsilon,\delta}^{\textup{DP}}}$ be a Gaussian noise amount which achieves \textbf{$(\epsilon, \delta)$-DP} for a query with $\ell_2$-sensitivity   $1$, where the expression of ${\sigma_{\epsilon,\delta}^{\textup{DP}}}$ can follow from classical ones \texttt{\upshape Dwork-2006} and \texttt{\upshape Dwork-2014}  of~\cite{dwork2006our,dwork2014algorithmic} (when $ \epsilon \leq 1$), the optimal one \texttt{\upshape DP-OPT} of Theorem~\ref{thm-DP-OPT}, or  our proposed mechanisms \texttt{\upshape Mechanism~1} of Theorem~\ref{thm-Mechanism-1} and \texttt{\upshape Mechanism~2} of Theorem~\ref{thm-Mechanism-2}. Then the above composition satisfies \textbf{$(\epsilon,\delta)$-DP} for $\epsilon$ and $\delta$ satisfying $\sigma_{*}\geq {\sigma_{\epsilon,\delta}^{\textup{DP}}}$ with $\sigma_{*}$ defined above.
\item  Result ii) of Lemma~\ref{Gaussiancomposition} implies the following. Let ${\sigma_{\epsilon,\delta}^{\textup{pDP}}}$ be a Gaussian noise amount which achieves \textbf{$(\epsilon, \delta)$-pDP} for a query with $\ell_2$-sensitivity   $1$, where the expression of ${\sigma_{\epsilon,\delta}^{\textup{pDP}}}$ can follow the optimal one, or  our proposed mechanisms. Then the above composition satisfies \textbf{$(\epsilon,\delta)$-pDP} for $\epsilon$ and $\delta$ satisfying $\sigma_{*}\geq {\sigma_{\epsilon,\delta}^{\textup{pDP}}}$ with $\sigma_{*}$ defined above.
\end{itemize}

\end{rem}

% We prove Lemma~\ref{Gaussiancomposition} in Appendix~\ref{appprf-Gaussiancomposition}.

% \section{Discussion} \label{sec-main-discussion}

% \subsection{Application to Local Differential Privacy}

%\section{Applications} \label{sec:Applications}

%Below we discuss   applications of differential privacy with the Gaussian mechanism to deep learning~\cite{abadi2016deep,wu2017bolt} and hypothesis testing~\cite{gaboardi2016differentially}, respectively.

%\section{Proof of Lemma~\protect\x} \label{appprf-Gaussiancomposition}

\noindent\textbf{Proof of Lemma~\ref{Gaussiancomposition}:
}

We consider $m$ queries $Q_1, Q_2, \ldots, Q_m$ with $\ell_2$-sensitivity $\Delta_1, \Delta_2, \ldots, \Delta_m$. The query result of $Q_i$ on dataset $D$ is added with independent Gaussian noise of standard deviation $\sigma_i$, in order to generate a noisy version $\widetilde{Q}_i(D)$.

We first state a result for a general query $Q$. Let the query result of $Q$ on dataset $D$ be added with Gaussian noise of standard deviation $\sigma$, in order to generate a noisy version $\widetilde{Q}(D)$. From Eq.~(\ref{defineStl2})~(\ref{defineGS2}) and~(\ref{DPeq14newqv3}), we obtain:
\begin{align}   \label{DP-eq-Gaussian}
\begin{array}{l}  \textup{with $y$ following  the probability distribution of $\widetilde{Q}(D)$}
\\ \textup{(i.e., a Gaussian distribution with mean $Q(D)$}
\\ \textup{and standard deviation $\sigma$),}
\\ \textup{the term
$\ln \frac{\fr{\widetilde{Q}(D) = y}}{\fr{\widetilde{Q}(D') = y }}$  obeys a Gaussian distribution}
\\ \textup{with mean $\frac{[\|Q(D) - Q(D')\|_{2}]^2}{2{\sigma}^2}$ and  variance $\frac{[\|Q(D) - Q(D')\|_{2}]^2}{{\sigma}^2}$.}
\end{array}
\end{align}

Let   $\widetilde{\boldsymbol{Q}}$ be the composition of mechanisms $\widetilde{Q}_1, \widetilde{Q}_2, \ldots, \widetilde{Q}_m$. Let $y_i$ follow the probability distribution of $\widetilde{Q}_i(D)$, and let $\boldsymbol{y}$ be the composition of $y_1, y_2, \ldots, y_m$, which means that $\boldsymbol{y}$ follow the probability distribution of $\widetilde{\boldsymbol{Q}}(D)$. Following Eq.~(\ref{eqn-L-Y-D-Dprime}),  the privacy loss function of $\boldsymbol{Q}$  on  neighbouring datasets $D$ and $D'$ can be defined as
\begin{align}
L_{\widetilde{\boldsymbol{Q}},D,D'}(\boldsymbol{y}) &=  \ln \frac{ \fr{   \widetilde{\boldsymbol{Q}}(D) = \boldsymbol{y}} }{  \fr{   \widetilde{\boldsymbol{Q}}(D') = \boldsymbol{y}  }}\nonumber \\ &  = \ln \frac{ \fr{\cap_{i=1}^m \left[\widetilde{Q}_i(D) = y_i\right]} }{  \fr{\cap_{i=1}^m \left[\widetilde{Q}_i(D') = y_i \right]}} . \label{privacylosscompose2}
\end{align}
Since $\widetilde{Q}_1, \widetilde{Q}_2, \ldots, \widetilde{Q}_m$ are independent, we further have
\begin{align}
L_{\widetilde{\boldsymbol{Q}},D,D'}(\boldsymbol{y}) = \sum_{i=1}^m \ln \frac{\fr{\widetilde{Q}_i(D) = y_i}}{\fr{\widetilde{Q}_i(D') = y_i }}. \label{privacylosscompose3}
\end{align}
From  (\ref{DP-eq-Gaussian}), $\ln \frac{\fr{\widetilde{Q}_i(D) = y_i}}{\fr{\widetilde{Q}_i(D') = y_i }}$ follows a Gaussian distribution with mean $\frac{[\|Q_i(D) - Q_i(D')\|_{2}]^2}{2{\sigma_i}^2}$ and  variance $\frac{[\|Q_i(D) - Q_i(D')\|_{2}]^2}{{\sigma_i}^2}$. Then from~(\ref{privacylosscompose3}), $L_{\widetilde{\boldsymbol{Q}},D,D'}(\boldsymbol{y})$ follows a Gaussian distribution with mean $\sum_{i=1}^m \frac{[\|Q_i(D) - Q_i(D')\|_{2}]^2}{2{\sigma_i}^2}$ and  variance $\sum_{i=1}^m \frac{[\|Q_i(D) - Q_i(D')\|_{2}]^2}{{\sigma_i}^2}$.

To account for the privacy level of $\widetilde{Q}$,
both $(\epsilon,\delta)$-differential privacy and $(\epsilon,\delta)$-probabilistic differential privacy can be given by conditions on $L_{\widetilde{\boldsymbol{Q}},D,D'}(\boldsymbol{y})$ for any pair of neighboring datasets $D$ and $D'$. In particular, from Theorem 5 of~\cite{balle2018improving}, $\widetilde{\boldsymbol{Q}}$   achieves $(\epsilon, \delta)$-differential privacy if and only if
\begin{align}
 & \left( \begin{array}{l}
    \mathbb{P}_{\boldsymbol{y}\sim\widetilde{\boldsymbol{Q}}(D)}[L_{\widetilde{\boldsymbol{Q}},D,D'}(\boldsymbol{y}) > \epsilon] \\[9pt] - e^{\epsilon}  \mathbb{P}_{\boldsymbol{y}\sim\widetilde{\boldsymbol{Q}}(D)}[L_{\widetilde{\boldsymbol{Q}},D,D'}(\boldsymbol{y}) < - \epsilon]
 \end{array} \right)  \leq \delta , \label{losseqleqdelta} \\ & \textup{for any pair of neighboring datasets $D$ and $D'$}. \nonumber
\end{align}
From Definition~\ref{defn-Prob-DP}, $\widetilde{\boldsymbol{Q}}$   achieves $(\epsilon, \delta)$-probabilistic differential privacy if and only if
\begin{align}
 &\mathbb{P}_{\boldsymbol{y}\sim\widetilde{\boldsymbol{Q}}(D)}[\big|L_{\widetilde{\boldsymbol{Q}},D,D'}(\boldsymbol{y})\big| > \epsilon] \leq \delta ,\label{losseqleqdeltab} \\ & \textup{for any pair of neighboring datasets $D$ and $D'$}.  \nonumber
\end{align}

Our analysis above shows that
$L_{\widetilde{\boldsymbol{Q}},D,D'}(\boldsymbol{y}) $ follows a Gaussian distribution with mean $A(D,D')$ and  variance $\frac{A(D,D')}{2}$ for
$A(D,D'): = \sum_{i=1}^m \frac{[\|Q_i(D) - Q_i(D')\|_{2}]^2}{2{\sigma_i}^2}$. Since $\|Q_i(D) - Q_i(D')\|_{2}$ is at most the $\ell_2$-sensitivity $\Delta_i$ of query $Q_i$, the term $A(D,D')$ is no greater than ${\sum_{i=1}^m \frac{  {\Delta_i}^2 }{{\sigma_i}^2} }$. Lemma 7 of~\cite{balle2018improving} proves that the left hand side of Eq.~(\ref{losseqleqdelta}) strictly increases when $A(D,D')$ increases. Hence, $\widetilde{\boldsymbol{Q}}$   achieves $(\epsilon, \delta)$-differential privacy if for $L^*$ obeying a Gaussian distribution with mean $A^*$ and  variance $\frac{A^*}{2}$ for
$A^*: = {\sum_{i=1}^m \frac{  {\Delta_i}^2 }{{\sigma_i}^2} }$, we have
\begin{align}
 \mathbb{P}[L^* > \epsilon] - e^{\epsilon}  \mathbb{P}[L^* < - \epsilon] \leq \delta . \label{losseqleqdelta2}
\end{align}

From~(\ref{DP-eq-Gaussian}) above and \cite{balle2018improving}'s Theorem~5, Inequality (\ref{losseqleqdelta2}) is also the  condition to ensure that answering a query with $\ell_2$-sensitivity $1$ and Gaussian noise amount $ \frac{1}{\sqrt{A^*}}$ satisfies $(\epsilon,\delta)$-differential privacy.

% For a query with $\ell_2$-sensitivity $1$, we use Theorem~\ref{thm-DP-OPT} to find that the minimal Gaussian noise amount to achieve $(\epsilon,\delta)$-differential privacy is $\sigma_{\texttt{\upshape DP-OPT}}$, which is given by Equations~(\ref{eqn-sigma-DP-OPT}) and~(\ref{eqn-sigma-DP-OPT}).
% In particular, from Theorem 8 of~\cite{balle2018improving}, $g(\epsilon,\delta)$ satisfies \begin{align}
% \Phi\left( \frac{1}{2\cdot g(\epsilon,\delta)} - \epsilon \cdot g(\epsilon,\delta) \right) - e^{\epsilon} \Phi\left( -\frac{1}{2\cdot g(\epsilon,\delta)} - \epsilon \cdot g(\epsilon,\delta) \right) = \delta, \label{gexpr}
% \end{align}
% where $\Phi(\cdot)$ is the cumulative distribution function of the standard univariate Gaussian distribution; i.e., $\Phi(x) = \frac{1}{\sqrt{2\pi}} \int_{-\infty}^{x} e^{-t^2/2} \text{d} t $.
%  Then the necessary and sufficient condition of Eq.~(\ref{losseqleqdelta2}) is $ \frac{1}{\sqrt{A^*}} \geq \sigma_{\epsilon,\delta}$, which implies
% $\sum_{i=1}^m \frac{  {\Delta_i}^2 }{{\sigma_i}^2} \leq \frac{1}{{\sigma_{\epsilon,\delta}}^2}$.

Similarly, $\widetilde{\boldsymbol{Q}}$   achieves $(\epsilon, \delta)$-probabilistic differential privacy if for $L^*$ obeying a Gaussian distribution with mean $A^*$ and  variance $\frac{A^*}{2}$ for
$A^*: = {\sum_{i=1}^m \frac{  {\Delta_i}^2 }{{\sigma_i}^2} }$, we have
\begin{align}
  \mathbb{P}[|L^*| > \epsilon]  \leq \delta . \label{losseqleqdeltapdp}
\end{align}
From~(\ref{DP-eq-Gaussian}) above, Inequality (\ref{losseqleqdeltapdp}) is also the condition to ensure that answering a query with $\ell_2$-sensitivity $1$ and Gaussian noise amount $ \frac{1}{\sqrt{A^*}}$ satisfies $(\epsilon,\delta)$-probabilistic differential privacy.

% Similar to the above analysis, the necessary and sufficient condition of Eq.~(\ref{losseqleqdeltapdp}) is $ \frac{1}{\sqrt{A^*}} \geq \sigma_{\texttt{\upshape pDP-OPT}}$ where $A^*: = {\sum_{i=1}^m \frac{  {\Delta_i}^2 }{{\sigma_i}^2} }$, which implies
% $\sum_{i=1}^m \frac{  {\Delta_i}^2 }{{\sigma_i}^2} \leq \frac{1}{{\sigma_{\texttt{\upshape pDP-OPT}}}^2}$. Hence,

With the above results and $ \frac{1}{\sqrt{A^*}}= \left(\sum_{i=1}^m \frac{  {\Delta_i}^2 }{{\sigma_i}^2}\right)^{-1/2} $,
Lemma~\ref{Gaussiancomposition} is proved.
 \qeda

\subsection{Proof of Lemma~\ref{lem-inverfc}} \label{sec-lem-inverfc}

\noindent\textbf{Lemma~\ref{lem-inverfc} (Restated).} For $0<y<1$, it holds that $\inverfc(y)<\sqrt{\ln \frac{2}{\sqrt{8y+1}-1}} $.

\noindent\textbf{Proof:}
We define a function $g(\cdot)$ as
\begin{align}
g(x) = \frac{1}{2} \exp(-2x^2) + \frac{1}{2} \exp(-x^2).\label{lem2fminusoney}
\end{align}
Then we derive for $0<y < 1$ that
\begin{align}
g^{-1}(y)=\sqrt{\ln \frac{2}{\sqrt{8y+1}\hspace{2pt}-\hspace{2pt}1}}.\label{lem2fminusoney}
\end{align}
We relate Lemma~\ref{lem-inverfc} with the result
\begin{align}
{\erfc}(x) < g(x) , \text{ for }x > 0 . \label{lem2erfcfx}
\end{align}
The rest of the proof includes two parts: i) using (\ref{lem2erfcfx}) to show Lemma~\ref{lem-inverfc}, and ii) proving (\ref{lem2erfcfx}).

\noindent\textbf{Using (\ref{lem2erfcfx}) to show Lemma~\ref{lem-inverfc}:}

We replace $x$ by $g^{-1}(y)$ in Eq.~(\ref{lem2erfcfx}), and thus obtain
\begin{align}
g(g^{-1}(y)) > {\erfc}(g^{-1}(y)). \label{lem2erfcfx2}
\end{align}
The term $g(g^{-1}(y))$ in Eq.~(\ref{lem2erfcfx2}) equals $y$ and can also be written as ${\erfc}(\inverfc(y))$; i.e., we can express Eq.~(\ref{lem2erfcfx2}) as follows:
\begin{align}
{\erfc}(\inverfc(y)) > {\erfc}(g^{-1}(y)). \label{lem2erfcfx3}
\end{align}
As ${\erfc}()$ is a decreasing function, Eq.~(\ref{lem2erfcfx3}) implies
\begin{align}
\inverfc(y) < g^{-1}(y).\label{lem2erfcfx4}
\end{align}
% For $0< \delta <1$, we can set $y$ as $\delta$ in Eq.~(\ref{lem2erfcfx4}). Then we have
% \begin{align}
% \inverfc(\delta) < g^{-1}(\delta).\label{lem2erfcfx5}
% \end{align}
From Eq.~(\ref{lem2fminusoney}), we know that $g^{-1}(y)$ in Eq.~(\ref{lem2erfcfx4})  equals $\sqrt{\ln \frac{2}{\sqrt{8y+1}\hspace{2pt}-\hspace{2pt}1}}$. Hence, Eq.~(\ref{lem2erfcfx4})   above means Lemma~\ref{lem-inverfc}.

\noindent\textbf{Proving (\ref{lem2erfcfx}):}

The complementary error function
${\erfc}(x)$ equals $\frac{2}{\sqrt{\pi}}\int_{x}^{\infty}e^{-{t}^2} \hspace{1pt} \text{d} t $. We will prove another form of the complementary error function for $x \geq 0$. Specifically, we will show
\begin{align}
{\erfc}(x) = \frac{2}{\pi}\int_{0}^{\frac{\pi}{2}}\exp\bigg(-\frac{x^2}{\sin^2 \theta}\bigg) \hspace{1pt}  \text{d} \theta, \text{ for }x \geq 0 . \label{eqCraigerf}
\end{align}
The right hand side of Eq.~(\ref{eqCraigerf})
is an alternative form of the complementary error function, and is known as Craig's formula~\cite{craig1991new} in the literature. Yet, to show Eq.~(\ref{eqCraigerf}), Craig~\shortcite{craig1991new} uses empirical arguments  and not many studies present a rigorous proof. Below we formally establish Eq.~(\ref{eqCraigerf}) for completeness.

Given $\frac{2}{\sqrt{\pi}}\int_{0}^{\infty}e^{-{s}^2} \hspace{1pt} \text{d} s = {\erfc}(0) = 1$, we now write ${\erfc}(x)$ (i.e., $\frac{2}{\sqrt{\pi}}\int_{x}^{\infty}e^{-{t}^2} \hspace{1pt} \text{d} t $) as follows:
\begin{align}
 & \frac{2}{\sqrt{\pi}}\int_{x}^{\infty}e^{-{t}^2} \hspace{1pt} \text{d} t\nonumber \\ &  = \frac{2}{\sqrt{\pi}}\int_{0}^{\infty}e^{-{s}^2} \hspace{1pt} \text{d} s \cdot \frac{2}{\sqrt{\pi}}\int_{x}^{\infty}e^{-{t}^2} \hspace{1pt} \text{d} t  \nonumber \\ &  =     \frac{4}{\pi}\int_{0}^{\infty}e^{-{s}^2} \hspace{1pt} \text{d} s \int_{x}^{\infty}e^{-{t}^2} \hspace{1pt} \text{d} t .  \label{eqCraig2}
\end{align}
We express the integral of Eq.~(\ref{eqCraig2}) in polar coordinates. Specifically, under $s   = r \cos \theta$ and $t = r \sin \theta$, the intervals $s \in [0, \infty)$ and $t \in [x, \infty)$ correspond to $r \in [x/\sin \theta, \infty)$ and $ \theta  \in [0, \frac{\pi}{2}]$. Also, it holds that~\mbox{$ \text{d} s\hspace{1pt}\text{d} t = r\text{d} r\hspace{1pt}\text{d} \theta$}. Then the right hand side  (RHS) of~Eq.~(\ref{eqCraig2}) is given by
\begin{align}
\text{RHS of Eq.~(\ref{eqCraig2})} &  = \frac{4}{\pi}\int_{0}^{\frac{\pi}{2}}  \hspace{1pt}\text{d} \theta \int_{x/\sin \theta}^{\infty} re^{-{r}^2} \hspace{1pt} \text{d} r \nonumber \\ & = \frac{4}{\pi}\int_{0}^{\frac{\pi}{2}}  \hspace{1pt}\text{d} \theta \left(-\frac{1}{2} e^{-y^2}\right) \bigg|_{y=x/\sin \theta}^{y=\infty}   \nonumber \\ & =  \frac{2}{\pi}\int_{0}^{\frac{\pi}{2}}\exp\bigg(-\frac{x^2}{\sin^2 \theta}\bigg) \hspace{1pt} \text{d} \theta . \label{eqCraig4}
\end{align}
Summarizing Eq.~(\ref{eqCraig2}) and Eq.~(\ref{eqCraig4}), we have proved Eq.~(\ref{eqCraigerf}).
To further bound ${\erfc}(x)$ based on Eq.~(\ref{eqCraigerf}), we obtain for $x>0$ that
\begin{align}
&\frac{2}{\pi}\int_{0}^{\frac{\pi}{2}}\exp\bigg(-\frac{x^2}{\sin^2 \theta}\bigg) \hspace{1pt} \text{d} \theta  \nonumber \\ &<  \frac{2}{\pi}\int_{0}^{\frac{\pi}{4}}\exp\bigg(-\frac{x^2}{\sin^2 \frac{\pi}{4}}\bigg) \hspace{1pt} \text{d} \theta + \frac{2}{\pi}\int_{\frac{\pi}{4}}^{\frac{\pi}{2}}\exp(-x^2) \hspace{1pt} \text{d} \theta  \nonumber \\ & =  \frac{2}{\pi} \cdot \frac{\pi}{4}  \cdot  \exp(-2x^2) + \frac{2}{\pi} \cdot \frac{\pi}{4}  \cdot  \exp(-x^2) \nonumber \\ & = g(x),\nonumber
%\label{lem2eqCraig5}
\end{align}
which along with Eq.~(\ref{eqCraigerf}) gives (\ref{lem2erfcfx}).

Since we have shown (\ref{lem2erfcfx}) and the result that (\ref{lem2erfcfx}) implies Lemma~\ref{lem-inverfc}, we complete proving Lemma~\ref{lem-inverfc}.
\qeda

\end{document}